\documentclass[twocolumn, twocolappendix]{aastex701}
\usepackage{natbib}
\usepackage{graphicx}	
\usepackage{color}
\usepackage{amsmath}	
\usepackage{amssymb}	
\usepackage{mathrsfs}
\usepackage{mathrsfs}
\usepackage{placeins}

\newcommand{\mBH}{M_{\mathrm{BH}}}
\newcommand{\msun}{M_\odot}

\newcommand{\edd}{L_{\rm bol}/L_{\rm Edd}}

\newcommand{\feii}{\ion{Fe}{2}}

\newcommand{\angstrom}{\textup{\angstrom}}

\defcitealias{Onoue19}{O19}

\graphicspath{{./}}

\shorttitle{BH Mass Measurements of SHELLQs Quasars at $6<z<7$}
\shortauthors{Onoue et al.}

\begin{document}

\title{SHELLQs. Black Hole Mass and Eddington Ratio Distributions of Intermediate-Luminosity Quasars  at $\mathbf{6< z< 7}$ 
\thanks{This research is based in part on data collected at the Subaru Telescope, which is operated by the National Astronomical Observatory of Japan. We are honored and grateful for the opportunity of observing the Universe from Maunakea, which has the cultural, historical, and natural significance in Hawaii.}} 

\correspondingauthor{Masafusa Onoue}
\email{masafusa.onoue@aoni.waseda.jp}

\author[0000-0003-2984-6803]{Masafusa Onoue}
\affiliation{Waseda Institute for Advanced Study (WIAS), Waseda University, 1-21-1, Nishi-Waseda, Shinjuku, Tokyo 169-0051, Japan}
\affiliation{Kavli Institute for the Physics and Mathematics of the Universe (Kavli IPMU, WPI), UTIAS, Tokyo Institutes for Advanced Study, University of Tokyo, Chiba, 277-8583, Japan}
\email{masafusa.onoue@aoni.waseda.jp}

\author[0000-0002-0000-6977]{John D. Silverman}
\affiliation{Kavli Institute for the Physics and Mathematics of the Universe (Kavli IPMU, WPI), UTIAS, Tokyo Institutes for Advanced Study, University of Tokyo, Chiba, 277-8583, Japan}
\affiliation{Department of Astronomy, School of Science, The University of Tokyo, 7-3-1 Hongo, Bunkyo-ku, Tokyo 113-0033, Japan}
\affiliation{Center for Data-Driven Discovery, Kavli IPMU (WPI), UTIAS, The University of Tokyo, Kashiwa, Chiba 277-8583, Japan}
\affiliation{Center for Astrophysical Sciences, Department of Physics \& Astronomy, Johns Hopkins University, Baltimore, MD 21218, USA}
\email{silverman@ipmu.jp}

\author[0000-0001-5063-0340]{Yoshiki Matsuoka}
\affiliation{Research Center for Space and Cosmic Evolution, Ehime University, Matsuyama, Ehime 790-8577, Japan}
\email{matsuoka.yoshiki.ld@ehime-u.ac.jp}

\author[0000-0001-8917-2148]{Xuheng Ding}
\affiliation{School of Physics and Technology, Wuhan University, Wuhan 430072, China}
\email{dingxh@whu.edu.cn}

\author[0000-0002-2099-0254]{Camryn L. Phillips}
\affiliation{Department of Astrophysical Sciences, Princeton University, 4 Ivy Lane, Princeton, NJ 08544, USA}
\email{cp5619@princeton.edu}

\author[0000-0002-0106-7755]{Michael A. Strauss}
\affiliation{Department of Astrophysical Sciences, Princeton University, 4 Ivy Lane, Princeton, NJ 08544, USA}
\email{strauss@astro.princeton.edu}

\author[0009-0007-0864-7094]{Junya Arita}
\affiliation{Department of Astronomy, School of Science, The University of Tokyo, 7-3-1 Hongo, Bunkyo-ku, Tokyo 113-0033, Japan}
\email{jarita@astron.s.u-tokyo.ac.jp}

\author[0000-0001-9452-0813]{Takuma Izumi}
\affiliation{National Astronomical Observatory of Japan, National Institutes of Natural Sciences, 2-21-1 Osawa, Mitaka, Tokyo 181-8588, Japan}
\affiliation{Department of Astronomy, The University of Tokyo, 7-3-1 Hongo, Bunkyo, Tokyo 113-0033, Japan}
\affiliation{Department of Astronomy, School of Science, Graduate University for Advanced Studies (SOKENDAI), Mitaka, Tokyo 181-8588, Japan}
\affiliation{Amanogawa Galaxy Astronomy Research Center, Kagoshima University, 1-21-35 Korimoto, Kagoshima 890-0065, Japan}
\email{takuma.izumi@nao.ac.jp}

\author[0009-0003-5438-8303]{Mahoshi Sawamura}
\affiliation{Department of Astronomy, The University of Tokyo, 7-3-1 Hongo, Bunkyo, Tokyo 113-0033, Japan}
\affiliation{National Astronomical Observatory of Japan, National Institutes of Natural Sciences, 2-21-1 Osawa, Mitaka, Tokyo 181-8588, Japan}
\email{mahoshi.sawamura@grad.nao.ac.jp}

\author[0000-0003-3954-4219]{Nobunari Kashikawa}
\affiliation{Department of Astronomy, School of Science, The University of Tokyo, 7-3-1 Hongo, Bunkyo-ku, Tokyo 113-0033, Japan}
\affiliation{Research Center for the Early Universe, Graduate School of Science, The University of Tokyo, 7-3-1 Hongo, Bunkyo-ku, Tokyo 113-0033, Japan}
\email{n.kashikawa@astron.s.u-tokyo.ac.jp}

\author[0000-0001-6102-9526]{Irham Andika}
\affiliation{Universit\"ats-Sternwarte M\"unchen, Fakult\"at f\"ur Physik, Ludwig-Maximilians-Universit\"at M\"unchen, Scheinerstr.~1, 81679 M\"unchen, Germany}
\email{irham.andika@lmu.de}

\author[0000-0003-4569-1098]{Kentaro Aoki}
\affiliation{Subaru Telescope, National Astronomical Observatory of Japan, 650 North A'ohoku Place, Hilo HI 96720 U.S.A.}
\email{kaoki@naoj.org}

\author[0000-0002-9850-6290]{Shunsuke Baba}
\affiliation{Institute of Space and Astronautical Science (ISAS), Japan Aerospace Exploration Agency (JAXA), 3-1-1 Yoshinodai, Chuo-ku, Sagamihara, Kanagawa 252-5210, Japan}
\email{sshunsuke.baba.astro@gmail.com}

\author[0000-0003-2895-6218]{Anna-Christina Eilers}
\affiliation{MIT Kavli Institute for Astrophysics and Space Research, Massachusetts Institute of Technology, Cambridge, MA 02139, USA}
\email{eilers@mit.edu}

\author[0000-0001-7201-5066]{Seiji Fujimoto}
\affiliation{David A. Dunlap Department of Astronomy and Astrophysics, University of Toronto, 50 St. George Street, Toronto, Ontario, M5S 3H4, Canada}
\affiliation{Dunlap Institute for Astronomy and Astrophysics, 50 St. George Street, Toronto, Ontario, M5S 3H4, Canada}
\email{seiji.fujimoto@utoronto.ca}

\author[0000-0002-6821-8669]{Tomotsugu Goto}
\affiliation{Department of Physics, National Tsing Hua University, 101, Section 2. Kuang-Fu Road, Hsinchu, 30013, Taiwan}
\affiliation{Institute of Astronomy, National Tsing Hua University, 101, Section 2. Kuang-Fu Road, Hsinchu, 30013, Taiwan}
\email{tomo@phys.nthu.edu.tw}

\author[0000-0001-6186-8792]{Masatoshi Imanishi}
\affiliation{National Astronomical Observatory of Japan, National Institutes of Natural Sciences, 2-21-1 Osawa, Mitaka, Tokyo 181-8588, Japan}
\affiliation{Department of Astronomy, School of Science, Graduate University for Advanced Studies (SOKENDAI), Mitaka, Tokyo 181-8588, Japan}
\email{masa.imanishi@nao.ac.jp}

\author[0000-0001-9840-4959]{Kohei Inayoshi}
\affiliation{Kavli Institute for Astronomy and Astrophysics, Peking University, Beijing 100871, People's Republic of China}
\email{inayoshi@pku.edu.cn}

\author[0000-0002-4923-3281]{Kazushi Iwasawa}
\affiliation{Institut de Ci\`encies del Cosmos (ICCUB), Universitat de Barcelona (IEEC-UB), Mart\'i i Franqu\`es, 1, 08028 Barcelona, Spain}
\affiliation{ICREA, Pg Llu\'is Companys 23, 08010 Barcelona, Spain}
\email{kazushi.iwasawa@icc.ub.edu}

\author[0000-0003-3804-2137]{Knud Jahnke}
\affiliation{Max Planck Institute for Astronomy, K\"onigstuhl 17, 69117 Heidelberg, Germany}
\email{jahnke@mpia.de}

\author[0009-0001-5295-5322]{Yuki Kaneko}
\affiliation{Graduate School of Advanced Science and Engineering, Waseda University,  Nishi-Waseda Campus, Waseda University, 3-4-1 Okubo, Shinjuku-ku, Tokyo 1699-8555 Japan}
\email{knkykwsd34@fuji.waseda.jp}

\author[0000-0002-3866-9645]{Toshihiro Kawaguchi}
\affiliation{Graduate School of Science and Engineering, University of Toyama, Gofuku 3190, Toyama 930-8555, Japan}
\email{kawaguti@eng.u-toyama.ac.jp}

\author[0000-0002-4052-2394]{Kotaro Kohno}
\affiliation{Institute of Astronomy, Graduate School of Science, The University of Tokyo, 2-21-1 Osawa, Mitaka, Tokyo 181-0015, Japan}
\affiliation{Research Center for the Early Universe, Graduate School of Science, The University of Tokyo, 7-3-1 Hongo, Bunkyo-ku, Tokyo 113-0033, Japan}
\email{kkohno@ioa.s.u-tokyo.ac.jp}

\author[0000-0003-1700-5740]{Chien-Hsiu Lee}
\affiliation{Hobby-Eberly Telescope, McDonald Observatory, The University of Texas at Austin, Fort Davis, TX 79734, USA}
\email{lchjoel1031@gmail.com}

\author[0000-0001-6106-7821]{Alessandro Lupi}
\affiliation{Como Lake Center for Astrophysics, DiSAT, Universit\`a degli Studi dell'Insubria, via Valleggio 11, I-22100 Como, Italy}
\affiliation{INFN, Sezione di Milano-Bicocca, Piazza della Scienza 3, 20126 Milano, Italy}
\email{alessandro.lupi@uninsubria.it}

\author[0000-0002-7402-5441]{Tohru Nagao}
\affiliation{Research Center for Space and Cosmic Evolution, Ehime University, 2-5 Bunkyo-cho, Matsuyama, Ehime 790-8577, Japan}
\affiliation{Amanogawa Galaxy Astronomy Research Center, Kagoshima University, 1-21-35 Korimoto, Kagoshima 890-0065, Japan}
\email{tohru@cosmos.phys.sci.ehime-u.ac.jp}

\author[0000-0002-3848-1757]{Dragan Salak}
\affiliation{Astronomical Institute, Tohoku University, 6-3 Aramaki, Aoba-ku, Sendai, Miyagi 980-8578, Japan}
\email{dragan@astr.tohoku.ac.jp}

\author[0000-0001-7825-0075]{Malte Schramm}
\affiliation{Universit\"at Potsdam, Karl-Liebknecht-Stra{\ss}e 24/25, D-14476 Potsdam, Germany}
\email{maschramm23@gmail.com}

\author[0000-0002-3531-7863]{Yoshiki Toba}
\affiliation{Department of Physical Sciences, Ritsumeikan University, 1-1-1 Noji-higashi, Kusatsu, Shiga 525-8577, Japan}
\affiliation{Academia Sinica Institute of Astronomy and Astrophysics, 11F of Astronomy-Mathematics Building, AS/NTU, No. 1, Section 4, Roosevelt Road, Taipei 10617, Taiwan}
\affiliation{Research Center for Space and Cosmic Evolution, Ehime University, Matsuyama, Ehime 790-8577, Japan}
\email{toba@fc.ritsumei.ac.jp}

\author[0000-0003-1937-0573]{Hideki Umehata}
\affiliation{Institute for Advanced Research, Nagoya University, Furocho, Chikusa, Nagoya 464-8602, Japan}
\email{umehata@a.phys.nagoya-u.ac.jp}

\author[0000-0002-3216-1322]{Marta Volonteri}
\affiliation{Institut d'Astrophysique de Paris, UMR 7095, CNRS and Sorbonne Universit\'e, 98 bis Boulevard Arago, 75014 Paris, France}
\email{martav@iap.fr}

\author[0000-0003-4793-7880]{Fabian Walter}
\affiliation{Max Planck Institute for Astronomy, K\"onigstuhl 17, 69117 Heidelberg, Germany}
\email{walter@mpia.de}

\author[0000-0002-7633-431X]{Feige Wang}
\affiliation{Department of Astronomy, University of Michigan, 1085 S. University Ave., Ann Arbor, MI 48109, USA}
\email{fgwang@umich.edu}

\author[0000-0001-5287-4242]{Jinyi Yang}
\affiliation{Department of Astronomy, University of Michigan, 1085 S. University Ave., Ann Arbor, MI 48109, USA}
\email{jyyangas@umich.edu}

\begin{abstract}
We present near-infrared spectroscopy of 21 quasars at $6.07\leq z \leq 6.90$ obtained with  JWST/NIRSpec and Subaru/MOIRCS.
These targets have absolute ultraviolet magnitudes of $-25 < M_{1450} < -22$  and are drawn from a sample of $z>6$ quasars identified in the \textit{Subaru High-z Exploration of Low-Luminosity Quasars} (SHELLQs) project.
These quasars occupy the intermediate-luminosity regime between luminous quasars and the faint high-redshift AGNs  uncovered by JWST.
We detect broad Balmer emission lines  and \ion{Mg}{2} $\lambda 2798$   with underlying continua from the NIRSpec and MOIRCS targets, respectively.
The virial black hole masses of this sample span a wide range of $7.2 < \log{(\mBH/\msun)} < 9.4$, with Eddington ratios of $-1.3 < \log{(\edd)} < 0.4$.
Combining these measurements with our previous mass estimates for other SHELLQs quasars,
we construct a sample of 27 quasars with $\mBH$ estimates at $6<z<7$ and derive the distributions of $\mBH$ and $\edd$.
When compared with luminosity-matched quasars at $z\approx1.3$, we find  median offsets of $\Delta \log{(\mBH/M_\odot)} = -0.4$  and  $\Delta \log{(\edd)} = 0.2$ in the high-redshift sample.
In addition, 11\% of the high-redshift quasars are accreting near or above the Eddington limit.
As a systematic check, adopting the Eddington-ratio-dependent calibration of single-epoch $\mBH$ estimates increases the inferred super-Eddington fraction to 22\%.
These results indicate that a representative population of distant supermassive black holes  undergo a rapid accretion phase, although not as extreme as that seen in the most luminous quasar population, consistent with the anti-hierarchical growth scenario of supermassive black holes.

\end{abstract}


\section{Introduction} \label{sec:Sec1}

High-redshift ($z>6$) quasars hold the key to unveiling the early growth pathways of supermassive black holes (SMBHs).
Extensive efforts have been made to search for these earliest-formed SMBHs based on $\gtrsim 1,000$ deg$^2$-scale ground-based wide-field surveys \citep[][for a recent review]{Fan23ARAA}, yielding hundreds of luminous (bolometric luminosity $\log L_\mathrm{bol}$ [erg\ s$^{-1}$]  $> 47$) and massive ($\log M_\mathrm{BH}$ [$M_\odot$] $ = 8$ -- $9$) SMBHs at $6 \leq z \leq 7.8$ \citep{Banados18, Yang20, Wang21, Yang26}.
New-generation infrared surveys are expected to push the redshift frontier even farther.
Euclid has begun its full extragalactic sky survey to search for luminous quasars  to $z\sim9$ \citep{Barnett19, Banados25, Yang26}.
The Nancy Grace Roman Space Telescope will  carry out a deeper infrared survey than Euclid, and will be able to identify relatively faint $z>7$ quasars over $\sim2,000$ deg$^2$ \citep{LiW23, Tee23}.

Ultradeep infrared observations by the James Webb Space Telescope \citep[JWST;][]{Rigby23} have proven effective in identifying low-luminosity active galactic nuclei (AGN), despite the limited field of view of its instruments ($\ll 1$ deg$^2$).
This new population of AGN, with luminosities similar to local Seyfert galaxies at $\log L_\mathrm{bol}~\mathrm{[erg\ s^{-1}]} = 44$ -- $45$, appears surprisingly abundant, suggesting that up to a few tens of percent of galaxies host accreting SMBHs at $z>4$ \citep[e.g.,][]{Onoue23,Kocevski23,Harikane23,Matthee24,Greene24,Juodzbalis26, Taylor25, Umeda25}.
Remarkably, many of these AGN show red optical colors \citep[e.g.,][]{Matthee24}, hinting at a scenario in which SMBHs are often surrounded by dense gas \citep{Inayoshi_Maiolino25, DeGraaff25b} or dust \citep{Vito18, Ni20, Matsuoka25} at high redshift.

A relatively unexplored parameter space lies between these two populations.
This intermediate-luminosity regime ($\log L_\mathrm{bol}~\mathrm{[erg\ s^{-1}]}  = 46$ -- $47$) holds the key to bridging the well-studied UV-selected luminous quasars and the optically-selected lower-luminosity AGN recently identified by JWST.
The 8.2-meter Subaru Telescope has made significant contributions through its wide-field imaging observations with Suprime-Cam \citep{Kashikawa15, Onoue17} and Hyper Suprime-Cam (HSC; \citealt{Miyazaki18, HSCSSP}).
The latter has led to a series of publications from \textit{the Subaru High-z Exploration of Low-Luminosity Quasars} (SHELLQs) project \citep{Matsuoka16, Matsuoka18a, Matsuoka18b, Matsuoka19, Matsuoka19b, Matsuoka22, Matsuoka24, Matsuoka25b}.
These observations are made possible by Subaru's unique capability of wide-field imaging at prime focus.
A ten-year survey with Vera C. Rubin Observatory \citep{LSST_2019} will soon explore a similar parameter space in the southern hemisphere \citep{Tee23}, albeit over a much larger area.

Spectroscopic follow-up observations are essential to constrain the properties of SMBHs in quasars, as quasar luminosity is determined by the product of SMBH mass and accretion rate.
The latter is typically expressed in terms of the Eddington ratio  (i.e., $L_\mathrm{bol}/L_\mathrm{Edd}$).
Beyond distances where direct SMBH mass estimates via stellar kinematics are possible \citep{Ghez08, Genzel10}, single-epoch spectroscopy of broad emission lines from the quasar broad-line region (BLR) is often used to derive virial SMBH masses, while infrared interferometry techniques have recently started to probe BLR kinematics up to $z\sim4$ \citep{Gravity26}.
The single-epoch method assumes that the BLR gas is gravitationally bound to the central SMBH, such that the broad-line width traces the virial velocity of the BLR gas.
The BLR radius ($R_\mathrm{BLR}$) is inferred from the $R_\mathrm{BLR}$--continuum luminosity relation calibrated against reverberation mapping \citep[e.g.,][]{Kaspi00, Vestergaard02, Bentz13}.

SMBH mass estimates have been obtained for dozens of luminous $z>6$ quasars, revealing systematically high Eddington ratios, near or above the Eddington limit (e.g., \citealt{Jiang07, Willott10a, Trakhtenbrot17, Yang21, Yang23, Farina22, Mazzucchelli23}; but see \citealt{Shen19}). 
The faint JWST AGNs are more commonly found in the sub-Eddington regime \citep[e.g.,][]{Matthee24, Juodzbalis26}, although the bolometric SEDs of this new population has still been elusive \citep{Greene26}.
Such observations, however, remain limited in the intermediate-luminosity range \citep{Willott10a, Kim18, Onoue19}.
This is partly due to the high atmospheric infrared sky background in ground-based observations, which effectively limits $\mBH$ estimates to quasars brighter than $\approx23$ mag in $y$-band.
Among these early studies,  \citet[][hereafter O19]{Onoue19} observed six of the first SHELLQs quasars at near-infrared wavelengths to estimate their SMBH masses from broad \ion{Mg}{2} $\lambda2798$ emission lines.
They showed that  the SHELLQs quasars generally accrete at  sub-Eddington rates, in contrast to the results of luminous quasars.

\citet{Takahashi24} matched the Ly$\alpha$ profiles of the SHELLQs objects in their discovery spectra with lower-redshift SDSS quasars \citep{SDSS_DR14Q}, predicting the SMBH mass distribution of the full SHELLQs quasar sample.
They found that the masses range from $10^7$ to $10^{10}\ \mBH$.
However, direct measurement of broad lines in the rest-frame near-UV and optical is preferable in order to better estimate their SMBH masses.
The unprecedented sensitivity of JWST now enables us to directly probe the SMBH masses of the faintest SHELLQs quasars beyond the ground-based limit ($y>23$ mag), as has been demonstrated in earlier JWST studies of the SHELLQs quasars  \citep{Ding23, Onoue25, Lyu25}.

In this paper, we present infrared spectroscopy (i.e., rest-frame near-ultraviolet or optical)
of an additional 21 SHELLQs quasars at $6.09\leq z \leq 6.90$ using both JWST and Subaru.
Spectroscopic redshifts and SMBH masses of some of these quasars have been published in the literature \citep{Ishimoto20, Wolf23, Ding23, Onoue25, Phillips25}.
Specifically, \citet{Silverman25} use the SMBH masses of the JWST targets in this paper to constrain the intrinsic distribution of $z=6$ quasars in the host stellar mass -- SMBH mass plane.

This paper is organized as follows: 
We first present our JWST/NIRSpec observations and spectral analyses in Section~\ref{sec:Sec2_NIRSpec}, and then
present our Subaru/MOIRCS observations and their spectral analyses in Section~\ref{sec:Sec2_MCS}.
BH mass measurements of these two samples are presented in Section~\ref{sec:BHmass}.
Section~\ref{sec:summary} gives the summary and our future prospects.

The magnitudes quoted in this paper are in the AB system \citep{Oke83}.
We correct for instrumental broadening when reporting the emission line widths.
We adopt a standard $\Lambda$CDM cosmology with $H_0=70$ km s$^{-1}$ Mpc$^{-1}$, $\Omega_m=0.3$, and $\Omega_\Lambda=0.7$.
Throughout this paper, $L_{3000}$ and $L_{5100}$ denote monochromatic luminosity $\lambda L_\lambda$ evaluated at rest-frame 3000~\AA\ and 5100~\AA, respectively, with $L_{\lambda_{\rm rest}} = 4\pi d_L^2 (1+z) f_\lambda[(1+z)\lambda_{\rm rest}]$.

\begin{deluxetable*}{lLCCClCCCCl}[p!]
\tablecaption{Targets \& Observations \label{tab:targets_all}}
\tablecolumns{11}
\tablewidth{0pt}
\tablehead{
\colhead{ID} &
\colhead{$z_\mathrm{Ly\alpha}$} &
\colhead{$y_\mathrm{HSC}$} &
\colhead{F150W/$K_s$} &
\colhead{$M_{1450}$} &
\colhead{Date} &
\colhead{Exp.~time} &
\colhead{$\alpha_{\rm \lambda, UV}$} &
\colhead{SNR} &
\colhead{Ref.} &
\colhead{Notes} \\
\colhead{} &
\colhead{} &
\colhead{[mag]} &
\colhead{[mag]} &
\colhead{[mag]} &
\colhead{} &
\colhead{[hour]} &
\colhead{} &
\colhead{} &
\colhead{} &
\colhead{}
}
\startdata
\multicolumn{11}{c}{\it JWST/NIRSpec} \\
\hline
J223644.58$+$003256.9 & 6.4  & 23.18\pm0.05 & 22.612\pm0.003 & -23.8 & 2022 Oct. 30 & 0.6 & -0.78 \pm 0.10 & 46 & 1 & $a$ \\
J225538.04$+$025126.6 & 6.34 & 22.94\pm0.08 & 22.837\pm0.003 & -23.9 & 2022 Oct. 28 & 0.6 & -1.78 \pm 0.17 & 32 & 2 &  \\
J114658.89$-$000537.7 & 6.30 & 24.77\pm0.29 & 24.775\pm0.019 & -21.5 & 2023 Jun. 22 & 2.4 & -2.01 \pm 0.63 & 42 & 2 & $b$ \\
J152555.79$+$430324.0 & 6.27 & 23.50\pm0.09 & 23.518\pm0.006 & -23.6 & 2023 Mar. 18 & 0.9 & -2.05 \pm 0.20 & 27 & 3 &  \\
J114648.42$+$012420.1 & 6.27 & 23.12\pm0.06 & 22.870\pm0.003 & -23.7 & 2023 Jun. 22 & 0.9 & -1.47 \pm 0.13 & 39 & 2 &  \\
J084431.60$-$005254.6 & 6.25 & 23.13\pm0.08 & 22.553\pm0.002 & -23.7 & 2022 Nov. 16 & 0.6 & -0.76 \pm 0.16 & 34 & 4 &  \\
J021721.59$-$020852.6 & 6.20 & 23.57\pm0.08 & 24.755\pm0.007 & -23.2 & 2023 Jul. 22 & 0.9 & -2.39 \pm 0.17 & 13 & 4 & $b$ \\
J091833.17$+$013923.4 & 6.19 & 23.19\pm0.04 & 23.982\pm0.004 & -23.7 & 2022 Dec. 25 & 0.6 & -1.57 \pm 0.09 & 29 & 2 &  \\
J084408.61$-$013216.5 & 6.18 & 23.68\pm0.15 & 23.976\pm0.009 & -23.7 & 2022 Nov. 28 & 0.9 & -2.64 \pm 0.32 & 39 & 2 & $b$ \\
J142517.72$-$001540.8 & 6.18 & 23.36\pm0.06 & 23.076\pm0.004 & -23.4 & 2023 Jan. 23 & 0.6 & -1.38 \pm 0.12 & 30 & 4 &  \\
J151248.71$+$442217.5 & 6.18 & 23.27\pm0.08 & 23.075\pm0.004 & -23.1 & 2023 Feb. 14 & 2.2 & -1.59 \pm 0.18 & 75 & 3 & $a,c$ \\
J091114.27$+$015219.4 & 6.07 & 24.33\pm0.13 & 23.962\pm0.009 & -22.1 & 2023 Nov. 28 & 2.4 & -1.20 \pm 0.28 & 23 & 4 &  \\
\hline
\multicolumn{11}{c}{\it Subaru/MOIRCS} \\
\hline
J221027.24$+$030428.5 & 6.90 & 22.91\pm0.06 & 21.76\pm0.08 & -24.4 & 2018 Jul. 7, 8 & 3.1 & -0.64 \pm 0.12 & 9.0 & 2 &  \\
J092120.56$+$000722.9 & 6.56 & 21.24\pm0.01 & 20.40\pm0.05 & -24.8 & 2019 Apr. 22   & 1.2 & -1.05 \pm 0.06 & 10.0 & 2 &  \\
J154505.62$+$423211.6 & 6.50 & 22.29\pm0.04 & 21.35\pm0.10 & -24.2 & 2018 Jul. 8, 9 & 4.5 & -1.07 \pm 0.12 & 8.3 & 2 &  \\
 &  &  &  &  & 2019 Apr. 22,23 &  &  &  &  &  \\
J135012.04$-$002705.2 & 6.49 & 22.89\pm0.05 & 22.48\pm0.14 & -24.4 & 2019 Apr. 23 & 3.7 & -1.53 \pm 0.17 & 4.0 & 3 & $d$ \\
J100401.37$+$023930.9 & 6.41 & 22.60\pm0.03 & 21.65\pm0.09 & -24.4 & 2021 Jan. 31 & 2.8 & -0.91 \pm 0.11 & 6.4 & 2 &  \\
J113753.64$+$004509.7 & 6.40 & 22.37\pm0.03 & 21.28\pm0.08 & -24.2 & 2019 Apr. 21, 23 & 3.7 & -0.77 \pm 0.10 & 7.3 & 3 &  \\
J230422.97$+$004505.4 & 6.36 & 22.73\pm0.05 & 21.79\pm0.14 & -24.3 & 2020 Aug. 11 & 2.3 & -0.87 \pm 0.18 & 5.3 & 2 &  \\
J140629.13$-$011611.1 & 6.29 & 21.90\pm0.03 & 21.69\pm0.09 & -25.0 & 2019 Apr. 22 & 1.9 & -1.71 \pm 0.12 & 4.5 & 2 &  \\
J121721.35$+$013142.5 & 6.20 & 20.94\pm0.01 & 20.54\pm0.06 & -25.4 & 2019 Apr. 22 & 0.9 & -1.54 \pm 0.07 & 6.4 & 5 &  \\
\enddata
\tablecomments{
The targets are grouped by instrument (JWST/NIRSpec at the top and  Subaru/MOIRCS at the bottom), and are ordered by decreasing Ly$\alpha$ redshift within each group.
The target IDs are based on coordinates, namely Jhhmmss.ss$+/-$ddmmss.s for a quasar at
(R.A., Decl.) = (hh:mm:ss.ss, $\pm$dd:mm:ss.s).
We use shorter IDs (Jhhmm$\pm$ddmm) elsewhere in the paper.
The $y$-band magnitudes, corrected for Galactic extinction \citep{Schlegel98}, are the PSF magnitudes from the Public Data Release 3 of the HSC-SSP catalog \citep{SSP_PDR3}.
The NIR photometry is reported, namely the F150W magnitudes for the JWST targets \citep{Ding25}, and $Ks$-band magnitudes for the MOIRCS targets.
The UV continuum slope $\alpha_{\rm \lambda, UV}$
($F_\lambda \propto \lambda^{\alpha_\lambda}$) is derived from the $y$ and the NIR band photometry.
The signal to noise ratio (SNR) column lists the median continuum SNR calculated over 300 km~s$^{-1}$, which approximately corresponds to the resolution element of the NIRSpec G395M grating.
Wavelengths around rest-frame  $5100$\,\AA\ is used for the NIRSpec targets, while the whole spectrum range is used for the MOIRCS targets.
The discovery references are as follows:
(1) \citet{Matsuoka16}, (2) \citet{Matsuoka18b}, (3) \citet{Matsuoka19b},  (4) \citet{Matsuoka18a}, and  (5) \citet{Banados16} and \citet{Wang17}, .
Specific spectroscopic properties of the  targets are indicated in the Notes column:
($a$) Balmer absorption lines; ($b$) strong narrow emission lines; ($c$) double-peaked H$\alpha$; ($d$) a candidate \ion{Mg}{2} BAL. }
\end{deluxetable*}

\section{JWST/NIRSpec}\label{sec:Sec2_NIRSpec}

We first present the rest-frame optical spectroscopy of our JWST targets.

\subsection{Target Selection}\label{sec:Sec2_NIRSpec_sample}

\begin{figure}
\centering
{\includegraphics[width=0.95\linewidth]{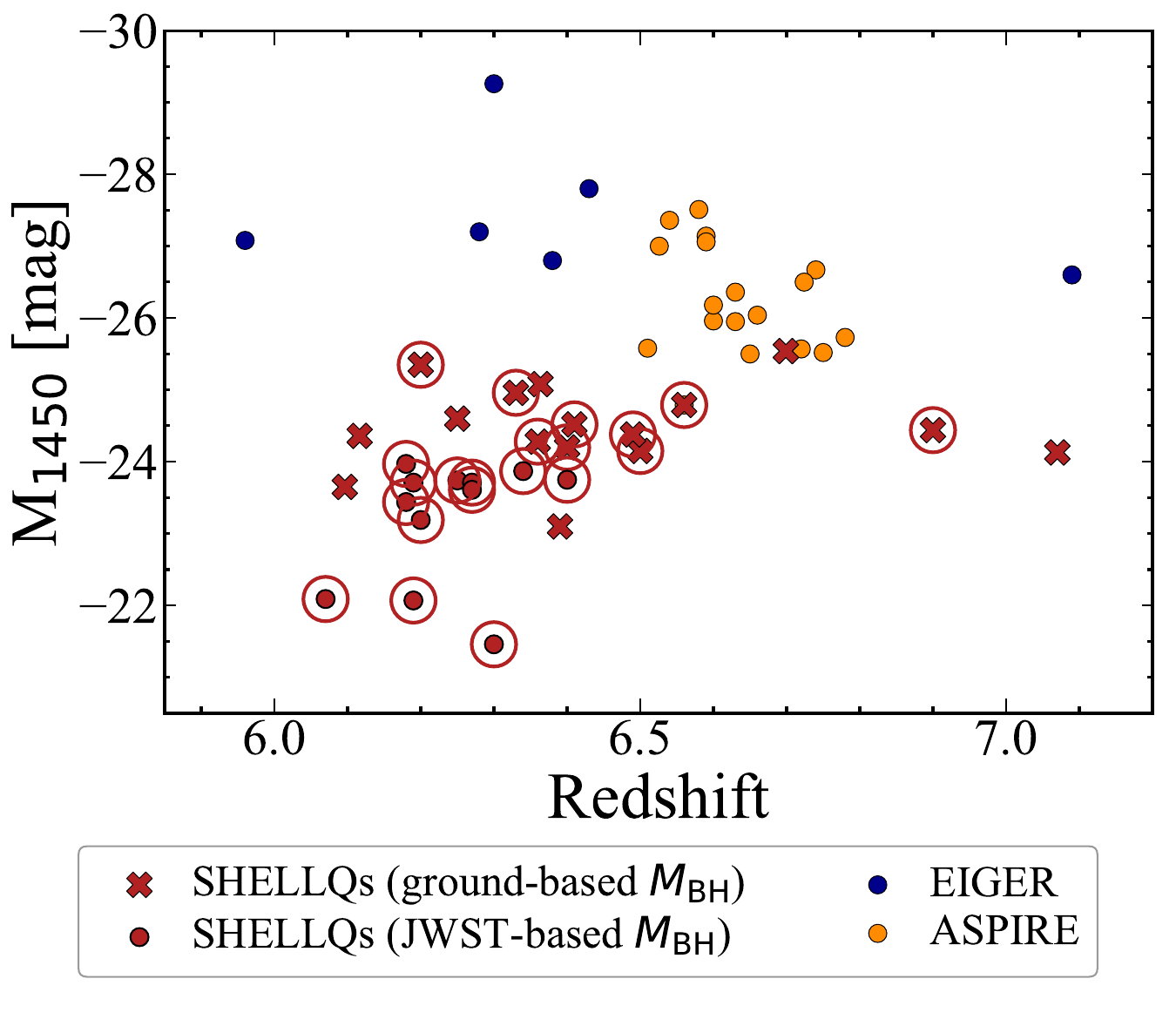}}
\caption{The redshift -  $M_{1450}$ distribution of the SHELLQs quasars with available BH mass estimates (red).
Those with Balmer-line-based $\mBH$ estimates available from JWST observations are shown as small circles, and those with \ion{Mg}{2}-based $\mBH$ estimates from ground-based observations are shown as crosses.
The SHELLQs quasars presented in this paper are highlighted with outer circles.
The remaining SHELLQs quasars, all with \ion{Mg}{2}-based $\mBH$ estimates, are taken from \citetalias{Onoue19}, \citet{Matsuoka19}, and \citet{Kato20}.
Luminous quasars targeted in the JWST EIGER program (blue; \citealt{Yue24}) and the ASPIRE program (orange; \citealt{Wang23, Wang26}) are also shown for comparison. 
}\label{fig:z_M1450}
\end{figure}

Our NIRSpec targets were observed as a part of a JWST Cy~1 program (GO 1967; PI: M.Onoue) between 2022 October 28 and 2023 November 28.
Ten quasars at $6.18\leq z \leq 6.4$ and $M_{1450}\geq -24$ mag were selected from a parent sample of $z\sim6$ quasars that were used to estimate the $z=6$ quasar luminosity function \citep{Matsuoka18c}.
Two fainter quasars, J1146$-$0005 and J0911$+$0152 with $M_{1450}\approx -22$ mag, were added to the sample in order to expand the luminosity range.
The basic properties of these 12 quasars are summarized in Table~\ref{tab:targets_all}.
Figure~\ref{fig:z_M1450} shows the $z$--$M_{1450}$ distribution of the SHELLQs quasars with available black hole mass estimates, including the JWST/NIRSpec and Subaru/MOIRCS targets presented in this paper, compared with targets of the JWST EIGER \citep{Yue24} and ASPIRE programs \citep{Wang23, Wang26}.

These quasars were also observed with NIRCam with two broadband filters (F150W and F356W).
In  \citet{Ding23} and \citet{Ding25}, we decompose these NIRCam images into a combination of 2D S\'ersic profile and point spread function (PSF) for separating host stellar emission from compact quasar emission.
One of the targets, J2236+0032, was observed with six additional broad- and medium-band NIRCam filters to constrain their post-starburst stellar population \citep{Onoue25}.
In Table~\ref{tab:targets_all}, we quote the decomposed F150W quasar magnitudes.

\subsection{G395M spectroscopy}\label{sec:Sec2_NIRSpec_obs}

The JWST spectroscopy was performed with the NIRSpec's 0\arcsec.2-wide S200A2 fixed slit \citep{Jakobsen22}.
We chose the G395M medium-resolution grating, which covers 2.87 -- 5.27 \micron.
This wavelength window corresponds to $\approx 4000 \text{--} 7300$~\AA\ in the rest frame of the target quasars.
Target acquisition was performed with the wide aperture target acquisition (WATA) method.
The on-source exposure times range from 0.6 to 2.4 hours with longer exposures assigned to UV-fainter targets.
We achieved these exposures with $3$-point primary dithering and no sub-pixel dithers.
The NRSIRS2RAPID readout pattern was used for those targets with $<1$ hour exposures, while NRSIRS2 for those with HSC $y$-band magnitude fainter than $24$ (J0911$+$0152, J1146$-$0005, J1512$+$4422).

\begin{figure*}[hp!]
\centering
 \includegraphics[width=0.95\linewidth]{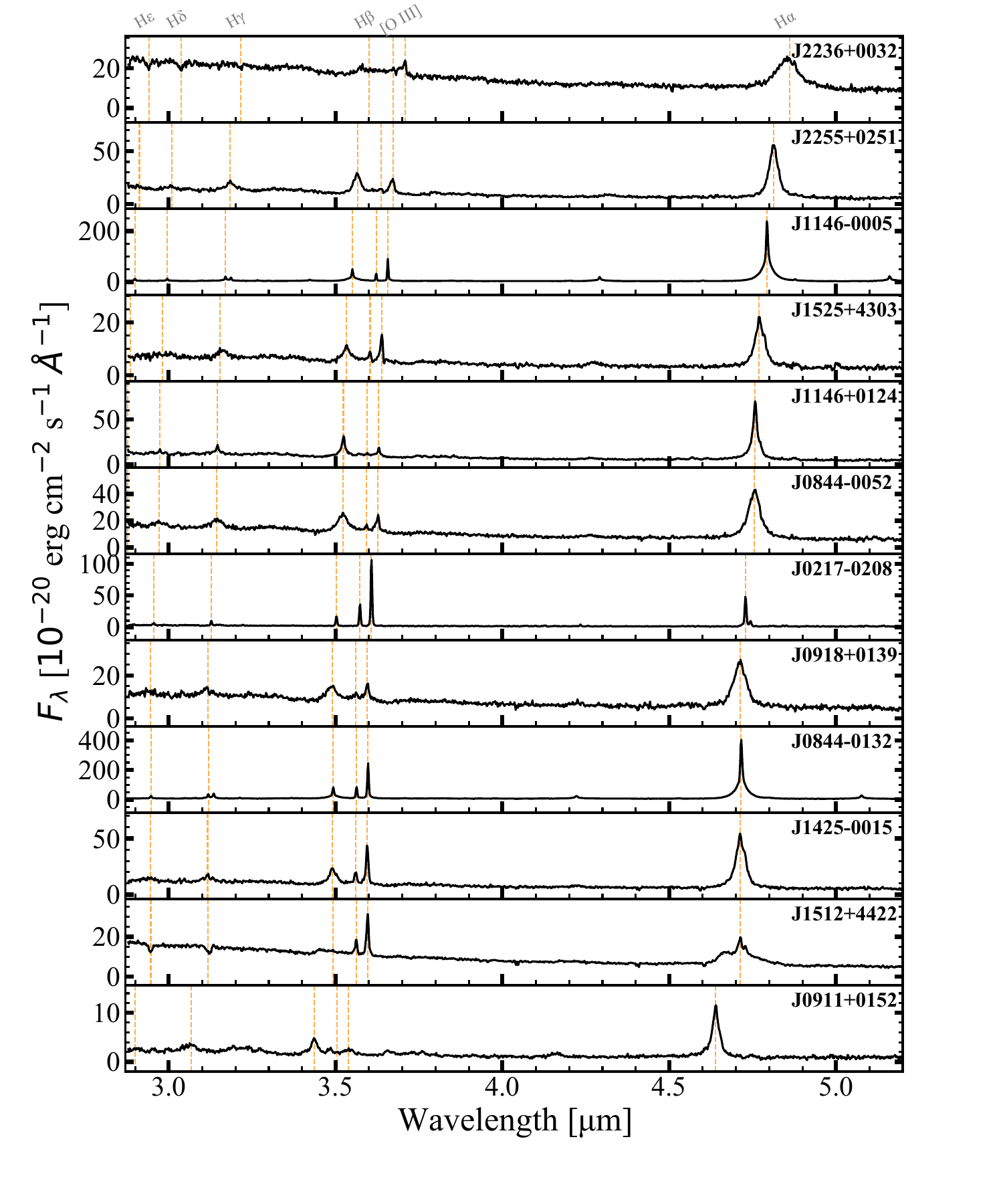}
\caption{The NIRSpec G395M spectra of the JWST targets. 
Flux densities are presented in units of $10^{-20}~\mathrm{erg}\,\mathrm{cm}^{-2}\,\mathrm{s}^{-1}\,\mathrm{\AA}^{-1}$.
The observed wavelengths of Balmer lines and [\ion{O}{3}] are indicated by vertical orange dashed lines. 
For J2236$+$0032 and J1512$+$4422, H$\gamma$, H$\delta$, and H$\epsilon$ are seen in absorption.
Weaker emission lines are also identified in the data, such as [\ion{O}{3}] $\lambda4364$ for some objects.
} \label{fig:spec_JWST}
\end{figure*}

The data reduction of these JWST data follows our earlier publications  \citep{Ding23, Onoue25}.
Briefly, the \textsf{uncal} images were downloaded from the MAST archive and processed with the JWST pipeline version 1.17.1 with the parameter reference file \textsf{jwst\_1322.pmap}, as registered in the JWST Calibration Reference Data System \footnote{\url{https://jwst-crds.stsci.edu}}.
Giant cosmic ray hits were masked during the Stage 1 pipeline by turning on the \textsf{expalnd\_large\_events} and \textsf{sat\_required\_snowball} functions implemented in the ``jump" step.
Smaller cosmic rays near the target traces were also flagged with visual inspection.
The $1/f$ noise stripes were subtracted before the Stage 2 pipeline based on a public code implemented in \textit{msaexp} \citep{msaexp}.
The point-source pathloss correction was applied during the Stage 2 pipeline reduction to take into account slit losses.
The processed two-dimensional spectra at each dither position were stacked using the Stage 3 pipeline with inverse-variance weighting. 
One-dimensional spectra were extracted with a 6-pixel-wide (0\arcsec.6 wide) box-car aperture. 
Figure~\ref{fig:spec_JWST} shows the full spectra of the JWST targets.
The median signal to noise ratios per pixel at rest-frame $5100\pm50$~\AA\ range from 10 -- 50 (Table~\ref{tab:targets_all}).

We note that \citet{Phillips25} use the Stage~2 spectra of the same data to model the 2D PSF and subtract the quasar component from each dither. 
This approach was taken to avoid the artificial broadening of the PSF that occurs during Stage 3 resampling and rectification, thereby ensuring the accurate extraction of extended host stellar and gas emission.

\subsection{Spectral Fitting} \label{sec:spec_model_JWST}

We model the rest-frame optical spectra of the 12 JWST targets using \textsf{QSOFitMORE} \citep[][version 1.2.0]{Fu21_QSOfitmore} with custom modifications to fit the continuum and emission lines simultaneously.
This is a wrapper package of \textsf{PyQSOFit} \citep{PyQSOFit, Shen19}, a public tool widely applied to large spectroscopic datasets such as SDSS \citep{Wu22}.
A single power-law function is adopted to model the quasar continuum, namely $F_\lambda\propto\lambda^{\alpha_\lambda}$, where $\alpha_\lambda$ is the slope index.
In \textsf{QSOFitMORE}, the pseudo-continuum of ionized iron is taken into account based on a spectrum of a local AGN by \citet{Boroson92}. 
For J2236$+$0032 and J1512$+$4422, which show Balmer absorption lines, we perform spectroscopic decomposition to separate the host stellar emission lines from the quasars, and derive the quasar continuum model. 
A full description of the analysis is presented in \citet{Onoue25}.
For J0217$-$0208 and J0844$-$0132, which show weak continuum and strong narrow emission lines, we model the continuum shape by taking the median flux density every 5000 km s$^{-1}$ in velocity space, with strong emission line regions masked, and interpolate with a third-order spline curve.
This smoothed continuum model is subtracted before fitting emission lines.
J1146$-$0005 is as faint as J0217$-$0208 and J0844$-$0132 in continuum, but a single power-law model returns a satisfactory fit.

\begin{figure*}[p!]
\centering
 \includegraphics[width=\linewidth]{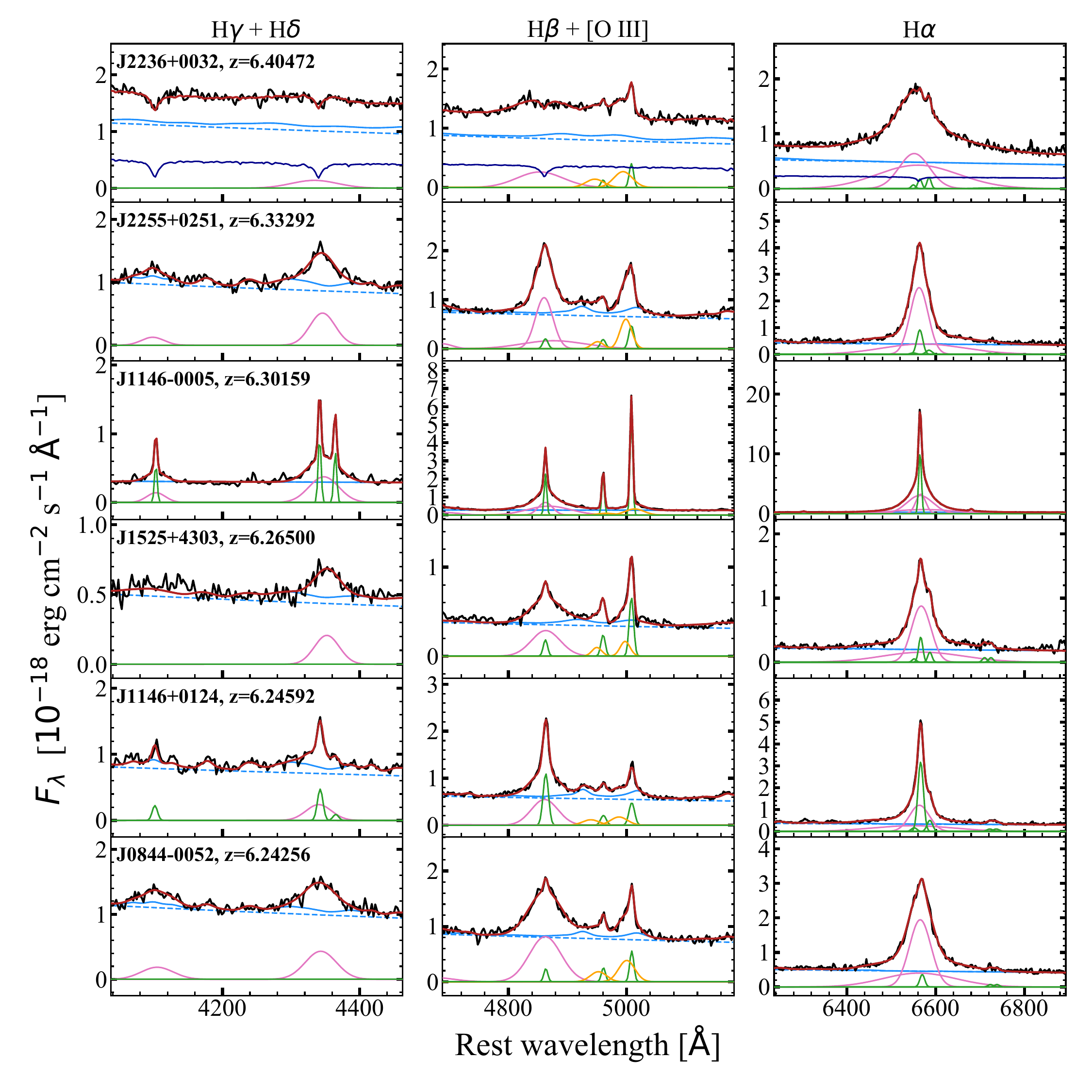}
\caption{Spectral models for the first half of the JWST targets in the rest frame.
Note the different $y$-axis scale from Figure~\ref{fig:spec_JWST}. 
For each target, the H$\gamma$+H$\delta$, H$\beta$+[\ion{O}{3}], and H$\alpha$ spectral regions are shown in the left, middle, and right panels, respectively.
The continuum-only and continuum+iron models are shown as blue dashed and blue solid lines, respectively.
The host-galaxy stellar continuum for J2236+0032 taken from \citet{Onoue25}, is shown as a dark blue line.
Emission line models are presented individually with broad components shown as  magenta lines and narrow components as green lines.
The blueshifted [\ion{O}{3}] ``wing'' components are shown as an orange line.
The sum of all line components is shown as a red line.
} \label{fig:specfit_JWST}
\end{figure*}

\begin{figure*}[p!]
\centering
 \includegraphics[width=\linewidth]{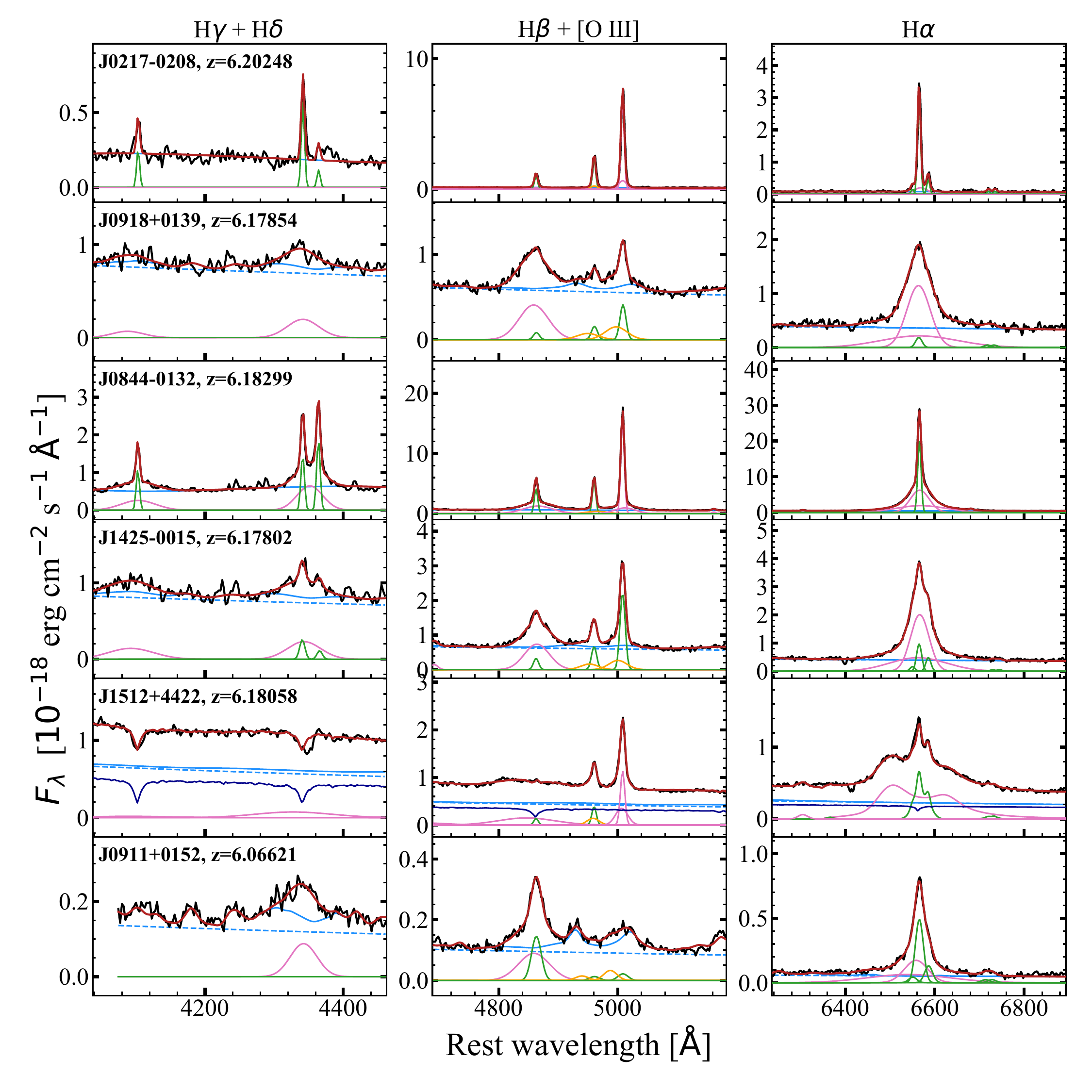}
\caption{Figure~\ref{fig:specfit_JWST}, but for the second half of the JWST targets.
1512+4422 shows Balmer absorption lines, similar to J2236+0032, and also exhibits a double-peaked H$\alpha$ profile \citep{Onoue25}.
Zoom-in panels of the Balmer emission lines of J0217$-$0208 are presented in Figure~\ref{fig:J0217_2comp}.
} \label{fig:specfit_JWST2}
\end{figure*}

A two-step approach was adopted to model the spectra.
First, we run \textsf{QSOFitMORE} to fit the H$\beta$+[\ion{O}{3}] spectral complex in addition to the continuum and iron emission, modeling each line with one or more Gaussian profiles.
For H$\beta$, we  adopt a combination of broad and narrow components, where we limit the line width of broad lines to $1,200 \leq \text{FWHM} < 10,000$ km s$^{-1}$, following \citet{Shen11}, and the widths of narrow lines to FWHM $<1,200$ km s$^{-1}$.
In most cases, a single broad component is sufficient to fit the broad components.
If multiple broad components are required, we use the total FWHM of the composite fit as their line widths.
The forbidden [\ion{O}{3}] $\lambda\lambda 4960,5008$ lines are modeled with one narrow and one broader component for each of the doublet lines.
The latter component is often blueshifted, commonly interpreted as a signature of outflowing ionized gas kinematics.
The narrow  component of the two [\ion{O}{3}] lines are required to have the same line widths without velocity offsets.
The line width of the broad [\ion{O}{3}]  component is left unconstrained  and is allowed to have a velocity offset of $|\Delta v| \leq 5,000$ km s$^{-1}$ from the systemic redshifts.
We repeat the H$\beta$+[\ion{O}{3}] fitting a few times, updating the redshift at each iteration based on the peak wavelength of the narrow [\ion{O}{3}] component.
The initial redshift is set to the Ly$\alpha$ redshift, and the final converged [\ion{O}{3}] redshift is adopted as the systemic redshift of each quasar.
We note that the [\ion{O}{3}] emission lines of J0911$+$0152 are weak, making it difficult to decompose them into two components, and thus their line flux measurements may be degenerate.

We then extend the line fitting to all emission lines visually identified in each spectrum.
Forbidden lines, such as [\ion{N}{2}] $\lambda\lambda 6549,6585$ and [\ion{S}{2}] $\lambda\lambda 6718,6732$, are constrained to have the same line width as the narrow [\ion{O}{3}] component.
Permitted lines, including the Balmer and Helium lines, are fitted with a combination of one broad and one narrow Gaussian profile, as was done for H$\beta$.
The broad components of individual Balmer lines are fitted independently.
In some cases, two or three broad Gaussian profiles are required to fit broad Balmer lines, similar to H$\beta$.
The line widths for the broad components are measured independently for different emission lines, including different Balmer transitions (Section~\ref{sec:Balmer_width}).
The narrow components of these permitted lines are fitted using the same FWHM as the narrow [\ion{O}{3}] component.
This reflects the assumption that the narrow-line region gas shares the same kinematics regardless of the line species.
We include [\ion{N}{2}] only when the line peak of [\ion{N}{2}] $\lambda6585$  is clearly visible in the spectra.
The final continuum and emission line models are presented in Figure~\ref{fig:specfit_JWST}.
Table~\ref{tab:emission_jwst} presents the continuum luminosity and line luminosities of representative emission lines.
Some of the diagnostic emission line ratios (i.e., [\ion{O}{3}]/H$\beta$ and [\ion{N}{2}]/H$\alpha$) are also presented in \citet{Phillips25}.

\begin{deluxetable*}{lCCCCCCCCCCC}[p!]
\rotate
\tabletypesize{\tiny}
\tablecaption{Spectral properties: JWST targets \label{tab:emission_jwst}}
\tablehead{
\colhead{ID} &
\colhead{$z_{\rm [OIII]}$} &
\colhead{$L_{5100}$} &
\colhead{$L_{\rm 5100, QSO}$} &
\colhead{$\alpha_{\rm opt}$} &
\colhead{FWHM H$\beta_{\rm b}$} &
\colhead{FWHM H$\alpha_{\rm b}$} &
\colhead{FWHM [O\,\textsc{iii}], c} &
\colhead{$L_{{\rm H}\beta,{\rm b}}$} &
\colhead{$L_{{\rm H}\beta,{\rm n}}$} &
\colhead{$L_{{\rm H}\alpha,{\rm b}}$} &
\colhead{$L_{{\rm H}\alpha,{\rm n}}$} \\
\colhead{} &
\colhead{} &
\colhead{[10$^{45}$ erg s$^{-1}$]} &
\colhead{[10$^{45}$ erg s$^{-1}$]} &
\colhead{} &
\colhead{[km s$^{-1}$]} &
\colhead{[km s$^{-1}$]} &
\colhead{[km s$^{-1}$]} &
\colhead{[10$^{43}$ erg s$^{-1}$]} &
\colhead{[10$^{43}$ erg s$^{-1}$]} &
\colhead{[10$^{43}$ erg s$^{-1}$]} &
\colhead{[10$^{43}$ erg s$^{-1}$]}
}
\startdata
J2236+0032 & 6.40472 $\pm$ 0.00087 & 2.550 $\pm$ 0.010 & 1.780 $\pm$ 0.012 & $-1.79 \pm 0.02$ & 5644 $\pm$ 741 & 4999 $\pm$ 88 & 412 $\pm$ 61 & 1.18 $\pm$ 0.14 & \nodata & 7.09 $\pm$ 0.10 & 0.09 $\pm$ 0.01 \\
J2255+0251 & 6.33292 $\pm$ 0.00096 & 1.450 $\pm$ 0.010 & 1.310 $\pm$ 0.009 & $-1.96 \pm 0.02$ & 2338 $\pm$ 44 & 2405 $\pm$ 36 & 552 $\pm$ 51 & 2.73 $\pm$ 0.06 & 0.10 $\pm$ 0.02 & 9.81 $\pm$ 0.12 & 0.69 $\pm$ 0.06 \\
J1146-0005 & 6.30159 $\pm$ 0.00001 & 0.612 $\pm$ 0.001 & 0.612 $\pm$ 0.001 & $-0.58 \pm 0.01$ & 2501 $\pm$ 67 & 1945 $\pm$ 349 & \nodata & 3.00 $\pm$ 0.01 & 0.56 $\pm$ 0.01 & 19.60 $\pm$ 1.46 & 3.61 $\pm$ 0.39 \\
J1525+4303 & 6.26500 $\pm$ 0.00054 & 0.720 $\pm$ 0.006 & 0.582 $\pm$ 0.005 & $-1.91 \pm 0.02$ & 3534 $\pm$ 81 & 2620 $\pm$ 33 & 420 $\pm$ 28 & 0.77 $\pm$ 0.02 & 0.07 $\pm$ 0.01 & 3.71 $\pm$ 0.05 & 0.21 $\pm$ 0.02 \\
J1146+0124 & 6.24591 $\pm$ 0.00045 & 1.180 $\pm$ 0.005 & 0.976 $\pm$ 0.004 & $-1.81 \pm 0.03$ & 3240 $\pm$ 97 & 2719 $\pm$ 110 & 590 $\pm$ 14 & 1.35 $\pm$ 0.01 & 0.55 $\pm$ 0.02 & 5.55 $\pm$ 0.21 & 2.15 $\pm$ 0.14 \\
J0844-0052 & 6.24256 $\pm$ 0.00047 & 1.630 $\pm$ 0.013 & 1.560 $\pm$ 0.012 & $-1.87 \pm 0.02$ & 3913 $\pm$ 162 & 2943 $\pm$ 56 & 377 $\pm$ 66 & 2.41 $\pm$ 0.11 & 0.09 $\pm$ 0.01 & 8.78 $\pm$ 0.09 & 0.22 $\pm$ 0.04 \\
J0217-0208 & 6.20248 $\pm$ 0.00001 & 0.278 $\pm$ 0.001 & 0.0858 $\pm$ 0.0004 & $-2.18 \pm 0.04$ & 1154 $\pm$ 1 & 1806 $\pm$ 119 & 230 $\pm$ 2 & 0.09 $\pm$ 0.01 & 0.31 $\pm$ 0.01 & 0.22 $\pm$ 0.01 & 1.23 $\pm$ 0.02 \\
J0918+0139 & 6.17854 $\pm$ 0.00139 & 1.160 $\pm$ 0.007 & 1.080 $\pm$ 0.006 & $-1.58 \pm 0.02$ & 3548 $\pm$ 121 & 3149 $\pm$ 39 & 643 $\pm$ 54 & 1.07 $\pm$ 0.03 & 0.04 $\pm$ 0.01 & 5.19 $\pm$ 0.03 & 0.14 $\pm$ 0.02 \\
J0844-0132 & 6.18299 $\pm$ 0.00001 & 1.170 $\pm$ 0.003 & 1.130 $\pm$ 0.003 & $-0.45 \pm 0.02$ & 3633 $\pm$ 43 & 2650 $\pm$ 17 & 259 $\pm$ 1 & 3.39 $\pm$ 0.01 & 1.30 $\pm$ 0.01 & 25.60 $\pm$ 0.06 & 8.23 $\pm$ 0.02 \\
J1425-0015 & 6.17801 $\pm$ 0.00047 & 1.250 $\pm$ 0.010 & 1.030 $\pm$ 0.008 & $-1.53 \pm 0.03$ & 3038 $\pm$ 51 & 2547 $\pm$ 32 & 537 $\pm$ 20 & 1.65 $\pm$ 0.03 & 0.15 $\pm$ 0.02 & 7.72 $\pm$ 0.06 & 0.59 $\pm$ 0.02 \\
J1512+4422 & 6.18058 $\pm$ 0.00063 & 1.520 $\pm$ 0.005 & 0.934 $\pm$ 0.004 & $-2.20 \pm 0.01$ & 4620 $\pm$ 97 & 8590 $\pm$ 181 & 440 $\pm$ 25 & 0.85 $\pm$ 0.02 & 0.055 $\pm$ 0.004 & 4.21 $\pm$ 0.03 & 0.51 $\pm$ 0.01 \\
J0911+0152 & 6.06621 $\pm$ 0.00028 & 0.180 $\pm$ 0.002 & 0.162 $\pm$ 0.001 & $-2.02 \pm 0.04$ & 3770 $\pm$ 583 & 3475 $\pm$ 70 & 1147 $\pm$ 411 & 0.24 $\pm$ 0.03 & 0.12 $\pm$ 0.05 & 1.06 $\pm$ 0.02 & 0.45 $\pm$ 0.02 \\
\enddata
\tablecomments{
The continuum slopes $\alpha_\mathrm{opt}$ are based on observed NIRSpec continuum without subtracting host contribution. 
The errorbars reported here do not include flux calibration uncertainties. 
}
\end{deluxetable*}

\subsection{Notes on Individual Objects}\label{sec:notes_on_indivial_JWST_targets}
Here we describe several remarkable properties of the JWST targets in detail.

\begin{description}
    \item[J2236$+$0032]  J2236$+$0032 is one of the two quasars in the sample, in which the spectrum of the host galaxy is immediately apparent. 
    Its high-order Balmer lines (H$\gamma$, H$\delta$, H$\epsilon$) are observed as absorption lines, unlike those of typical broad-line quasars. 
    We interpret these Balmer absorption lines as the spectroscopic signature of the post-starburst stellar population of the host \citep{Onoue25}. 
    We also highlight that J2236$+$0032 and J1512$+$4422 (see below) are hosted by the two most massive galaxies in our sample \citep{Ding23, Ding25}.
    This is consistent with the picture that quasar-mode feedback becomes important as massive galaxies mature and transition toward quiescence, as described in \citet{Onoue25}.
    
    \item[J1512$+$4422]  J1512$+$4422 is the other quasar in our sample that shows Balmer absorption lines in H$\gamma$ and H$\delta$.
    We also observed double-peaked broad H$\alpha$ emission, in addition to narrow H$\alpha+$[\ion{N}{2}] and [\ion{S}{2}].
    \citet{Onoue25} model this line multiplet with a disk BLR model, where the emission arises from relativistic Keplerian motion of  gas in the BH accretion disk \citep{Eracleous09, Ward24}.
    \citet{Liu26} recently obtained a high-resolution spectrum of this object, where they resolve the Balmer absorption lines and detected a spatially extended ionized gas outflow.
    
    \item[J0844$-$0132 \& J1146$-$0005] These two quasars are characterized by strong narrow emission lines in the Balmer series on top of their broad components. We detect several other weak emission lines, especially [\ion{O}{3}]~$\lambda4362$,  [\ion{O}{1}]~$\lambda6302$, and various transitions of \ion{He}{1}. 
    They are distinguished from the other JWST targets by both their continuum slopes and their large H$\alpha$/H$\beta$ flux ratios, and rather reminiscent of ``Little Red Dot” objects (see Sec~\ref{sec:spec_properties}).
    
    \item[J0217$-$0208]  This quasar shows the narrowest Ly$\alpha$ line among our targets and is classified as a narrow-line quasar in the series of the SHELLQs papers \citep{Matsuoka18b}.
    The JWST medium-resolution spectrum of J0217$-$0208 shows that its optical lines are also relatively narrow (FWHM (H$\alpha_{\rm broad}=1,810\pm120$~km s$^{-1}$). 
    Its line modeling and nature (an AGN or a star-forming galaxy) will be extensively discussed in Section~\ref{sec:J0217_comp}.
\end{description}

\subsection{Spectral Properties}\label{sec:spec_properties}

\subsubsection{Continuum}\label{sec:continuum_slopes}
\begin{figure*}[tbp]
    \centering
    \includegraphics[width=0.65\linewidth]{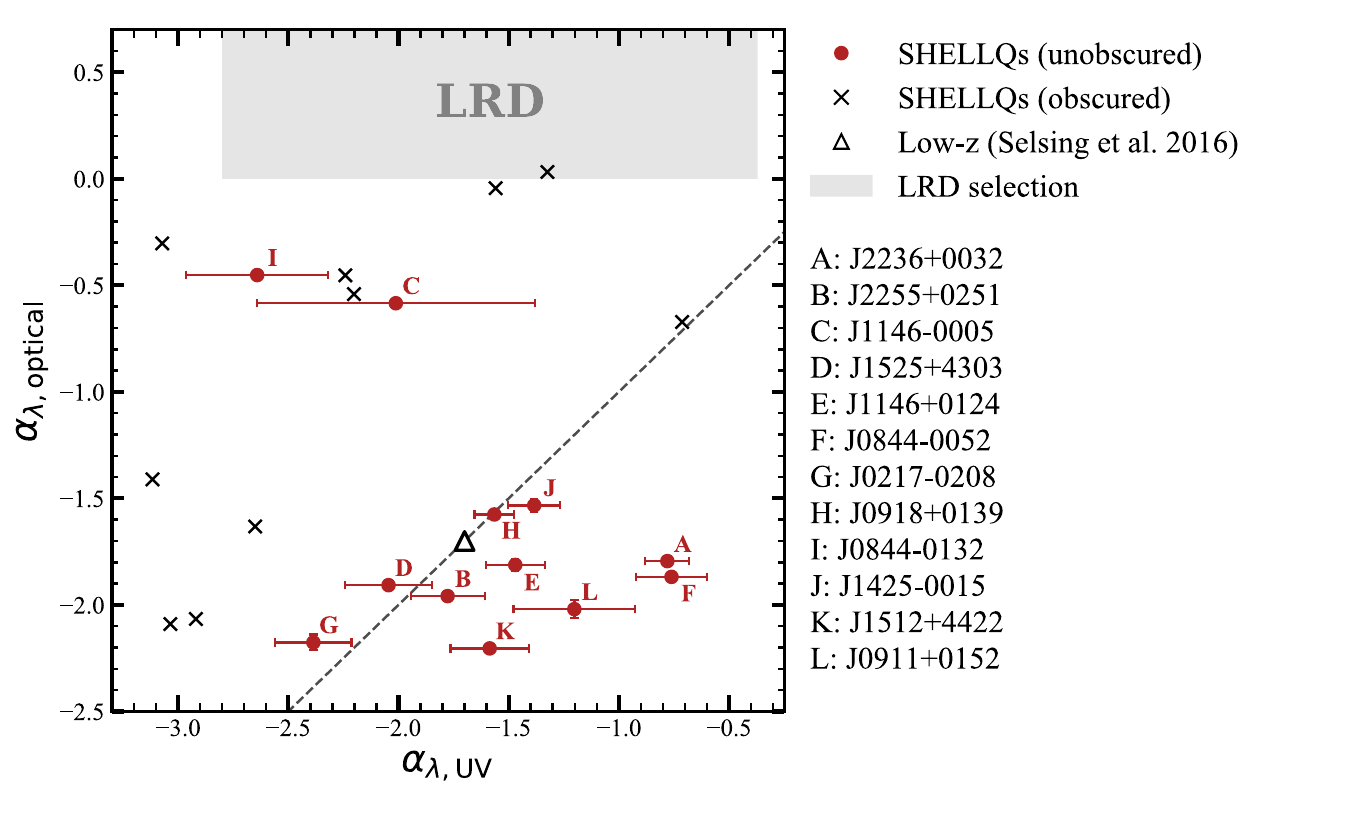}
    \caption{
    UV ($\alpha_{\rm \lambda, UV}$) and optical ($\alpha_{\rm \lambda, optical}$) continuum slopes. 
    The JWST targets in this work are shown as red circles with each object labeled by a letter as indicated in the lower right.
    The dust-obscured broad-line SHELLQs objects observed by \citet{Matsuoka25} are shown as black crosses.
    The continuum slopes of typical $1<z<2$ quasars measured by \citet{Selsing16} are shown as an open triangle.
    The color window used for the selection of Little Red Dots in \citet{Kocevski25} is indicated by the shaded region.
    The dashed line indicates the 1:1 relation.
    }
    \label{fig:continuum}
\end{figure*}

The clear detection of rest-frame optical continuum emission of the NIRSpec targets enables us to investigate their continuum shapes.
Figure~\ref{fig:continuum} compares UV and optical slopes of the NIRSpec targets, where the optical slopes $\alpha_\mathrm{\lambda, optical}$ are derived from the NIRSpec data.
The UV continuum slopes $\alpha_\mathrm{\lambda, UV}$ are obtained from HSC $y$-band photometry and F150W photometry of the targets \citep{Ding25}.
We note that total continuum emission and photometry are used here in order to obtain the observed continuum shape of each object, and part of the measured flux may be contributed by host-galaxy emission.

Figure~\ref{fig:continuum} shows that the NIRSpec targets generally have blue UV continuum shapes, consistent with typical unobscured quasars at $1<z<2$ \citep[e.g., $\alpha_\mathrm{\lambda, UV}\approx-1.7$;][]{VB01, Selsing16}.
Most targets also show blue optical slopes with $\alpha_\mathrm{\lambda, optical} < -1.5$, again comparable to those of low-redshift quasars.
These similarities in the UV and optical continuum slopes suggest that the accretion-disk emission in the majority of these $z>6$ quasars is  similar to that in lower-redshift quasars.

On the other hand, there are two objects, J0844$-$0132 and J1146$-$0005, which stand out with relatively redder optical slopes with $\alpha_\mathrm{\lambda, optical}\approx-0.5$. 
The SED of these objects lie between those of normal quasars and Little Red Dot (LRD) objects, which exhibit ``V-shape" continuum shapes \citep{Matthee24, Greene24, Kocevski25}.
A plausible explanation of the characteristic continuum emission of LRDs is that their optical continuum is  thermal emission from a dense gas cocoon surrounding a young black hole \citep{Inayoshi_Maiolino25, deGraaff25, Naidu25}.
In this context, J0844$-$0132 and J1146$-$0005 may trace an intermediate phase
in the evolutionary sequence from LRD-like young accreting black holes to more mature unobscured quasars.

The location of J0844$-$0132 and J1146$-$0005 in Figure~\ref{fig:continuum} is also similar to a sub-sample of the SHELLQs quasars, for which their Ly$\alpha$ emission is narrow (${\rm FWHM} < 500$~km~s$^{-1}$) and strong (line luminosity $L_{\rm Ly\alpha} > 10^{44}~{\rm erg~s^{-1}}$).
\citet{Matsuoka25} show that seven out of ten targets exhibit broad Balmer emission lines in their JWST/NIRSpec observations, indicating that they are modestly dust-obscured quasars.
J1146$-$0005 is among the same population (FWHM $=330\pm 100$~km~s$^{-1}$), while 
J0844$-$0132 shows a broader Ly$\alpha$ (FWHM $=1610\pm 280$~km~s$^{-1}$). 
This similarity in continuum shapes therefore suggests that J0844$-$0132 and J1146$-$0005 represent a related population to the  \citet{Matsuoka25}'s obscured quasar sample bridging LRDs and normal quasars.

\subsubsection{Balmer Line Widths}\label{sec:Balmer_width}
Lower-redshift quasar studies have shown that the broad components of high-order Balmer emission lines exhibit systematically broader profiles than H$\alpha$  \citep{GH05, Schulze10, Schulze18}.
These lines also show shorter response times to continuum flux variations in reverberation mapping \citep{Bentz10, Grier17}, suggesting that higher-order Balmer lines originate from inner regions of the BLR gas.

\begin{figure*}[tbp]
    \centering
    \includegraphics[width=0.85\linewidth]{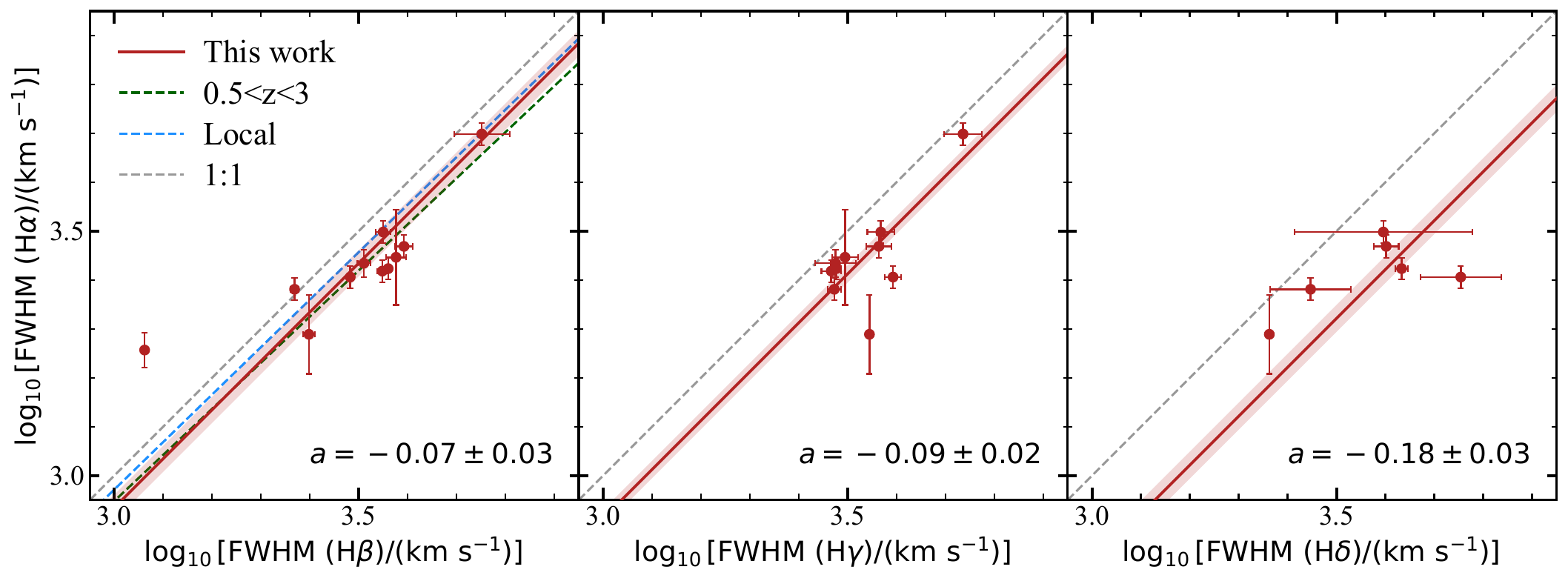}
    \caption{
    Comparison of the FWHMs of broad Balmer emission lines. 
    Red solid lines show the line correlation where the slope is fixed to 1. 
    Grey dashed lines show the one-to-one relation.
    In the H$\alpha$ vs H$\beta$ plot, the local relation of \citet{GH05} and the relation at $0.5<z<3$ from \citet{Schulze18} are indicated by cyan and green dashed lines, respectively.
    }
    \label{fig:Balmer}
\end{figure*}

In Figure~\ref{fig:Balmer}, we show the relation between line FWHMs of different Balmer series transitions in our sample.
We perform orthogonal distance regression (ODR) in logarithmic space to fit a linear relation for each line combination (H$\alpha$ vs H$\beta$, H$\gamma$, and H$\delta$), namely 
$\log_{10} {\rm FWHM}_{\rm H\alpha} = \log_{10} {\rm FWHM}_{\rm line} + a$, where $a$ is the intercept.
We chose ODR to take into account measurement errors in both variables.
We here only consider broad components of the Balmer emission lines and remove J1512+4422, which exhibits a double-peak profile in H$\alpha$ \citep{Onoue25}.
We impose a 5\% error floor on the measurement errors to prevent the fitting from being dominated by specific objects with small errors.
The slope of each relation is fixed to 1, given the limited dynamic range and the small sample size.

\begin{deluxetable}{lCC}[tb]
\tabletypesize{\small}
\tablecaption{Correlations of Balmer emission line widths \label{tab:Balmer_fwhm}}
\tablehead{
\colhead{Line } &
\colhead{Intercept \tablenotemark{$\dagger$}} &
\colhead{${\rm FWHM}_{\rm H\alpha}/{\rm FWHM}_{\rm line}$}
}
\startdata
H$\beta$  & $-0.066 \pm 0.027$ & $0.86 \pm 0.05$ \\
H$\gamma$ & $-0.087 \pm 0.018$ & $0.82 \pm 0.03$ \\
H$\delta$ & $-0.18 \pm 0.03$   & $0.66 \pm 0.04$ \\
\enddata 
\tablecomments{$\dagger$ The intercept $a$ in the relation
$\log_{10} {\rm FWHM}_{\rm H\alpha}
= \log_{10} {\rm FWHM}_{\rm line} + a$
is obtained using ODR regression.
The intercept for each line combination is also presented in the form of
${\rm FWHM}_{\rm H\alpha}/{\rm FWHM}_{\rm line}\ (=10^a)$
in the third column.
}
\end{deluxetable}

The best-fit parameters of the fits are reported in Table~\ref{tab:Balmer_fwhm}. 
We find that the line widths of high-order transitions are generally broader than H$\alpha$. 
The majority are consistent with the relation of lower-redshift quasars \citep{GH05, Schulze18} with H$\beta$ broader than H$\alpha$ by  16\%. 
For higher-order transitions, the linear regression finds that H$\gamma$ and H$\delta$ are 22\%\ and 51\%\ broader than H$\alpha$, respectively. 
Our results therefore indicate that the low-redshift trends among different Balmer transitions extend to $z\sim6$.

On the other hand, J0217$-$0208 and J0911$+$0152 are outliers with $\mathrm{FWHM(H\alpha)} > \mathrm{FWHM(H\beta)}$, which would bring the mean relation closer to the one-to-one relation.
A larger sample would therefore be required to firmly establish the high-redshift line width relation.
Also note that the weak H$\delta$ emission line is sensitive to continuum and iron emission uncertainties.

\subsubsection{Continuum -- Balmer Line Relation}\label{sec:Balmer_cont}

\begin{deluxetable}{lcccc}[tbp]
\tabletypesize{\small}
\tablecaption{Correlations between $L_{\rm 5100,QSO}$ and $L_{\rm Balmer}$
\label{tab:Balmer_luminosity}}
\tablehead{
\colhead{Relation} &
\colhead{$a$} &
\colhead{$10^{\,a}$} &
\colhead{$b$} &
\colhead{$\sigma$}
}
\startdata
H$\alpha_{\rm broad}$ &
$0.69 \pm 0.08$ & $4.89 \pm 0.90$ & $1.05 \pm 0.08$ & $0.10$ \\
H$\alpha_{\rm total}$ &
$1.14 \pm 0.08$ & $13.8 \pm 2.6$ & $0.66 \pm 0.09$ & $0.11$ \\
H$\beta_{\rm broad}$ &
$0.13 \pm 0.12$ & $1.35 \pm 0.37$ & $0.96 \pm 0.12$ & $0.15$ \\
H$\beta_{\rm total}$ &
$0.54 \pm 0.13$ & $3.47 \pm 1.01$ & $0.60 \pm 0.13$ & $0.17$ \\
H$\gamma_{\rm broad}$ &
$0.21 \pm 0.15$ & $1.63 \pm 0.57$ & $0.53 \pm 0.15$ & $0.13$ \\
H$\gamma_{\rm total}$ &
$0.21 \pm 0.09$ & $1.62 \pm 0.34$ & $0.58 \pm 0.10$ & $0.12$ \\
H$\delta_{\rm broad}$ &
$-0.20 \pm 2.08$ & $0.63 \pm 3.01$ & $0.63 \pm 1.90$ & $0.22$ \\
H$\delta_{\rm total}$ &
$-0.21 \pm 0.27$ & $0.62 \pm 0.38$ & $0.56 \pm 0.27$ & $0.26$ \\
\hline
H$\beta$ -- H$\alpha$ (broad) &
$0.64 \pm 0.03$ & $4.35 \pm 0.26$ & 1 (fixed) & $0.08$ \\
H$\beta$ -- H$\alpha$ (total) &
$0.64 \pm 0.02$ & $4.41 \pm 0.23$ & 1 (fixed) & $0.07$ \\
\enddata
\tablenotetext{}{
The results of linear least-$\chi^2$ fits after excluding the two objects
exhibiting strong narrow Balmer emission (J0844$-$0132 and J1146$-$0005; see text) are presented in this table.
For the continuum--Balmer line relations, the fitting form is
$\log (L_{\rm Balmer}~\mathrm{[erg\ s^{-1}]}/10^{42}) =
    a \log (L_{\rm 5100, QSO}~\mathrm{[erg\ s^{-1}]}/10^{44}) + b$,
where $L_{\rm 5100,QSO}$ is the QSO-only continuum luminosity derived from
our NIRCam image decomposition analysis presented in \citet{Ding25}.
The first column lists the Balmer line used as $L_{\rm Balmer}$.
We show the correlation for both broad and total Balmer emission lines.
We also introduce $\sigma$, which represents the intrinsic scatter.
For the H$\alpha$--H$\beta$ relation, the slope is fixed to unity, namely
$\log_{10} L_{\rm H\alpha} = a + \log_{10} L_{\rm H\beta}$.
}
\end{deluxetable}

\begin{figure*}[tbp]
    \centering
    \includegraphics[trim=0cm 0 0 0, clip, width=0.99\linewidth]{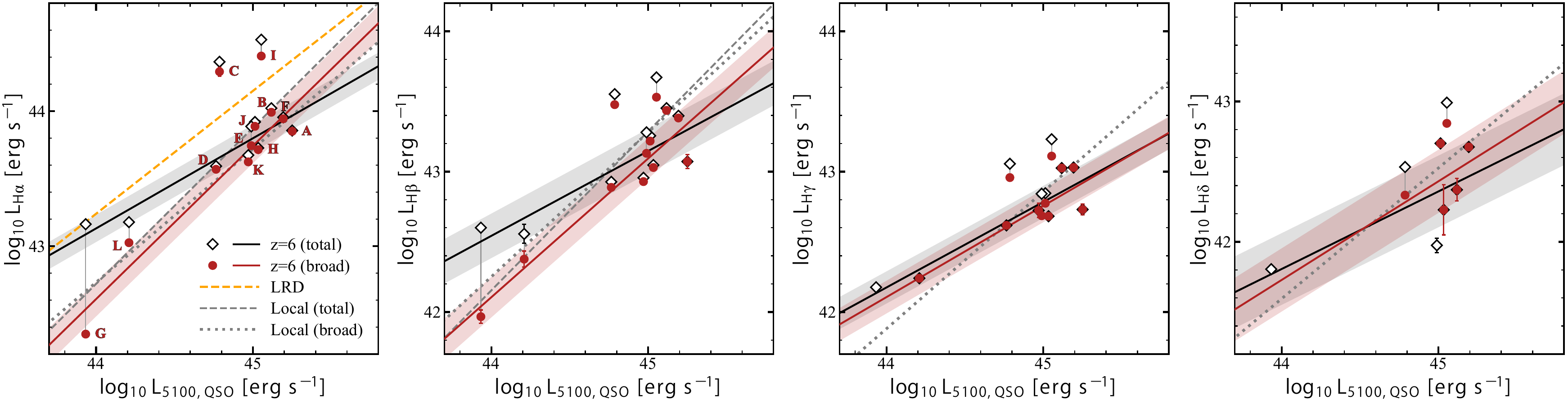}
    \caption{
    Correlations between 5100 \AA\ luminosity and Balmer line luminosities from H$\alpha$ to H$\delta$. 
    Here we use the host-galaxy–subtracted 5100~\AA\ luminosity ($L_{\rm 5100, QSO}$) based on JWST/NIRCam F356W imaging decomposition \citep{Ding25}.
    In each panel, the luminosities of the broad-line components are shown as red circles, while those including both broad and narrow components are shown as open black diamonds.
    For each line pair, the best-fit linear relation is plotted as a solid line with the shaded area indicating the $1\sigma$ scatter of the residuals relative to the best-fit relation.
    The gray dashed and dotted lines represent the correlations observed in
    the local universe for the total \citep{GH05} and broad-only
    \citep{LaMura07} luminosities, respectively.
    For the $L_{\rm 5100, QSO}$-$L_{\rm H\alpha}$ panel, the relation for $z<4.5$ LRDs from \citet{DeGraaff25b} is shown with an orange dashed line.
    The two outliers in our sample (J0844$-$0132 and J1146$-$0005) fall at the upper left in all panels.
    }
    \label{fig:Balmer_luminosity}
\end{figure*}

Local AGN show a tight correlation between the AGN continuum and Balmer line luminosities \citep{GH05, LaMura07}.
This relation enables single-epoch BH mass estimates solely from Balmer emission lines without needing to determine the AGN continuum, which may be affected by possible contribution of host stellar emission.
Using our JWST sample, we now investigate whether the local continuum--Balmer luminosity relation holds for high-redshift quasars.

The rest-frame 5100~\AA\ continuum luminosity $L_{5100}$ is measured from the best-fit continuum model to the observed NIRSpec spectra.
We then subtract the host stellar emission by leveraging our NIRCam imaging data analysis presented in  \citet{Ding23} and \citet{Ding25}.
Specifically, we refer to the fraction of stellar emission in the NIRCam F356W images of the JWST targets, which trace rest-frame $4,300$--$5,500$~\AA\ at the redshifts of our targets, and derive the host-subtracted 5,100~\AA\ continuum luminosity $L_{\rm 5100, QSO}$.
For the two post-starburst quasars, J2236+0032 and J1512+4422, we use the spectroscopically-decomposed quasar spectra presented in \citet{Onoue25} to evaluate the quasar-only continuum luminosity.

We fit the quasar continuum--Balmer luminosity relations to the following form:
\begin{equation}
    \log (L_{\rm Balmer}~\mathrm{[erg\ s^{-1}]}/10^{42}) =
    a \log (L_{\rm 5100, QSO}~\mathrm{[erg\ s^{-1}]}/10^{44}) + b,
    \label{eq:cont_Balmer}
\end{equation}%
where $a$ and $b$ represent the slope and intercept, respectively.
We exclude J0844$-$0132 and J1146$-$0005 from this fitting, as their LRD-like continuum shapes and high H$\alpha$/H$\beta$ ratios indicate that they are a distinct population from the other objects studied here  (Sections~\ref{sec:continuum_slopes}, \ref{sec:Balmer_Decrement}).
We note that these two objects lie far from the regression of the other objects in the continuum -- Balmer emission line luminosity plane.
The fitting results are summarized in Table~\ref{tab:Balmer_luminosity}.

Figure~\ref{fig:Balmer_luminosity} shows the best-fit correlations between $L_{\rm 5100,QSO}$ and $L_{\rm Balmer}$.
For comparison, Figure~\ref{fig:Balmer_luminosity} also shows the local total-line relation of \citet{GH05} and the broad-line-only relation of \citet{LaMura07}.
In the H$\alpha$ panel, we also show the correlation for $z<4.5$ LRDs obtained in \citet{DeGraaff25b};  their sample is fainter than ours in the quasar continuum ($L_{\rm 5100} < 10^{44}~{\rm erg~s^{-1}}$).
We find that the best-fit relation between quasar optical continuum and H$\alpha$, effectively equivalent to H$\alpha$ equivalent widths, is similar to the local relations.
At $L_{\rm 5100,QSO} = 10^{45}~{\rm erg~s^{-1}}$, the $z=6$ relation is slightly lower than the \citet{GH05} relation by 0.08 dex in H$\alpha$ total luminosity, and is still consistent within the object-to-object scatter (0.10 dex).
The luminosity of broad H$\alpha$  is well aligned with the \citet{LaMura07} relation at the same luminosity range.

For H$\beta$ and the higher-order Balmer transitions, we also do not find substantial differences between the local relations  and those measured for our $z=6$ sample.
For H$\beta$, the $z=6$ relations are offset by  $-0.14$ dex for the total luminosity and $-0.19$ dex for the broad-component luminosity at $L_{5100,\mathrm{QSO}}=10^{45}\ {\rm erg\ s^{-1}}$.
These modest offsets have little impact on single-epoch black hole mass
estimates because the inferred mass depends only weakly on the continuum
luminosity, scaling as $M_{\rm BH}\propto L_{5100}^{0.5}$, and therefore
changes $M_{\rm BH}$ by less than $0.1$ dex.
This is well below the typical systematic uncertainty of $\sim0.4$ dex associated with single-epoch virial mass estimates \citep{Vestergaard06,Shen13_review}.
We emphasize that a larger sample with a wider luminosity coverage would be required to firmly establish the high-redshift continuum -- Balmer emission line relation.

\subsubsection{Balmer Decrement}\label{sec:Balmer_Decrement}

\begin{figure}[tb]
    \centering
    \includegraphics[width=0.9\linewidth]{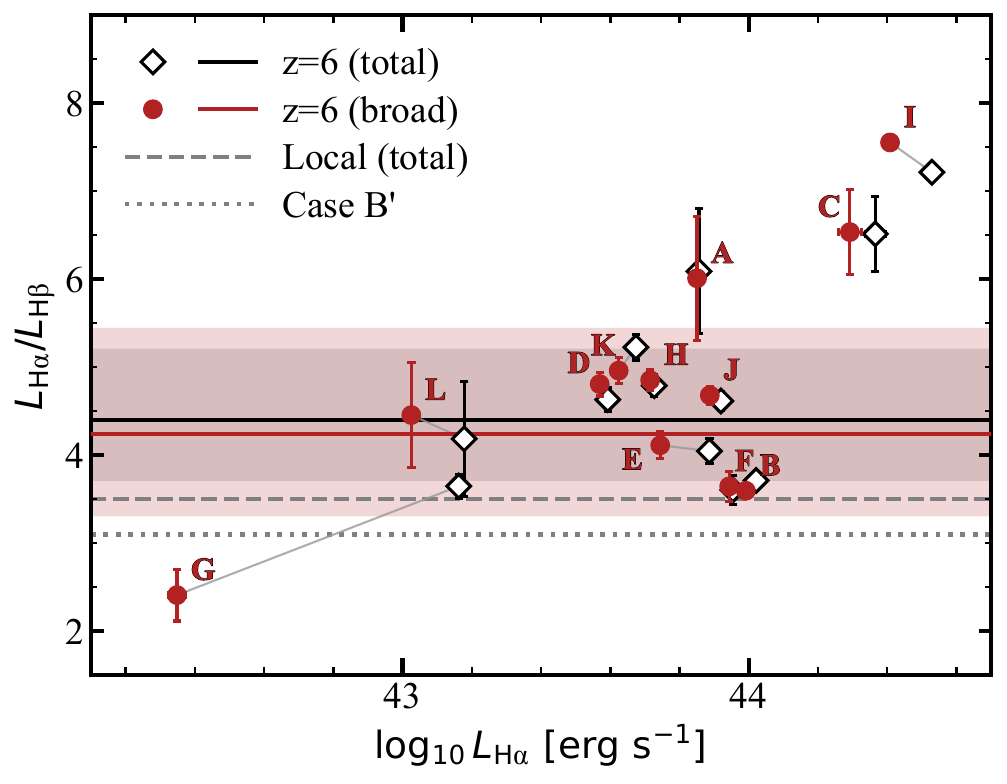}
    \caption{
    Balmer decrements, $L_{\rm H\alpha}/L_{\rm H\beta}$, as a function of the H$\alpha$ luminosity.
    The symbols are the same as in Figure~\ref{fig:Balmer_luminosity}, and the gray lines connect the total and broad-component measurements for the same object.
    The solid black and red lines represent the linear regression for the total and  broad emission-line luminosities, respectively, with the shaded regions indicating their uncertainties.
    The dashed line represents the local relation for the total luminosities \citep{GH05}, while the gray dotted line indicates the fiducial Case B$'$ ratio for AGN NLR gas \citep[$L_{\rm H\alpha}/L_{\rm H\beta}=3.1$;][]{Halpern83, Osterbrock06}.
    }
    \label{fig:Balmer_luminosity_HaHb}
\end{figure}

We show in Figure~\ref{fig:Balmer_luminosity_HaHb} the luminosity ratios of H$\alpha$ and H$\beta$ of the JWST targets as a function of H$\alpha$ luminosity.
We perform linear regression for both broad-line luminosity and total line luminosity, namely
\begin{equation}
    \log_{10} L_{\rm H\alpha} = a + \log_{10} L_{\rm H\beta},
\end{equation}%
where $10^a$ represents the H$\alpha$/H$\beta$ luminosity ratio.
We fix the slope to unity, because our JWST sample spans a limited luminosity range and the sample size is small.
The best-fit H$\alpha$/H$\beta$ luminosity ratios are $4.35\pm0.26$ for the broad components, and  $4.41\pm0.23$ for the total luminosity.
These values exceed the intrinsic ratio commonly assumed for AGN narrow-line-region (NLR) gas ($=3.1$; \citealt{Halpern83, Osterbrock06}) as well as the typical broad-line or total H$\alpha$/H$\beta$ ratios of low-redshift  quasars ($\approx3.5$; e.g.,  \citealt{GH05}, \citealt{LaMura07}, \citealt{Mejia-Restrepo22}, \citealt{Son25}).

These elevated H$\alpha$/H$\beta$ ratios are difficult to explain solely by dust attenuation. 
The mean broad-line ratio implies a $V$-band extinction of $\approx 1$ mag when attributed to dust reddening.
Under the extinction curve of the Small Magellanic Cloud (SMC) commonly adopted for quasars \citep{Richards03, Hopkins04, Pei_SMC}, the corresponding attenuation at Ly$\alpha$ wavelength is $\approx 5$ mag, which is difficult to reconcile  with the detection of broad Ly$\alpha$ emission and the high Ly$\alpha$ luminosities of these objects ($L_{\rm Ly\alpha}\gtrsim10^{44}\ {\rm erg\ s^{-1}}$).
Recent JWST studies have suggested that high-redshift sources may have lower $A_\lambda/A_V$ at UV wavelengths than predicted by local extinction curves \citep{Markov25}, which could reduce the inferred attenuation at Ly$\alpha$.
Nevertheless, an extinction as large as $A_V\approx1$ mag would remain inconsistent with the blue continuum slopes of these objects discussed in Section~\ref{sec:continuum_slopes}.
Dust attenuation is therefore unlikely to be the sole origin of the elevated Balmer decrements.

\begin{figure}
    \centering
    \includegraphics[width=\linewidth]{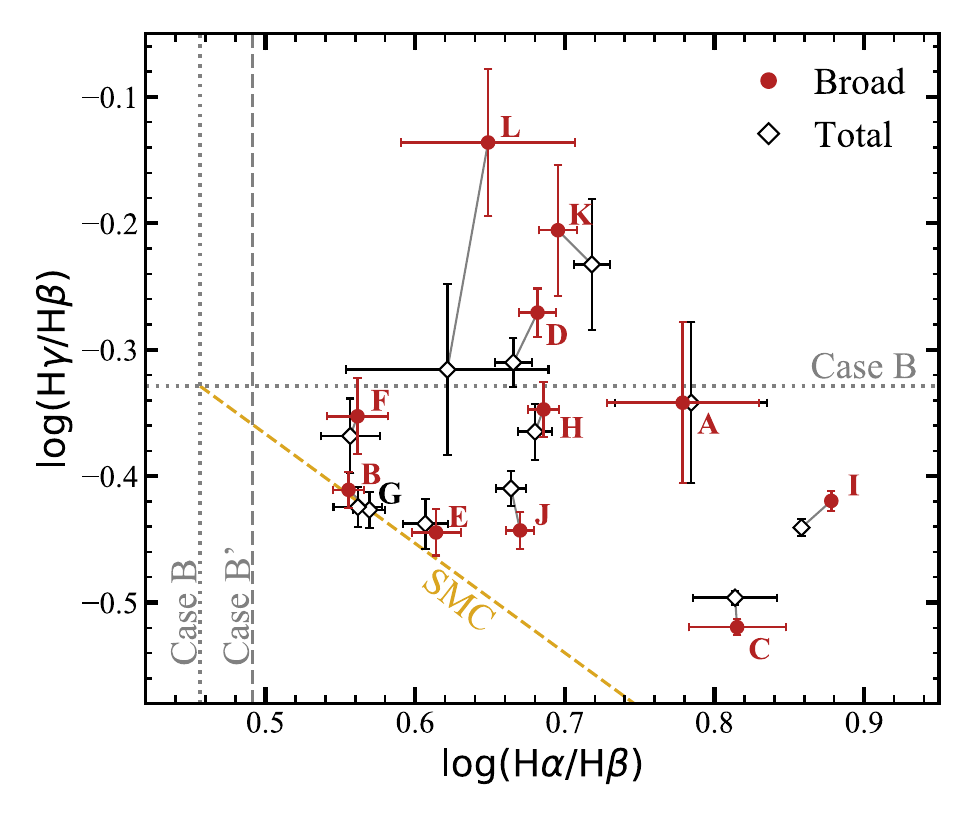}
    
    \caption{H$\alpha$/H$\beta$ and H$\gamma$/H$\beta$ ratios.
    Red filled circles represent the broad emission lines, while black open diamonds show the total emission lines. 
    Gray dashed lines connect the two measurements for each object.
    Each object is annotated as in Figure~\ref{fig:continuum}.
    The vertical and horizontal dotted lines indicate the Case~B values, while the vertical dashed line shows the intrinsic NLR Balmer decrement of H$\alpha$/H$\beta=3.1$.
    The orange dashed line indicates the expected line ratios under the SMC extinction curve \citep{Pei_SMC}.
    }
    \label{fig:BD}
\end{figure}

We further illustrate this point in Figure~\ref{fig:BD}, where we compare Balmer decrements measured from different line combinations.
The broad H$\alpha$/H$\beta$ ratios are generally higher, at a given H$\gamma$/H$\beta$ ratio, than predicted by the SMC extinction curve.
This offset suggests that the enhanced H$\alpha$/H$\beta$ ratios may reflect non-recombination processes in the dense BLR gas.
For example, collisional excitation from the $n=2$ to the $n=3$ level can preferentially enhance H$\alpha$ emission \citep{Osterbrock06}.
\citet{Son25} also argue that H$\alpha$ responds less rapidly to continuum variability than H$\beta$  because of its higher optical depth.
The enhanced H$\alpha$/H$\beta$ ratios of these intermediate-luminosity quasars could partly arise from a decline in the AGN continuum luminosity from a more luminous phase.

The most extreme Balmer decrements are observed in J0844$-$0132 and J1146$-$0005, the two LRD-like objects discussed in Section~\ref{sec:continuum_slopes}, which both exhibit H$\alpha$/H$\beta\approx7$.
These objects are also outliers in the $L_{5100}$ versus $L_{\rm Balmer,broad}$ relations.
Their Balmer decrements are reminiscent of those measured for LRDs, which typically exhibit  H$\alpha$/H$\beta \approx 8.7$  \citep{DeGraaff25b}, although J0844$-$0132 and J1146$-$0005 do not follow the correlation between $L_{5100}$ and H$\alpha$/H$\beta$ reported for their LRDs.
Several models have been proposed to explain the extremely high Balmer decrements of LRDs, especially the ``black hole star'' model, in which young black holes are surrounded by dense gaseous envelopes \citep{Inayoshi_Maiolino25, Kido25, YanZ25, DeGraaff25b, Naidu25}.
In this framework, H$\beta$ photons in the envelope undergo repeated resonant scattering  and can be converted into H$\alpha$ and Pa$\alpha$ photons, resulting in an enhanced H$\alpha$/H$\beta$ ratio.
Further investigation, including follow-up observations of reprocessed dust emission, is required to address the origin of their high Balmer decrements and to assess both their similarity to the LRD population and a possible evolutionary connection between the two populations.
With these uncertainties in mind, we do not apply dust attenuation corrections to the observed broad-line luminosities of the JWST targets in the following discussion.

Finally, we note that a few objects show H$\gamma$/H$\beta$ ratios higher than the intrinsic value under the Case~B recombination.
While H$\gamma$ is clearly detected in these objects, the lower signal-to-noise ratios of these higher-order transitions, together with the difficulty in modeling the continuum and pseudo-iron emission over a wide wavelength range, may introduce additional systematic uncertainties in the measured line fluxes.
We therefore do not overinterpret these elevated H$\gamma$/H$\beta$ ratios.
Higher-quality and higher-resolution spectroscopy is required to robustly model weak Balmer lines and decompose them into broad and narrow components.

\subsection{Notes on narrow-line quasars} \label{sec:NLQ}

Three of the JWST targets analyzed in this paper, J0844$-$0132, J1146$-$0005, and J0217$-$0208, are also included in \citet{Matsuoka25}, who present JWST/NIRSpec spectroscopy of a subset of the SHELLQs quasars that show relatively narrow Ly$\alpha$ (FWHM $<500$ km~s$^{-1}$).
About two-thirds of their targets exhibit clear broad components in Balmer emission lines, securing their nature as dust-obscured quasars, while the others are likely dominated by galaxy emission.

We note that the fitting methodology of the NIRSpec spectra is different  between this work and \citet{Matsuoka25}.
In their analysis, permitted lines are modeled using three components: narrow, intermediate, and broad. 
Forbidden lines are modeled using only the narrow and intermediate components, with the line widths and velocity shifts tied across lines within each component.
In this framework, the intermediate component is interpreted as galactic outflowing gas, and is seen in [\ion{O}{3}] and H$\alpha$ for all objects, and marginally in [\ion{N}{2}]. 
For J0844$-$0132 and J1146$-$0005, this treatment assigns part of the apparently broad Balmer emission to the intermediate component, resulting in narrower broad-line widths   by 40--80\%\ than in our fits.
\citet{Matsuoka25} also correct the quasar emission for dust extinction using the Balmer decrements of the broad-line components, assuming an intrinsic H$\alpha$/H$\beta$ ratio of 2.86, when estimating the line luminosities and SMBH properties (see Section~\ref{sec:Balmer_Decrement}).
The different fitting methods also affect whether J0217$-$0208 is interpreted as exhibiting broad emission from the BLR gas. 
We revisit this model-dependent interpretation of the J0217$-$0208 spectrum in the next section.

\subsection{Balmer Line Decomposition for J0217$-$0208} \label{sec:J0217_comp}

\begin{figure*}[bt]
\centering
\includegraphics[width=\linewidth]{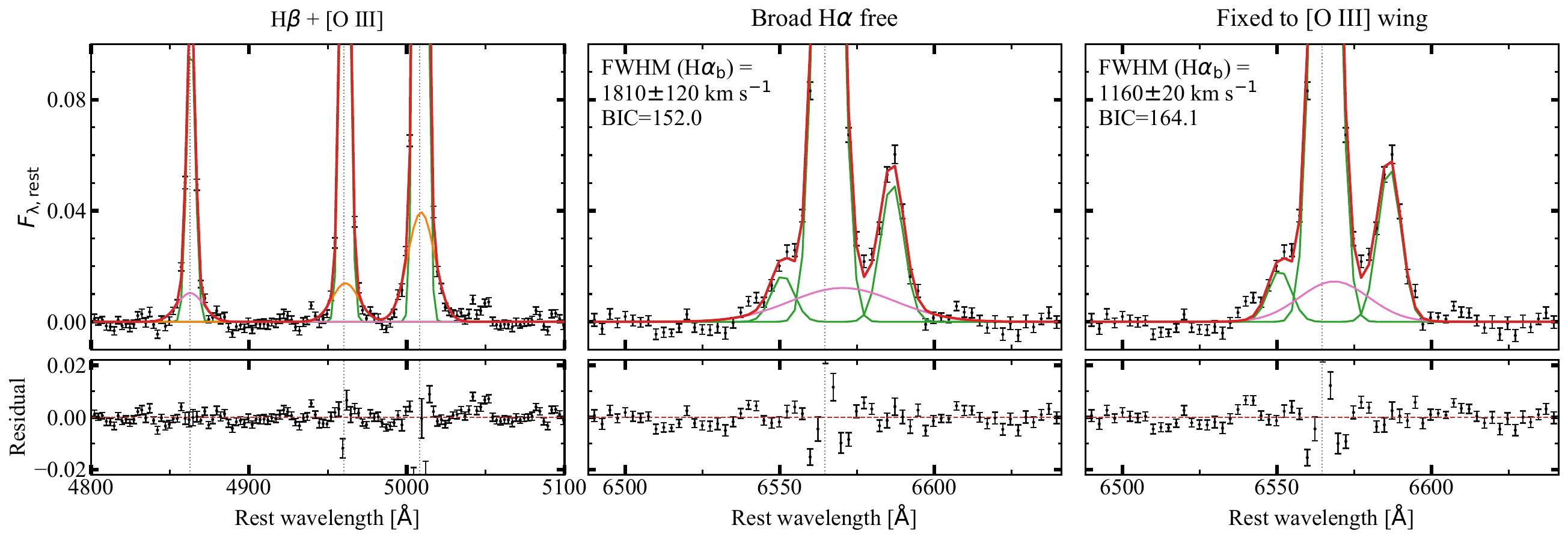}
\caption{Balmer emission line fitting for J0217$-$0208. 
(\textit{Left:}) H$\beta$+[\ion{O}{3}] complex. 
The line colors are the same as in Figure~\ref{fig:specfit_JWST}.
(\textit{Middle \& Right:})
H$\alpha$+[\ion{N}{2}] complex. 
The middle panel shows the fit in which broad H$\alpha$ is modeled without any constraint on its FWHM.
The right panel shows the fit in which the FWHM of broad H$\alpha$ is fixed to that of the [\ion{O}{3}] wing.
In each panel, the continuum-subtracted spectrum is shown in black, with error bars indicating the $1\sigma$ flux uncertainty in each pixel.
The broad H$\alpha$ FWHM and BIC are indicated at the top left of each panel.
The bottom panels show the residuals.
}\label{fig:J0217_2comp}
\end{figure*}

The line decomposition of J0217$-$0208 is uncertain because its Balmer emission-line profiles are dominated by narrow lines compared to those of the other JWST targets.
Figure~\ref{fig:J0217_2comp} shows zoomed-in panels of the decomposed H$\beta$+[\ion{O}{3}] and H$\alpha$+[\ion{N}{2}] emission-line complexes.
Both the H$\beta$ and [\ion{O}{3}] doublet emission lines exhibit narrow and broader components. 
The broader components have ${\rm FWHM}~({\rm H\beta_{broad}})=1150~{\rm km~s^{-1}}$ and ${\rm FWHM}~({\rm [O~III]{wing}})=1160~{\rm km~s^{-1}}$, and contribute 23\% and 14\% of the total line fluxes, respectively.
These similar line widths may suggest that the broad H$\beta$ component traces the same outflowing ionized gas as the [\ion{O}{3}] wing component.

The middle panel of Figure~\ref{fig:J0217_2comp} shows that a broad H$\alpha$ component is required in our fiducial model.
Its ${\rm FWHM}({\rm H\alpha_{broad}})=1,810~{\rm km~s^{-1}}$ is derived without any constraints on its line profile.
The right panel of Figure~\ref{fig:J0217_2comp} shows an alternative model in which we fix the line width of the broader H$\alpha$ component to that of the [\ion{O}{3}] wing (${\rm FWHM}=1160~{\rm km~s^{-1}}$), which is a factor of 1.5 narrower than in the fiducial case.
We compare the two models using the Bayesian Information Criterion (BIC; \citealt{Schwarz78}) within $\Delta v=\pm3,000~{\rm km~s^{-1}}$ around H$\alpha$ and find that the first model yields a smaller BIC ($\Delta {\rm BIC}=-12.1$), favoring a BLR origin rather than an outflow for the broad component.
We therefore adopt the broad H$\alpha$ component as BLR emission in our fiducial interpretation. For consistency with our spectral decomposition, we also use the broad H$\beta$ component to estimate the SMBH properties in the following analysis.

However, the classification of J0217$-$0208 as a broad-line quasar remains marginal, given the complexity of its line profile and the modest BIC difference between the BLR and outflow models.
\citet{Matsuoka25} and \citet{MC25} reached different interpretations of the same spectrum using fitting frameworks that include additional H$\alpha$ and [\ion{N}{2}] components associated with outflowing gas, which are not included in our fiducial model.
In these frameworks, part of the flux attributed to broad H$\alpha$ in our model can instead be assigned to an outflow component, making its interpretation model dependent.
We nevertheless retain J0217$-$0208 in our broad-line quasar sample based on our fiducial fitting scheme, while acknowledging that it may instead be a star-forming galaxy with outflowing gas.
We also note that this object is included in the analysis of \citet{Silverman25}.
As discussed in Section~\ref{sec:BHmass}, the H$\beta$- and H$\alpha$-based $\mBH$ estimates differ by only 7\%, so the choice of mass estimator does not significantly affect the inferred SMBH properties.

\section{Subaru/MOIRCS}\label{sec:Sec2_MCS}

In addition to the quasars targeted with JWST, we observed nine SHELLQs quasars in the $K_s$ band with the Multi-Object Infrared Camera and Spectrograph \citep[MOIRCS;][]{Ichikawa06, Suzuki08}, a Cassegrain instrument on the 8.2\,m Subaru telescope.
These near-infrared observations enable single-epoch SMBH mass measurements based on the redshifted \ion{Mg}{2} emission line.
Combined with the sample presented in \citetalias{Onoue19}, there are 15 SHELLQs quasars with ground-based near-infrared spectroscopy.

\subsection{Target Selection}\label{sec:Sec_MCS_sample}

We selected our MOIRCS targets from the SHELLQs sample available at the time of our observations in the S18A--S20B semesters \citep{Matsuoka16, Matsuoka18a, Matsuoka18b}.
The targets were selected from a sub-sample of the SHELLQs quasars, for which \ion{Mg}{2}-based SMBH mass measurements are feasible from the ground.
The targets were required to have Ly$\alpha$ redshifts of $z \geq 6.2$ in order to avoid severe atmospheric absorption at the observed wavelength of the \ion{Mg}{2}\ peak.
We also imposed an absolute magnitude cut of $M_{1450} \leq -24$, corresponding roughly to the HSC $y$-band magnitude of $y \leq 23$.
This magnitude cut ensures sufficient signal-to-noise ratios for detecting the \ion{Mg}{2}\ emission line with the 8m-class Subaru telescope.
We excluded SHELLQs quasars that had already been observed with near-infrared spectroscopy in previous studies \citep{Onoue19, Matsuoka19}.

During our observing runs,  we prioritized $6.2\leq z \leq 6.5$ SHELLQs quasars that were used in the study of the $z\sim6$ quasar luminosity function \citep{Matsuoka18c}.
One of the targets, J1217$+$0131 at $z=6.20$, was independently discovered by Pan-STARRS1 \citep{Banados16} and the DECam Legacy Survey \citep{Wang17}, and later recovered by our HSC photometric selection \citep{Matsuoka18b}.

In all, we observed nine SHELLQs quasars with Subaru/MOIRCS. 
These targets are presented in the second half of Table~\ref{tab:targets_all}. 
They occupy the brighter part of this distribution, as shown in Figure~\ref{fig:z_M1450}.
In Table~\ref{tab:targets_all}, we report the PSF magnitudes of these targets in the HSC $y$-band from the internal DR4 HSC-SSP catalog.

This new sample, together with the previous studies of \citet{Onoue19}, \citet{Willott10a}, and \citet{Shen19}, completes the near-infrared spectroscopic coverage of the luminosity-function sample of \citet{Matsuoka18c} at $6.2 \leq z \leq 6.5$ and $M_{1450} \leq -24$. 
Follow-up near-infrared spectroscopy remains incomplete at $z > 6.5$.
The JWST sample presented in Section~\ref{sec:Sec2_NIRSpec} covers the fainter SHELLQs quasars in \citet{Matsuoka18c}.
We will use this well-defined $z\sim6$ sample with black hole mass estimates for obtaining the $z\sim6$ black hole mass function in our future work (Onoue et al. in prep.).

\subsection{$K$-band spectroscopy}\label{sec:Sec_MCS_obs}

Our MOIRCS observations, the upgraded 2k $\times$ 2k Hawaii 2RG detectors of which were installed in 2015 \citep{Walawender16, Fabricius16}, were carried out over multiple observing runs: 2018 July 7--8 (Program ID: S18A-061), 2019 April 21--23 (S19A-015), 2020 August 11 (S20B-114) and 2021 January 31 (S20B-114).
The observations were performed in the multi-object spectroscopy (MOS) mode for secure acquisition of our faint targets in the narrow MOS slits, guided by bright alignment stars.
We chose the VPH-$K$ grism \citep{Ebizuka11} to cover 2.0 -- 2.4 \micron\  at a spectral resolution of $R=1700$ with a slit width of 0\arcsec.8. 
We made a custom MOS mask for each target quasar with filler sources and alignment stars.
The wavelength coverage of the obtained spectrum of each target depends on the location on the MOS mask.
The observations of J1217$+$0131 were carried out with a preset long-slit mask, for which the target acquisition was performed a blind offset from a nearby star.
The total integration time for each target was 0.9--4.5 hours, depending on the faintness and the visibility of the targets during the observing runs.
Those integration times were divided into pairs of 240 seconds (S18A and S19A runs) or 300 seconds (S20B) exposures to subtract the time-varying sky background with the standard ABBA nodding offsets. 
The MOS masks were re-aligned every $\sim1.5$ hours to make sure that the targets were at the centers of the target slits during integration.

We also took 5--10 minute $Ks$-band imaging data of each target during the same run as the spectroscopic observations.
We use the $Ks$-band $2\arcsec$-diameter aperture photometry to flux-calibrate the obtained spectrum of each target, so the flux calibration is not affected by potential quasar variability between the spectroscopic and imaging observations.
The data reduction of the imaging data used the {\it MCSALL} task of {\it MCSRED}\footnote{The software was developed by Ichi Tanaka: \url{https://www.naoj.org/staff/ichi/MCSRED/mcsred.html}}, the IRAF-based pipeline for MOIRCS imaging data. 
For each target the photometric zeropoint was determined with field 2MASS stars after converting their 2MASS $Ks$ magnitudes to the Maunakea Observatory filter system based on the prescription of \citet{Leggett06}.
The obtained $Ks$-band photometry as well as the rest-frame UV continuum slope $\alpha_{\rm \lambda, UV}$ based on HSC $y$-band and MOIRCS $Ks$-band photometry are reported in Table~\ref{tab:targets_all}.

The raw spectroscopic data were reduced in the standard manner using NOAO/IRAF \citep{Tody86}. 
We rectified the 2D spectra to correct for spectral tilt, and extracted 1D spectra using a Gaussian-weighted aperture.
The profile was determined by fitting a Gaussian  to each two-dimensional spectrum collapsed along the dispersion direction.
At each wavelength bin, the profile center and width were fixed, and only the normalization was scaled to derive the extracted flux.
The corresponding error spectra were estimated from the mean sky spectra.

\begin{figure*}[ht!]
\centering
 \includegraphics[width=\linewidth, trim=0 4cm 0 0, clip]{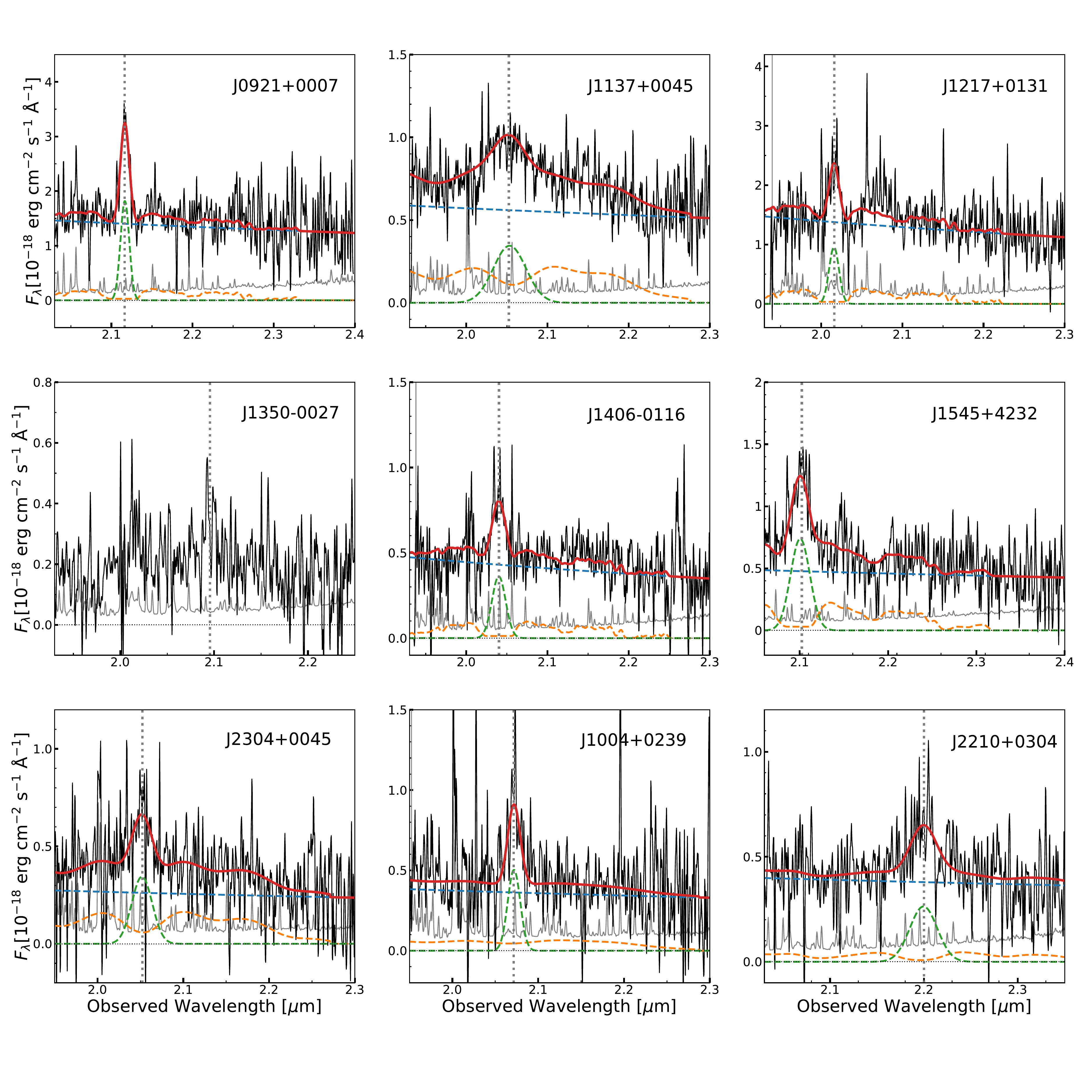}
 \caption{The $K$-band spectra of the 9 MOIRCS targets (black). 
For visualization purposes, the spectrum in each panel is smoothed using a Gaussian kernel with $\sigma = 2$ pixels.
The noise spectrum is shown in grey.
The best-fit power-law continuum, iron, and \ion{Mg}{2} emission line models are shown as blue, orange, and green dashed curves, respectively, and their sum is shown as solid red curves.
The iron model in this figure is based on the I Zw 1 template by \citet{Vestergaard01}.
The vertical black dotted lines show the expected \ion{Mg}{2} centers from their systemic redshifts.
The fitting model for J1350$-$0027, which shows an absorption feature at the line peak, is presented in Figure~\ref{fig:spec_J1350}.
} \label{fig:spec}
\end{figure*}

Telluric absorption was corrected using spectra of A-type standard stars observed immediately before or after each science exposure through the same MOS slit as the target quasar.
The telluric-corrected one-dimensional spectra were then flux-calibrated by scaling to the MOIRCS $K_s$-band magnitudes.
The spectra were subsequently resampled with a 4~{\rm \AA} step, corresponding to approximately two pixels.
Spectra obtained on different observing dates were stacked at this stage.
The final reduced spectra are presented in Figure~\ref{fig:spec}.

\begin{deluxetable*}{lCCCC}[tb]
\tabletypesize{\small}
\tablecaption{\ion{Mg}{2} line properties for the MOIRCS targets \label{tab:MgII_VO09}}
\tablehead{
\colhead{ID} &
\colhead{$z_{\rm MgII}$} &
\colhead{$L_{3000}$} &
\colhead{{\rm FWHM}(Mg\,\textsc{ii})} &
\colhead{$L_{\rm MgII}$} \\
\colhead{} &
\colhead{} &
\colhead{[10$^{45}$ erg s$^{-1}$]} &
\colhead{[km s$^{-1}$]} &
\colhead{[10$^{43}$ erg s$^{-1}$]}
}
\startdata
J2210+0304 & 6.859_{-0.005}^{+0.006} & 4.7 \pm 0.4 & 4700_{-900}^{+800} & 5.3_{-0.7}^{+0.9} \\
J0921+0007 & 6.562_{-0.001}^{+0.001} & 14.5 \pm 0.5 & 1800_{-100}^{+100} & 12.1_{-0.8}^{+1.3} \\
J1545+4232 & 6.505_{-0.002}^{+0.002} & 4.9 \pm 0.3 & 3700_{-200}^{+200} & 9.8_{-0.6}^{+0.7} \\
J1350$-$0027 & 6.49                  & 2.3 \pm 0.1 & 370_{-100}^{+60}     & 0.73_{-0.12}^{+0.10} \\
J1004+0239 & 6.402_{-0.003}^{+0.003} & 3.5 \pm 0.2 & 2500_{-300}^{+1100} & 4.3_{-0.4}^{+1.1} \\
J1137+0045 & 6.347_{-0.003}^{+0.004} & 3.1 \pm 0.2 & 6800_{-300}^{+200} & 12.0_{-0.5}^{+0.4} \\
J2304+0045 & 6.331_{-0.004}^{+0.005} & 2.5 \pm 0.1 & 4100_{-600}^{+300} & 4.6_{-0.4}^{+0.3} \\
J1406$-$0116 & 6.289_{-0.003}^{+0.003} & 3.7 \pm 0.1 & 2900_{-300}^{+200} & 3.4_{-0.3}^{+0.4} \\
J1217+0131 & 6.202_{-0.003}^{+0.003} & 11.5 \pm 0.2 & 2100_{-200}^{+200} & 6.3_{-0.8}^{+0.8} \\
\enddata
\tablecomments{
The line properties reported here are based on \citet{Vestergaard01} iron template.
The same table for the case when we use \citet{Tsuzuki06} iron template is reported in Appendix~\ref{sec:BHmass_T06}.
$L_{3000}$ is the monochromatic luminosity at rest-frame 3000~\AA\ in the unit of $10^{45}\ {\rm erg\ s^{-1}}$. 
The \ion{Mg}{2} emission line of J1350$-$0027 is fitted with two Gaussian profiles (see main text for details).
}
\end{deluxetable*}

\subsection{Spectral Fitting}\label{sec:Sec_MCS_fit}

\subsubsection{Continuum + Iron}\label{sec:Sec_MCS_fit_cont}
We modeled the obtained $K$-band spectra of the MOIRCS targets with three components: power-law continuum, the iron emission line forest and the \ion{Mg}{2} emission line, following the methodology described in \citetalias{Onoue19}.
The rest-frame UV continuum from the accretion disk is modeled with a power-law:
\begin{equation}
 F_\mathrm{\lambda, cont}=F_0 \ \lambda^{\alpha_\lambda},
\end{equation}%
where $F_0$ is the amplitude and $\alpha_\lambda$ is the power-law index.
The slope of the continuum cannot be well constrained from the MOIRCS spectra alone owing to their narrow wavelength coverage.
We instead estimated the slopes from the broad-band photometry at the HSC $y$-band and the MOIRCS $Ks$-band, assuming that the broad-band magnitudes are dominated by the continuum.
The estimated slope values are reported in Table~\ref{tab:targets_all}.

Rest-frame iron emission lines are also taken into account as a pseudo-continuum component at rest-frame wavelengths 2000--3000 \AA.
We use two iron templates in this work: the first from \citet{Vestergaard01} and the second from \citet{Tsuzuki06}.
Both iron templates are based on a high-resolution spectrum of I Zw 1, a local narrow-line Seyfert 1 galaxy.
The difference of the two templates and its effect on the \ion{Mg}{2}\ and \feii\ measurements have been widely discussed in the literature \citep[e.g.,][]{Sameshima17, Woo18, Onoue20, Schindler20, Yang21}.
In short, the \ion{Mg}{2}-based single-epoch BH mass measurements are calibrated with the \citet{Vestergaard01} template, while the template from \citet{Tsuzuki06} is superior in terms of the accuracy of the \feii\ flux beneath the \ion{Mg}{2}\ line and thus the total \ion{Mg}{2}\ flux.
As discussed in Section~\ref{sec:BHmass}, we use the iron template from  \citet{Vestergaard01} in our primary models for BH mass measurements, while we also report our models with the \citet{Tsuzuki06} template to report accurate \ion{Mg}{2} properties.
Following \citet{Kurk07}, we added 20\% of the mean continuum flux density at $\lambda_\mathrm{rest}=2930$--2970 \AA\ to the \citet{Vestergaard01} template to effectively compensate for the \feii\ flux in the \ion{Mg}{2}\ region.

The iron emission was fitted using a grid of  broadened iron templates.
We convolved the original template with Gaussian kernels to produce templates with FWHM values ranging from 500 to 10,000 km s$^{-1}$ in steps of 500 km s$^{-1}$.
Each template was shifted to the systemic redshift of the quasar based on measurements of [\ion{C}{2}] 158~$\micron$ when available.
J1137+0045, J2304+0045, J1406$-$0116, and J1217+0131 were observed and analyzed in Sawamura, M. et al. (\textit{in prep.}), and J0921+0007 was observed in \citet{Wang24}.
If [\ion{C}{2}] observations were not available (J2210+0304, J1545+4232, J1350$-$0027, J1004+0239), we used Ly$\alpha$ redshifts as an initial guess of the systemic redshifts and iteratively updated them using the \ion{Mg}{2} redshifts.
For each assumed iron template, we alternately fitted the continuum model and the iron emission until the scale factor of the iron template converged ($<1\%$), following the description of \citet{Vestergaard01}.
We then adopted the model with the minimum $\chi^2$ among the converged fits as the best-fit iron model.
We do not consider the iron component for J1350$-$0027, which shows a noisy spectrum, and instead use a constant continuum model.

\subsubsection{\ion{Mg}{2} Line Fitting}\label{sec:Sec_MCS_fit_MgII}
The \ion{Mg}{2} emission line properties of each quasar were measured after subtracting the best-fit power-law+iron continuum model.
We detected significant flux excess for all targets at the wavelengths close to where \ion{Mg}{2} are expected based on the optical and [C{\sc ii}] redshifts.
We fitted a single Gaussian profile for each spectrum to measure the amplitude, central wavelength and line width.
The \ion{Mg}{2} line of J1350$-$0027 was modeled as resolved \ion{Mg}{2} $\lambda\lambda2797,2803$ doublet emission using two Gaussian components with a common FWHM (See Section~\ref{sec:MCS_targets_comments}).

The uncertainties of the continuum and emission line parameters are measured by Monte Carlo resampling.
From each unsmoothed flux-calibrated spectrum, $1,000$ mock spectra are generated by adding random noise to each spectral pixel based on its noise vector.
The mock spectrum is then smoothed by the same Gaussian kernels and the power-law+iron+\ion{Mg}{2} fitting was repeated
with the procedure described in Section~\ref{sec:Sec_MCS_fit_cont}.
The $1\sigma$ uncertainty is given by $16$\% and $84$\% percentiles of the distribution of the best-fit values.
The derived emission line properties including the redshift measurements are shown in Table~\ref{tab:MgII_VO09}.

We tested with the five objects with available [\ion{C}{2}] to fit the spectra with and without fixing the iron templates to [\ion{C}{2}] redshifts.
We find that the resulting \ion{Mg}{2} line widths change by only $< 1$\%: therefore, we confirm that the availability of [\ion{C}{2}]  redshift does not significantly affect our line measurements.

\subsection{Notes on Individual Objects} \label{sec:MCS_targets_comments}

\begin{figure}[tb!]
\centering
 \includegraphics[width=\linewidth]{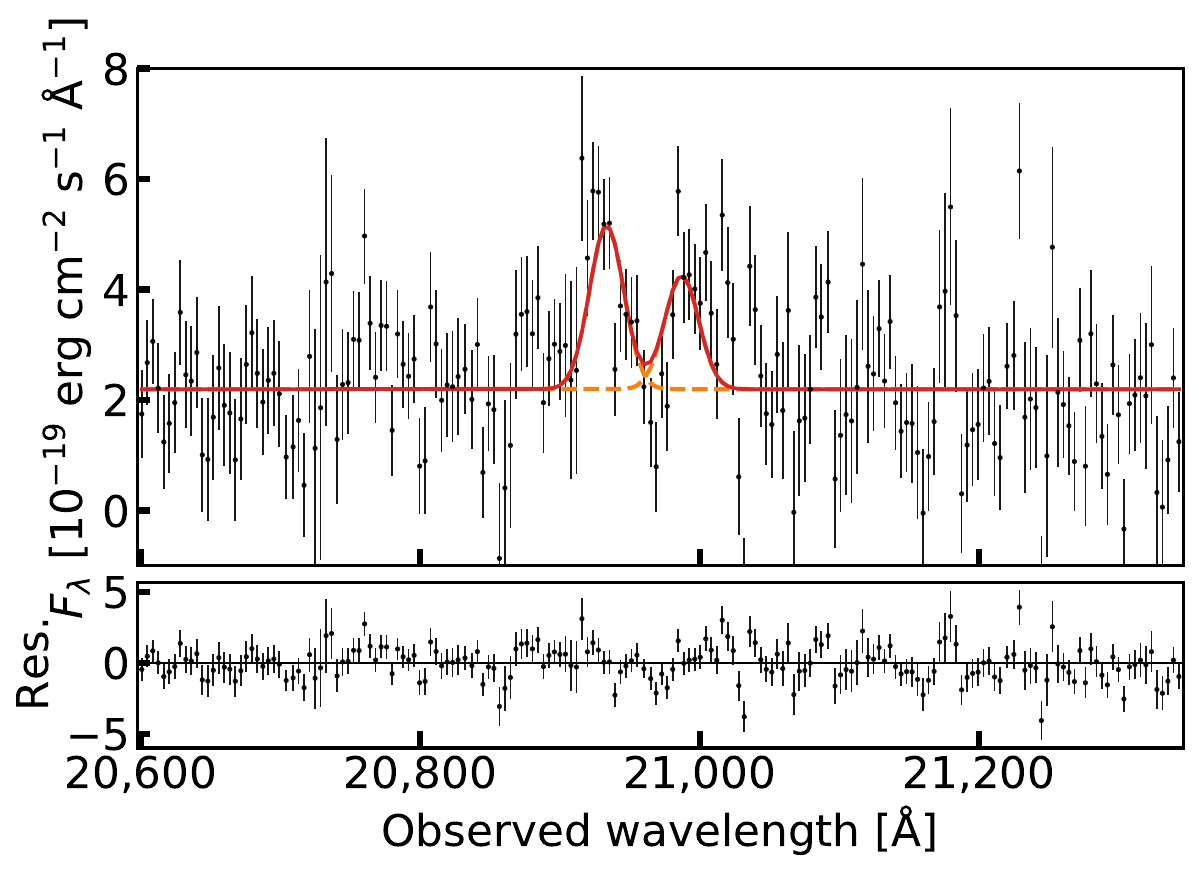}
\caption{The unsmoothed spectrum of J1350$-$0027. 
The spectrum is modeled assuming that the observed double-peaked feature corresponds to resolved \ion{Mg}{2} $\lambda\lambda2797,2803$ doublet emission.
Orange dashed curves indicate the two Gaussian components, and the red solid curve shows the combined model including a constant continuum.
The residuals are shown in the bottom panel.
} \label{fig:spec_J1350}
\end{figure}

\begin{description}
\item[J0921+0007]  This quasar exhibits a relatively narrow \ion{Mg}{2} emission line with FWHM $=1700\pm100~{\rm km\ s^{-1}}$. 
    It is marginally detected in rest-frame hard X-ray observations, $L_{\rm 2-10~kev} = (2.96 \pm 1.71) \times 10^{45}~{\rm erg~s^{-1}}$, by the extended ROentgen Survey with an Imaging Telescope Array (eROSITA), later confirmed by deeper Chandra observations  \citep{Wolf23}. 
    Its X-ray loudness relative to rest-frame ultraviolet luminosity makes it a high-redshift analog of a local Seyfert 1 galaxy. 
    The Eddington ratio of J0921+0007 is among the highest among our SHELLQs sample (See Section~\ref{sec:MBH_distribution}).
    
\item[J1350$-$0027]  This quasar is the faintest in  $Ks$-band among our MOIRCS targets ($22.48\pm0.14$ mag). 
    Figure~\ref{fig:spec} shows that J1350$-$0027 exhibits a double-peaked feature at \ion{Mg}{2}.
    This line profile may arise from either resolved \ion{Mg}{2} $\lambda\lambda2797,2803$ doublet emission or broad \ion{Mg}{2} absorption imprinted on an underlying broad emission component \citep[e.g.,][]{Hall02, Choi22}.

    Figure~\ref{fig:spec_J1350} shows  unbinned $Ks$-band spectrum of this source, in which \ion{Mg}{2} is reasonably well modeled with a resolved doublet emission model.
    The measured line FWHM of $370_{-100}^{+60}$ km~s$^{-1}$ is $40$\% narrower than that of Ly$\alpha$ ($620\pm200$ km~s$^{-1}$; \citealt{Matsuoka19b}).
    Within the SHELLQs sample, this object lies at the boundary of the definition of ``narrow-line'' quasars, namely Ly$\alpha$ FWHM $<500$ km~s$^{-1}$ \citep{Matsuoka19b}.
    \citet{Matsuoka25} showed from JWST rest-optical spectroscopy that this sub-population consists of a mixture of dust-reddened broad-line quasars and galaxies.
    These objects may also be related to the LRD population identified by JWST.

    It is therefore possible that J1350$-$0027 is a dust-reddened quasar, similar to those reported by \citet{Matsuoka25}, in which only narrow emission lines from ionized gas in NLR are visible in the rest-frame ultraviolet \citep{Onoue21}.
    However, given the low signal-to-noise ratio of the \ion{Mg}{2} spectrum and the absence of JWST spectroscopy currently available for this object,
    we cannot reliably distinguish between the dust-obscured quasar scenario and the low-ionization broad absorption line quasar scenario.
    We therefore exclude J1350$-$0027 from the subsequent statistical analysis of the SHELLQs broad-line quasars.
\end{description}

\section{BH mass and Eddington Ratio} \label{sec:BHmass}

Assuming that the BLR gas is gravitationally bound to the central black hole, the virial SMBH mass is given by:
\begin{equation}
M_\mathrm{BH}=f G^{-1}  v^2_\mathrm{BLR} R_\mathrm{BLR},
\end{equation}%
where $G$ is the gravitational constant, $v_\mathrm{BLR}$ is the circular velocity of the BLR gas, and $R_\mathrm{BLR}$ is its distance from the black hole.
The amplitude $f$ takes into account the geometry of the BLR gas.
In practice, these relations are implemented using the FWHM or velocity dispersion of broad emission lines as a proxy for $v_{\rm BLR}$, together with the continuum luminosity to estimate $R_{\rm BLR}$ via the BLR size-luminosity relation calibrated against reverberation mapping studies \citep{Kaspi00, Vestergaard02, Vestergaard06, Shen24}.

We use Balmer emission lines to estimate SMBH masses for our JWST targets. 
\citet{Vestergaard06} calibrate the H$\beta$-based $\mBH$ as
\begin{equation} \label{eq:V06}
M_{\rm BH} = 8.1 \times 10^6 \left(\frac{{\rm FWHM~{\rm (H\beta)}}}{10^3\ {\rm [km\ s^{-1}]}} \right)^{2} \left( \frac{L_{5100}}{10^{44}\ {\rm [erg\ s^{-1}]}}\right)^{0.5} ~\msun,
\end{equation}%
where FWHM~(H$\beta$) is the full width at half maximum of the broad H$\beta$ line, and $L_{5100}$ is the monochromatic luminosity at rest-frame $5100$~\AA.
We note that the observed 5100~\AA~ continuum luminosities of our JWST targets ($\sim10^{45}$ erg s$^{-1}$) are well covered by the local sample of \citet{Vestergaard06}.
The measurement uncertainties of the virial black hole masses are derived by propagating the measurement errors of the  line widths and the continuum model.
These uncertainties are smaller than the systematic uncertainty of $\sim0.4$~dex inherent to single-epoch virial mass estimates \citep{Shen13_review, Shen24}.

Alternatively, we  apply the virial SMBH mass recipe of \citet{GH05}, which is based on the broad H$\alpha$ line width and luminosity:
 \begin{equation} \label{eq:GH05}
 M_{\rm BH} = 2.0 \times 10^6 \left(\frac{{\rm FWHM~{\rm (H\alpha)}}}{10^3\ {\rm [km\ s^{-1}]}} \right)^{2.06} \left( \frac{L_{\rm H\alpha}}{10^{42}\ {\rm [erg\ s^{-1}]}}\right)^{0.55}   M_\odot.
 \end{equation} %
This widely-used $\mBH$ recipe leverages the tight correlation between $L_{5100}$ and $L_{\rm H\alpha}$ of their local sample, which makes it possible to estimate $\mBH$ solely from H$\alpha$.
These H$\beta$- and H$\alpha$-based $\mBH$ are summarized in the bottom half of Table~\ref{tab:mbh_jwst}.

\begin{figure}[bt]
\centering
\includegraphics[width=0.97 \linewidth]{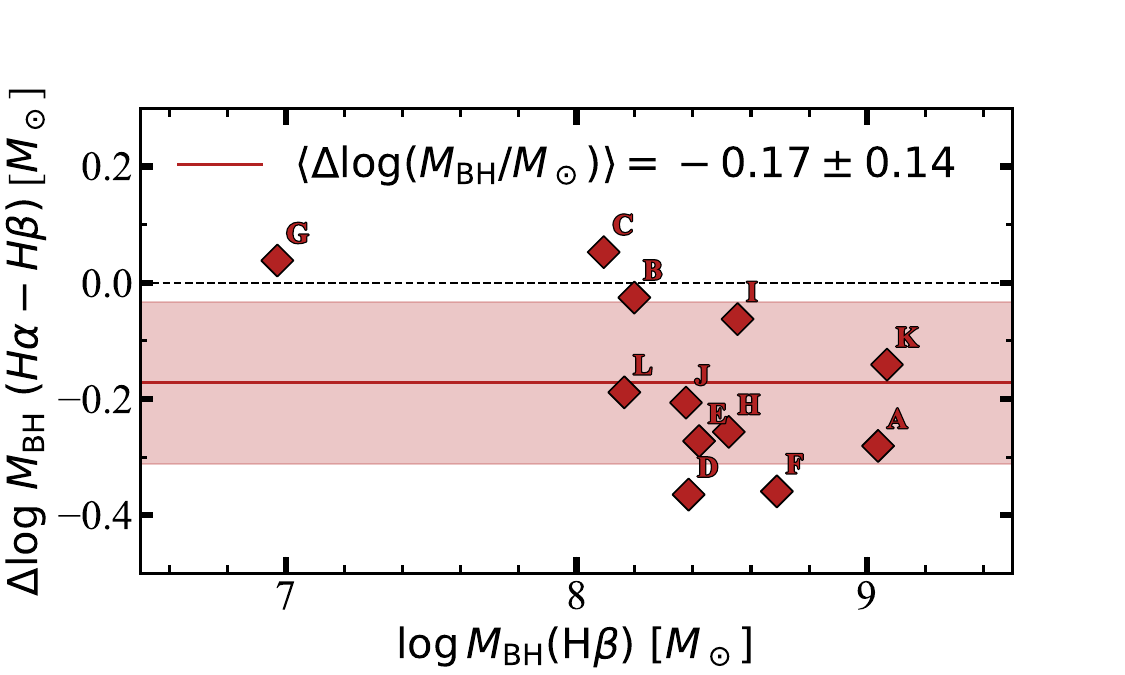}
\caption{Comparison of the H$\beta$-based and H$\alpha$-based $\mBH$ for the JWST targets.
Each object is annotated as in Figure~\ref{fig:continuum}.
The y-axis represents the difference between the H$\alpha$- and H$\beta$-based $\mBH$. 
The mean offset and the 1$\sigma$ scatter are shown by the red line and shaded region, respectively.
}\label{fig:delta_MBH}
\end{figure}

Figure~\ref{fig:delta_MBH} compares the  $\mBH$ measurements based on H$\beta$ (Equation~\ref{eq:V06}) and H$\alpha$ (Equation~\ref{eq:GH05}).
We find modest offsets of the H$\alpha$-based $\mBH$ with $\Delta \log \mBH = \log \mBH (H\beta) - \log \mBH (H\alpha) = -0.17 \pm0.14$.
This systematic offset is explained by the difference in the $L_{5100}$--$L_{\rm H\alpha}$ correlation between $z\sim6$ and low redshift (Figure~\ref{fig:Balmer_luminosity}), as well as by the different BLR geometries assumed in the two $\mBH$ recipes \citep[see, e.g.,][for detailed discussions]{Woo15, Mejia-Restrepo22}.
Nevertheless, the estimated masses are broadly consistent within the typical systematic uncertainty of single-epoch estimates ($\sim0.4$~dex; \citealt{Shen13_review}), and this small offset does not alter our key findings (Section~\ref{sec:MBH_Woo}).

For the \ion{Mg}{2}-based SMBH mass estimates of the MOIRCS targets, we adopt the prescription of \citet{Vestergaard09}:
\begin{equation}
M_\mathrm{BH} = 7.2\times 10^{6}
\left(\frac{\text{FWHM}~(\mathrm{Mg}\,\textsc{ii})}{10^3\ \mathrm{[km\ s^{-1}]}}\right)^2
\left(\frac{L_{3000}}{10^{44}\ \mathrm{[erg\ s^{-1}]}}\right)^{0.5} [M_\odot], \label{eq:VO09}
\end{equation}%
where FWHM~(\ion{Mg}{2}) is the full width at half maximum of the \ion{Mg}{2} line, and $L_{3000}$ is the monochromatic luminosity at rest-frame $3000$~\AA.

We use the \ion{Mg}{2} line width measurements from the line fitting with the \citet{Vestergaard01} iron template, to be consistent with \citetalias{Onoue19} and other similar works in the literature.
These $\mBH$ estimates based on \ion{Mg}{2}  emission lines are summarized in the first half of  Table~\ref{tab:mbh_jwst}.
The case when the \citet{Tsuzuki06} iron template is used is presented in Appendix~\ref{sec:BHmass_T06}.
In addition,  we will discuss the impact of the choice of different $\mBH$ calibration in Section~\ref{sec:MBH_Woo}.

The luminosity-based Eddington ratios $\edd$ of our NIRSpec and MOIRCS targets are presented in Table~\ref{tab:mbh_jwst}.
For this purpose, the Eddington luminosity is given by $L_{\rm Edd} = 1.3\times 10^{38} \left(\mBH / M_\odot\right)$ erg s$^{-1}$.
The H$\beta$-based $\mBH$ is used for the JWST targets.
The bolometric luminosity $L_{\rm bol}$ is estimated by scaling the UV or optical continuum luminosity of each target, adopting the standard bolometric corrections for quasars presented in \citet{Richards06a}.
For the JWST targets, we use the host-subtracted 5100~\AA\ luminosity,
\begin{equation}
L_\mathrm{bol} = 9.26~ L_{\rm 5100, QSO}~{\rm [erg\ s^{-1}]}.  \label{eq:Lbol_Hb}
\end{equation}
For the MORICS targets, we use the 3000~\AA\ luminosity,
\begin{equation}
L_\mathrm{bol} = 5.15~ L_{3000}~{\rm [erg\ s^{-1}]}. \label{eq:Lbol_MgII}
\end{equation}

Finally, we note that our SHELLQs quasars have either \ion{Mg}{2} or Balmer line observations, but none of them has both.
This is partly because the JWST targets in this study are too faint for ground-based \ion{Mg}{2} spectroscopy.
Early high-redshift JWST studies have shown overall agreement between \ion{Mg}{2}- and H$\beta$-based $\mBH$ estimates, although offsets of up to $\sim0.5$ dex have been reported for individual objects \citep{Eilers23, Yang23, Lyu25}.
Such differences are, however, comparable to the typical uncertainties of single-epoch mass estimates \citep{Shen13_review}.
It therefore remains unclear whether these BLR emission lines trace the same BLR kinematics as in the local universe and whether locally calibrated single-epoch $\mBH$ estimators can be reliably applied at high redshift.

\begin{deluxetable*}{lCCCCCC}[p!]
\tabletypesize{\small}
\tablecaption{Black Hole Mass Estimates\label{tab:mbh_jwst}}
\tablewidth{0pt}
\tablehead{
\colhead{ID} &
\colhead{$M_{\rm BH, H\beta}$} &
\colhead{$M_{\rm BH, H\alpha}$} &
\colhead{$M_{\rm BH, Mg\,II}$} &
\colhead{$L_{\rm bol}/L_{\rm Edd}$} &
\colhead{$L_{\rm bol}$} \\
\colhead{} &
\colhead{[$10^8 M_\odot$]} &
\colhead{[$10^8 M_\odot$]} &
\colhead{[$10^8 M_\odot$]} &
\colhead{} &
\colhead{[10$^{45}$ erg s$^{-1}$]}
}
\startdata
\multicolumn{6}{c}{\it JWST/NIRSpec} \\
\hline
J2255+0251 & $1.58 \pm 0.06$ & $1.49 \pm 0.05$ & \nodata & $0.590 \pm 0.023$ & $12.12 \pm 0.08$ \\
J2236+0032 & $10.90 \pm 2.86$ & $5.71 \pm 0.21$ & \nodata & $0.116 \pm 0.031$ & $16.47 \pm 0.11$ \\
J0844$-$0132 & $3.58 \pm 0.09$ & $3.10 \pm 0.04$ & \nodata & $0.225 \pm 0.005$ & $10.46 \pm 0.03$ \\
J0844$-$0052 & $4.89 \pm 0.41$ & $2.14 \pm 0.09$ & \nodata & $0.227 \pm 0.019$ & $14.46 \pm 0.11$ \\
J0918+0139 & $3.34 \pm 0.23$ & $1.85 \pm 0.05$ & \nodata & $0.231 \pm 0.016$ & $10.02 \pm 0.06$ \\
J1425$-$0015 & $2.38 \pm 0.08$ & $1.48 \pm 0.04$ & \nodata & $0.307 \pm 0.011$ & $9.51 \pm 0.07$ \\
J1525+4303 & $2.43 \pm 0.11$ & $1.05 \pm 0.03$ & \nodata & $0.171 \pm 0.008$ & $5.39 \pm 0.05$ \\
J1146$-$0005 & $1.24 \pm 0.07$ & $1.40 \pm 0.53$ & \nodata & $0.352 \pm 0.019$ & $5.66 \pm 0.01$ \\
J1146+0124 & $2.64 \pm 0.16$ & $1.41 \pm 0.12$ & \nodata & $0.263 \pm 0.016$ & $9.04 \pm 0.04$ \\
J0217$-$0208 & $0.0934 \pm 0.0002$ & $0.10 \pm 0.01$ & \nodata & $0.654 \pm 0.004$ & $0.794 \pm 0.004$ \\
J1512+4422 & $11.70 \pm 0.49$ & $8.46 \pm 0.03$ & \nodata & $0.057 \pm 0.002$ & $8.65 \pm 0.04$ \\
J0911+0152 & $1.46 \pm 0.45$ & $0.95 \pm 0.04$ & \nodata & $0.079 \pm 0.025$ & $1.50 \pm 0.01$ \\
\hline
\multicolumn{6}{c}{\it Subaru/MOIRCS} \\
\hline
J2210+0304 & \nodata & \nodata & $10.90_{-4.95}^{+5.52}$ & $0.17_{-0.06}^{+0.14}$ & $24.0 \pm 1.6$ \\
J0921+0007 & \nodata & \nodata & $2.81_{-0.31}^{+0.41}$ & $2.05_{-0.34}^{+0.29}$ & $74.9 \pm 2.8$ \\
J1545+4232 & \nodata & \nodata & $6.86_{-0.82}^{+0.72}$ & $0.28_{-0.04}^{+0.04}$ & $25.1 \pm 1.5$ \\
J1004+0239 & \nodata & \nodata & $2.65_{-0.87}^{+1.82}$ & $0.52_{-0.21}^{+0.26}$ & $18.1 \pm 0.9$ \\
J1137+0045 & \nodata & \nodata & $25.40_{-3.71}^{+2.38}$ & $0.08_{-0.01}^{+0.01}$ & $27.5 \pm 1.0$ \\
J1406$-$0116 & \nodata & \nodata & $3.79_{-0.70}^{+0.61}$ & $0.39_{-0.06}^{+0.09}$ & $19.2 \pm 0.5$ \\
J2304+0045 & \nodata & \nodata & $6.03_{-1.65}^{+1.77}$ & $0.16_{-0.04}^{+0.06}$ & $12.7 \pm 0.6$ \\
J1217+0131 & \nodata & \nodata & $3.55_{-0.73}^{+0.69}$ & $1.29_{-0.20}^{+0.34}$ & $59.4 \pm 1.2$ \\
\enddata
\tablecomments{
The uncertainties of the SMBH mass estimate reported here do not include 0.4 dex of intrinsic scatter of the virial $\mBH$ estimates \citep{Shen13_review}.
For JWST targets, H$\beta$-based $\mBH$ is based on the host subtracted 5100 \AA\ luminosity $L_{\rm 5100, QSO}$.
The Eddington ratio $\edd$ are based on the H$\beta$-based $\mBH$.
The bolometric luminosities $L_{\rm bol}$ are from $L_{\rm 5100, QSO}$ and $L_{\rm 3000}$ for JWST and MOIRCS targets, respectively.
}
\end{deluxetable*}

\begin{figure*}[htb]
\centering
\includegraphics[width=0.48\textwidth,trim=20 0 20 0,clip]{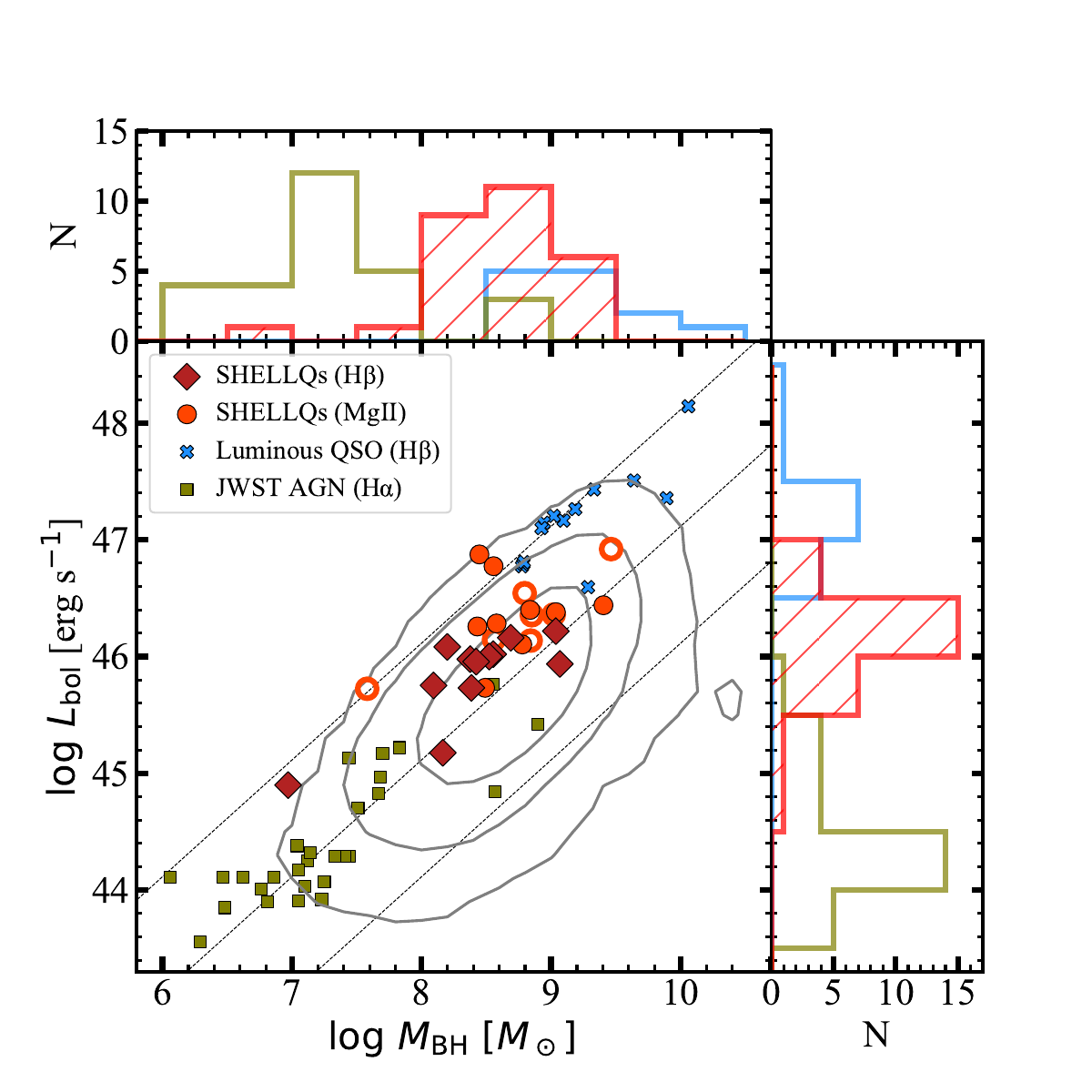}
\includegraphics[width=0.48\textwidth,trim=20 0 20 0,clip]{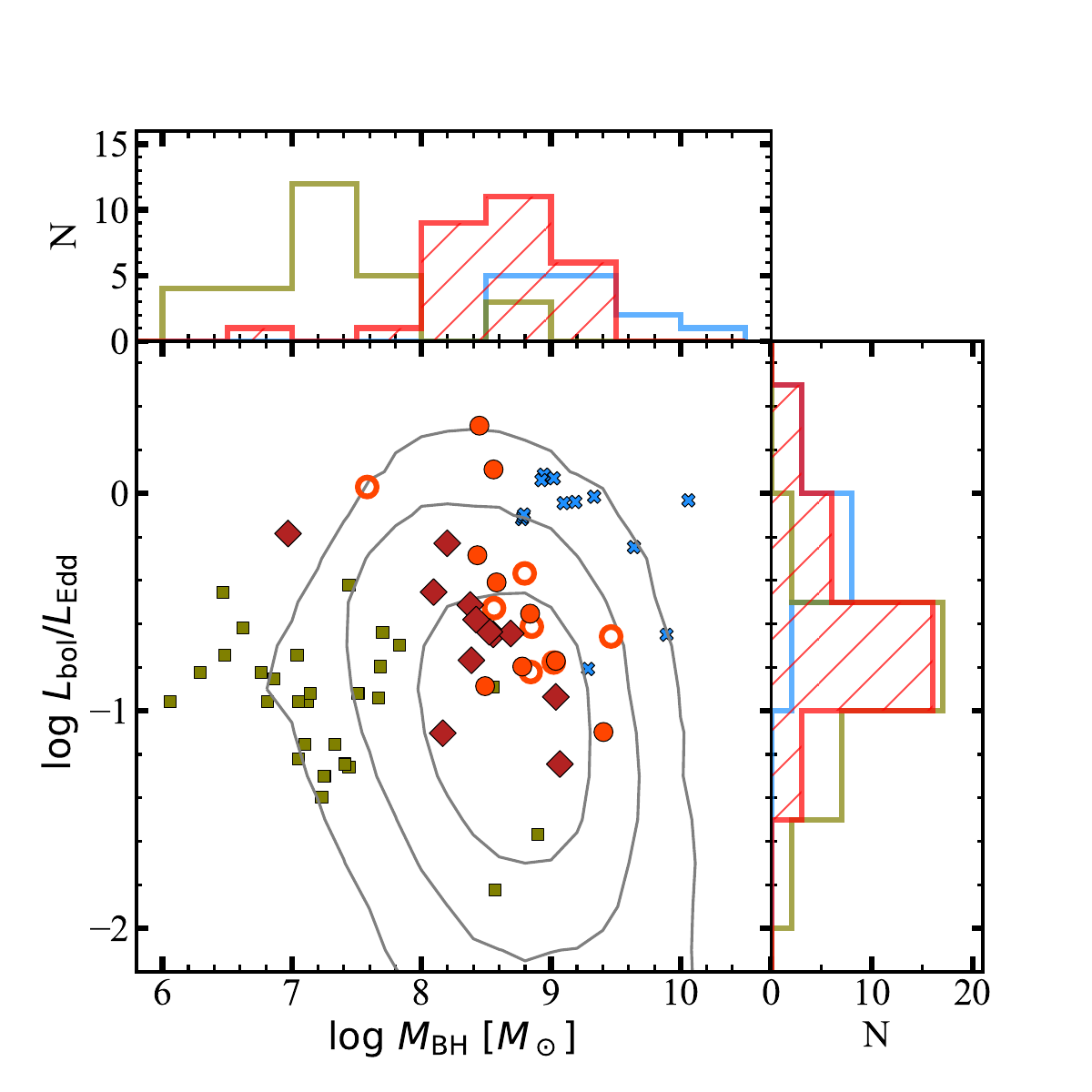}
\caption{(\textit{Left}) Black hole mass–bolometric luminosity distribution at $z\sim6$--7.
SHELLQs quasars with JWST-based H$\beta$ $\mBH$ measurements are shown as red diamonds, while those with \ion{Mg}{2}-based $\mBH$ are shown as orange circles. 
Filled symbols indicate the SHELLQs quasars observed by MOIRCS in this work, whereas open symbols are taken from \citetalias{Onoue19} and \citet{Kato20}.
For the JWST targets in this work, the host-galaxy contribution to the 5100~\AA\ luminosity has been subtracted based on the NIRCam decomposition analysis of \citet{Ding25}.
H$\beta$-based $\mBH$ measurements for $z>6$ luminous quasars from the literature are shown as blue crosses \citep{Yang23, Eilers23, Yue24} .
H$\alpha$-based faint JWST AGN from \citet{Juodzbalis26} are shown as green squares.
The background contours show the density distribution of $z\approx1$--2 SDSS DR16Q quasars \citep{Wu22}, enclosing 68\%, 95\%, and 99.7\% of the population, restricted to objects with \ion{Mg}{2}-based black hole mass estimates~(Section~\ref{sec:comp_lowz}).
Their bolometric luminosities and black hole masses are derived using the same prescriptions as those adopted for the high-redshift samples.
The diagonal lines indicate  Eddington ratios of $\edd=$ 1, 0.1, and 0.01 from left to right.
The upper and right panels show the black hole mass and bolometric luminosity distributions, respectively.
The orange histograms show the distributions of the full SHELLQs targets, combining the JWST and ground-based samples.
(\textit{Right}) The corresponding SMBH mass -- Eddington ratio  distribution for the same samples.
}\label{fig:MBH}
\end{figure*}

\subsection{Distribution of $\mBH$ and $\edd$} \label{sec:MBH_distribution}

The redshift dependence of BH mass and Eddington ratio distributions provides insight into the growth history of SMBHs over  cosmic time.
The fraction of SMBHs accreting at super-Eddington rates holds the key to explaining the existence of billion-solar-mass BHs at high redshift \citep[e.g.,][]{Kawaguchi04, Volonteri15, Shirakata19, LiW23, Lupi24, Inayoshi_Maiolino25}.

There are 27 SHELLQs quasars in total, including 12 objects with H$\beta$-based measurements obtained in this work and 15 objects with \ion{Mg}{2}-based black hole mass measurements from this work and previous near-infrared observations  \citep{Onoue19, Matsuoka19, Kato20}.
These SHELLQs quasars span a bolometric luminosity range of $46 \lesssim \log L_{\rm bol}\ {\rm [erg\ s^{-1}]} \lesssim 47$ and occupy an intermediate regime between classical luminous quasars \citep[e.g.,][]{Yang23, Yue24} and faint broad-line AGN recently identified by JWST \citep[e.g.,][]{Maiolino24, Juodzbalis26}.

The left panel of Figure~\ref{fig:MBH} shows the distribution of black hole mass and bolometric luminosity for the 27 SHELLQs quasars.
The right panel shows the corresponding distribution of Eddington ratios.
For comparison, we also show in both panels luminous $z>6$ quasars with Balmer-line-based $\mBH$ measurements \citep{Yang23, Eilers23, Yue24} and faint broad-line AGNs identified by the JWST JADES program \citep{Juodzbalis26}.

The single-epoch black hole masses of the SHELLQs quasars span a wide range of $7.0 \leq \log (M_{\rm BH}/M_\odot) \leq 9.5$, with a median of $\left<\log (M_{\rm BH}/M_\odot)\right> = 8.6$.
The Eddington ratios of the SHELLQs quasars range from $-1.2 \leq \log (L_{\rm bol}/L_{\rm Edd}) \leq 0.3$, with a median of $\left<\log (L_{\rm bol}/L_{\rm Edd})\right> = -0.6$.
The side histograms in Figure~\ref{fig:MBH} show that the $\mBH$ and $\edd$ of the SHELLQs quasars are systematically shifted toward lower $\mBH$ and higher $\edd$ than those of the luminous quasars.
This trend also holds when the SHELLQs sample is compared to the \ion{Mg}{2}-based $\mBH$ measurements available in the literature at $z\gtrsim6$ \citep[e.g.,][]{Willott10a, Mazzucchelli17, Mazzucchelli23, Shen19, Yang21, Farina22}.
For example, \citet{Farina22} show that the median $\mBH$ and $\edd$ of their XQR-30 sample at $L_{\rm bol}>2.9\times10^{46}~{\rm erg~s^{-1}}$ are $\left<\log (M_{\rm BH}/M_\odot)\right> = 9.5$ and $\left<\log (L_{\rm bol}/L_{\rm Edd})\right> = -0.32$, respectively.
Thus, compared to the more luminous quasar population, the SHELLQs quasars trace SMBHs in an earlier and more active phase of their growth in the early universe.

While we are now identifying sub-Eddington SMBHs that are as massive as those powering typical luminous quasars ($\log \mBH / M_\odot \sim 9$), thanks to the deep imaging of the HSC-SSP,
we do not find any at $\log \mBH / M_\odot \sim 10$, although the depth of the HSC-SSP should be able to detect such SMBHs if they accrete at $\gtrsim1$\% Eddington.
Such ultramassive SMBHs are known to exist at $z\sim6$ \citep{Shemmer04, Netzer07, Wu15}, although they are intrinsically rare.
Their absence from the HSC-SSP sample may therefore be primarily due to the smaller sky coverage of HSC-SSP ($1{,}100$ deg$^2$) compared to SDSS ($15{,}000$ deg$^2$), together with the incomplete SMBH mass measurements for the SHELLQs quasars.
Also, such rarest $\log \mBH / M_\odot \sim 10$ systems will likely be found in the $\gtrsim2,000~\rm{deg^2}$ wide-area surveys of Euclid and Roman.

At the opposite end of the Eddington-ratio distribution,
three out of 27 (i.e., 11\%) of the $z\sim6$ SHELLQs quasars show accretion rates higher than the Eddington limit, albeit with systematic uncertainties of $0.4$~dex in individual $\mBH$ estimates.
Those are J0921$+$0007 and J1217$+$0131 in this study, and J0859$+$0052 in \citetalias{Onoue19}, which all have Mg{\sc ii}-based $\mBH$. 
Remarkably, one of these objects, J0921$+$0007 is detected in X-rays \citep{Wolf23} despite the fact that its rest-frame ultraviolet luminosity is $\approx1$ dex fainter than other X-ray bright $z>6$ quasars \citep{Nanni17}.
Existence of super-Eddington SMBHs at high redshift has been reported in the literature \citep{Jiang07, Willott10a, Banados21, Wang21, Yang21, Farina22, Mazzucchelli23}.
Specifically, \citet{Willott10a} reported that about half of their nine quasars at comparable $M_{1450}$ to our MOIRCS targets accrete at super-Eddington rates ($\edd>1$).
In Section~\ref{sec:MBH_Woo}, we discuss the possibility that the fraction of super-Eddington SMBHs in our sample may be even higher.

\subsection{Comparison with  $\mBH$ and $\edd$ at Lower redshifts} \label{sec:comp_lowz}

\begin{figure*}[tb]
\centering
\includegraphics[  trim=0.3in 0.25in 0.3in 0.25in, clip,
width=0.95 \linewidth]{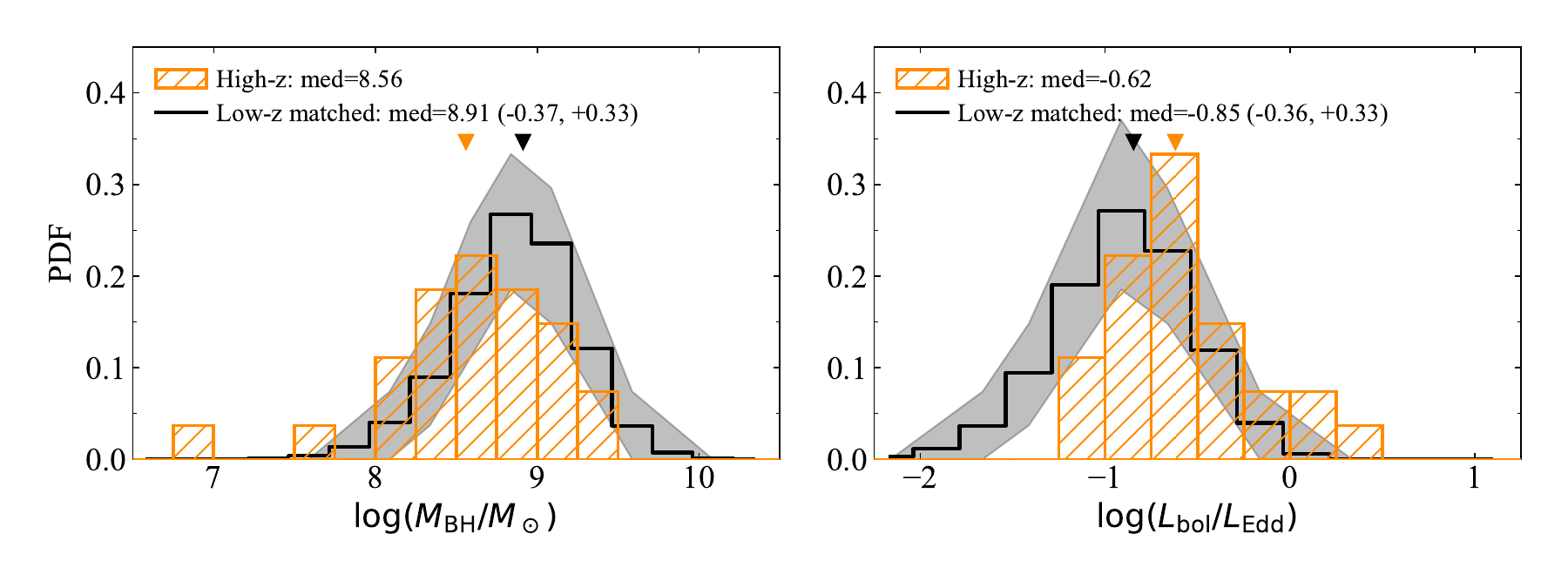}
\caption{Distribution of black hole mass (left) and Eddington ratio (right) of the 27 SHELLQs quasars at $z\sim6$--7  (orange) and the luminosity-matched SDSS DR16 quasars at $z\approx1.3$ (black).
The low-redshift sample is extracted from Monte Carlo resampling to reproduce the  $L_{\rm bol}$ distribution of the high-redshift sample.
The black histogram shows the median probability distribution function (PDF) in each $\log\mBH/M_\odot$ or $\log \edd$ bin over the Monte Carlo realizations, while the grey shaded area indicates the 16--84 percentile.
The median $\log(M_{\rm BH}/M_\odot)$ and $\log(L_{\rm bol}/L_{\rm Edd})$ values for the high-redshift and low-redshift samples, together with the mean 16th and 84th percentile ranges for the low-redshift sample, are indicated in the inset of each panel.
}\label{fig:comp_lowz}
\end{figure*}

Previous studies have investigated the distributions of BH mass and Eddington ratio for luminous quasars at $\log L_\mathrm{bol}~\mathrm{[erg\ s^{-1}]} \gtrsim 47$ \citep[e.g.,][]{Willott10a,Yang21,Farina22,Mazzucchelli23}; however, such luminous quasar samples are naturally biased toward most massive and  rapidly accreting SMBHs.
Here we compare the BH mass and Eddington ratio distributions of the intermediate-luminosity SHELLQs sample with those of a low-redshift control sample to investigate whether the accretion properties of SMBHs evolve with redshift.

For the control sample at low redshift, we use the spectral measurements for the sixteenth data release of Sloan Digital Sky Survey quasar catalog \citep{Lyke20}.
We select quasars for which \ion{Mg}{2}-based black hole mass estimates are available in \citet{Wu22}.
The mean redshift of this low-redshift sample is $z=1.3$.
We do not use quasars with H$\beta$-based black hole masses for the control sample, because only a small number of such objects exist at bolometric luminosities $L_{\rm bol} > 10^{46.5}\ \mathrm{erg\ s^{-1}}$.

We perform Monte Carlo resampling of the low-redshift quasars.
In each realization, 27 objects are randomly drawn with replacement, such that each low-redshift object is matched in $L_{\rm bol}$ to one of the high-redshift quasars within $\pm0.1$ dex.
We calculate $\mBH$, $L_{\rm bol}$, and $\edd$  for the extracted low-redshift quasars following the same method as for the high-redshift sample (Equations~\ref{eq:VO09}, \ref{eq:Lbol_MgII}). 
We then compare the luminosity-matched low-redshift distribution with the high-redshift sample.
An ensemble of 10,000 realizations is used to derive the average distributions, thereby accounting for sample variance and uncertainties in the matching procedure.

Figure~\ref{fig:comp_lowz} presents the black hole mass and Eddington ratio distributions of both the high-redshift sample and the luminosity-matched low-redshift quasar sample described above.
We obtain a median black hole mass of $\log (M_{\rm BH}/M_\odot) = 8.91_{-0.37}^{+0.33}$ for the low-redshift quasars, where the quoted uncertainties represent the 16th–84th percentile range.
The scatter of the median $M_{\rm BH}$ between different realizations is $0.08$ dex, which is smaller than the spread of the distribution.
The corresponding median Eddington ratio is $\log (L_{\rm bol}/L_{\rm Edd}) = -0.85_{-0.36}^{+0.33}$.
We find that the median black hole mass of the high-redshift sample is offset by $\Delta \log (M_{\rm BH}/M_\odot) = -0.35$ with respect to the low-redshift sample.
Likewise, the high-redshift sample has a higher accretion efficiency, with a mean offset of $\Delta \log (L_{\rm bol}/L_{\rm Edd}) = 0.23$.
Therefore, SMBH accretion activity is systematically more efficient at $6<z<7$ than at lower redshifts, suggesting that the trend previously found for luminous quasars extends to the more representative high-redshift SMBH population.

It is also remarkable that the high-redshift sample exhibits an extended tail toward the super-Eddington regime ($\edd\gtrsim1$) and a relative deficit of  Eddington ratios below 5\% Eddington.
This trend is unlikely to be driven by observational biases, as both the high- and low-redshift quasar samples discussed here were selected based on luminosity, and not by Eddington ratio.
In order to address this apparent trend, we apply the Anderson–Darling (AD) test \citep{AD52}, which is sensitive to differences in distribution tails, to test the null hypothesis that the black hole mass and Eddington ratio distributions between the low- and high-redshift samples are drawn from the same underlying parent distribution. 
We perform the AD test for each Monte Carlo realization of the low-redshift sample.
We obtain a median $p$-value of 0.007, with the null hypothesis rejected ($p<0.05$) in 87.2\% of the random realizations, indicating that the difference between the high- and low-redshift quasar samples persists for the majority of the luminosity-matched low-redshift samples.
This result suggests that the enhanced fraction of rapidly accreting SMBHs at high redshift contributes to the observed difference from the low-redshift population.
This result suggests that the enhanced fraction of rapidly accreting SMBHs at high redshift contributes to the observed difference from the low-redshift population.

The excess of super-Eddington objects and the overall enhancement of the SMBH accretion activity in the high-redshift sample is relevant to the so-called AGN downsizing discussed in X-ray AGN studies, in which the most rapidly accreting and luminous black holes reach their peak activity at higher redshifts \citep{Ueda03, Merloni04}.
Such a picture has also been indicated by theoretical studies of distant SMBHs \citep{Madau14, Shirakata19, Inayoshi20, Lupi24}.
In this context, our result  extends the downsizing picture from the most luminous quasar population to a more representative population at the epoch of reionization.

\subsection{Possible Bias in MOIRCS $\mBH$ Estimates} \label{sec:comp_lowz_noise}

We assess the impact of the modest  quality of the MOIRCS spectra on the \ion{Mg}{2}-based virial SMBH mass estimates by performing a noise-degradation test using SDSS DR16Q quasar spectra \citep{Lyke20}. 
We retrieved SDSS spectra of $z=$1--2 counterparts of our MOIRCS sample, the broad \ion{Mg}{2} emission lines of which are clearly detected with S/N $>20$ and FWHM $<5000$~km~s$^{-1}$ in the DR16Q catalog.
We also applied a quality cut of S/N $>15$ per 300~km~s$^{-1}$ at rest-frame 3000~\AA, which yields 73 objects.
We artificially redshifted these spectra to $z=6.4$ using their SDSS redshifts, and resampled both the flux and variance spectra with a 4~\AA\ step per pixel, matching that of the MOIRCS spectra. 
We then degraded the spectra to S/N $=6$ per 300~km~s$^{-1}$ by adding random noise, comparable to the typical quality of our MOIRCS data (Table~\ref{tab:targets_all}). 
For both the original and noise-degraded spectra, we measured the \ion{Mg}{2} line width and the continuum flux density at rest-frame 3000~\AA\ by fitting a single Gaussian profile plus a power-law continuum model.

We find that the noise degradation introduces scatters of $\pm$1.9\% in the continuum flux density and $\pm$19\% in \ion{Mg}{2} FWHM, while the median offsets from the original measurements are only 0.09\% and 1.2\%, respectively.
We therefore conclude that the modest quality of the MOIRCS spectra does not introduce a significant systematic bias in the \ion{Mg}{2}-based SMBH mass estimates.
It instead adds an uncertainty of $\pm0.17$ dex when propagated through the single-epoch $\mBH$ recipe (Eq.~\ref{eq:V06}), with the error budget dominated by the FWHM term.
This additional scatter is neither large enough to explain the observed tails in the $\mBH$ and Eddington ratio distributions relative to their mean values, nor large compared with the 0.4 dex systematic uncertainty inherent in single-epoch $\mBH$ estimates \citep{Shen13_review}.

\subsection{Impact of Single-epoch $\mBH$ Calibration} \label{sec:MBH_Woo}

Single-epoch $\mBH$ estimates rely on the $R_\mathrm{BLR}$--luminosity relation calibrated for local AGNs, which follows $R_\mathrm{BLR}\propto L_{5100}^{0.5}$ \citep{Bentz13}.
Observations have shown that high-Eddington SMBHs with $\edd\gtrsim1$ exhibit systematically reduced BLR sizes (i.e., shorter time lags) compared to sub-Eddington AGNs at a given luminosity \citep{Grier17, Du19}.
A possible explanation for this trend is that the BLR size depends not only on luminosity but also on  accretion physics.
This effect can potentially affect current interpretations of the early growth history of SMBHs, because many of the most luminous and/or highest-redshift quasars at $z\gtrsim6$ show high Eddington ratios \citep[e.g.,][]{Mortlock11, Wu15, Banados18, Yang20, Wang21}.

Previous studies attempted to correct this effect using observables such as rest-frame ultraviolet \ion{Fe}{2} emission as empirical proxies for the Eddington ratio \citep[e.g.,][]{Du19, Pan25}.
A possible dependence of single-epoch $\mBH$ estimates on the Eddington ratio has also been discussed by \citet{Bonta20}.
\citet{Woo26} recently introduced the explicit Eddington-ratio dependence into the $R_\mathrm{BLR}$--luminosity relation by defining a three-parameter fundamental plane.
This modified single-epoch $\mBH$ calibration minimizes the intrinsic scatter of the H$\beta$-based $\mBH$ estimates to 0.21 dex.

\begin{figure*}[tb]
\centering
\includegraphics[trim=0.3in 0.25in 0.3in 0.25in, clip, width=0.95 \linewidth]{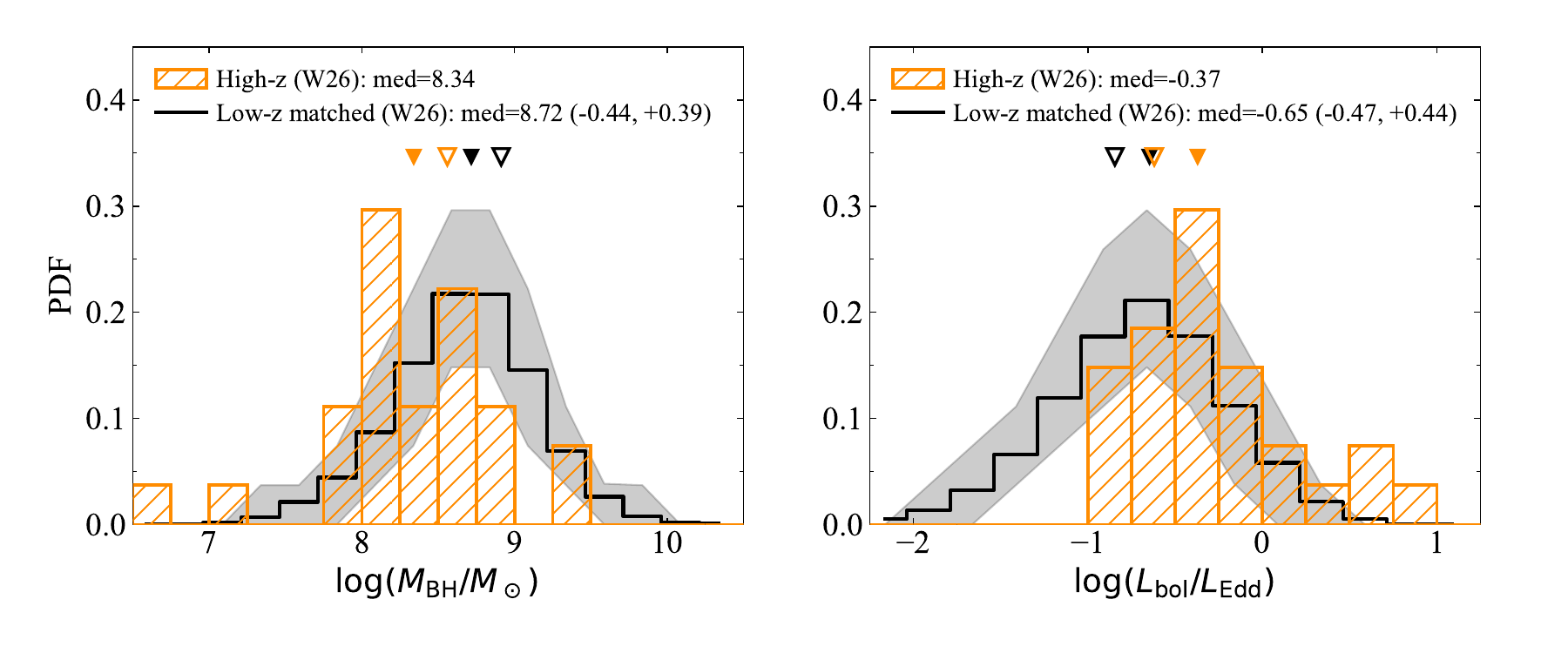}
\caption{Same as Figure~\ref{fig:comp_lowz}, but with $\mBH$ re-estimated using the calibration of \citet{Woo26}.
The median $\log(M_{\rm BH}/M_\odot)$ and $\log(L_{\rm bol}/L_{\rm Edd})$ values in Figure~\ref{fig:comp_lowz} are indicated with open symbols to be compared with those based on the \citet{Woo26}'s calibration.
}\label{fig:comp_lowz_W26}
\end{figure*}

Figure~\ref{fig:comp_lowz_W26} shows the $\mBH$ and $\edd$ distributions of our $z>6$ sample when the \citet{Woo26} calibration is applied to the H$\beta$- and \ion{Mg}{2}-based $\mBH$ estimates.
The distributions of the low-redshift SDSS quasars are also re-measured using the same calibration.
Both the high- and low-redshift samples shift toward lower $\mBH$ and correspondingly to higher $\edd$.
For the high-redshift sample, which has systematically higher accretion rates than the matched low-redshift quasars, the median object-by-object shifts relative to our fiducial calibration are $\Delta \log{(\mBH/M_\odot)} = -0.23_{-0.22}^{+0.05}$  and $\Delta \log{(\edd)} = 0.23_{-0.05}^{+0.21}$, where the quoted ranges represent the 16th--84th percentiles of the object-to-object offset distribution.
The asymmetric ranges reflect the fact that the largest corrections occur for high-Eddington sources; for example, J0921$+$0007, J1217$+$0131 and J0859$+$0052, which have $\edd\gtrsim1$ under the \citet{Vestergaard09} calibration, show the largest offsets with $\Delta \log (\mBH/M_\odot) \simeq -0.5$ dex and $\Delta \log{(\edd)} \simeq 0.5$ dex.

A notable consequence of this application is that the fraction of super-Eddington SMBHs at high redshift increases from 11\% to 22\%.
Reflecting the stronger corrections for the higher-Eddington-ratio high-redshift quasars described above, the differences in the median BH mass and Eddington ratio between the high- and low-redshift samples also become slightly larger, with $\Delta \log (\mBH/M_\odot){\rm high-low} = -0.38$ dex and $\Delta \log{(\edd)}{\rm high-low} = 0.28$ dex, respectively, compared to $-0.35$ dex and $0.23$ dex for the fiducial calibrations.
These results are consistent with the possible importance of frequent super-Eddington episodes in the growth of distant SMBHs, as suggested by other observations \citep{Collin04, Kawaguchi04, Davies19, Arita23, Eilers24}.
The reduced SMBH masses and higher inferred accretion rates, if real,  partly alleviate the long-standing tension between the rapid emergence of massive black holes and the short cosmic time available within the first billion years of the universe.
The exact frequency and duty cycle of the super-Eddington phase in the intermediate-luminosity high-redshift quasar population will be addressed in future work, where sample completeness will be taken into account.

Finally, we note that the high fraction of super-Eddington SMBHs at high redshift is not significantly affected by the choice between H$\alpha$- and H$\beta$-based $\mBH$ estimates, because the H$\alpha$-based $\mBH$ is systematically lower than the H$\beta$-based estimate  (Figure~\ref{fig:delta_MBH}) and  therefore yields a higher Eddington ratio.

\section{Summary} \label{sec:summary}

We present the results of JWST/NIRSpec and Subaru/MOIRCS  near-infrared spectroscopy of 21 intermediate-luminosity quasars at $6.07 \leq z \leq 6.90$ drawn from the SHELLQs project.
These SHELLQs quasars occupy the luminosity range of $\log L_{\rm bol}\ \mathrm{[erg~s^{-1}]} \sim 46$--47, which lies between classical luminous $z>6$ quasars and the faint broad-line AGN recently uncovered by JWST.
The JWST/NIRSpec data from our Cycle~1 project (GO 1967) enable Balmer-line-based measurements for 12 fainter quasars beyond the sensitivity limit of ground-based spectroscopy, while the MOIRCS spectra provide \ion{Mg}{2}-based constraints for 9 targets brighter than $M_{1450}=-24$ mag ($\approx23$ mag in $y$ band).
We perform single-epoch black hole mass estimates using Balmer and \ion{Mg}{2} emission lines, and construct a $M_{\rm BH}$ sample of 27 objects combining the new data with earlier studies of the SHELLQs quasars.
This large sample enables us to constrain the black hole mass and Eddington ratio distributions of $z>6$ quasars in the intermediate-luminosity regime, $\log L_{\rm bol} \sim 46$--47.

Our main results are as follows.

\begin{enumerate}

\item
The JWST/NIRSpec data reveal a variety of rest-frame optical emission lines in the 12 targets.
Among them, two targets (J2236+0032 and J1512+4422) exhibit stellar absorption features in high-order Balmer lines, which we interpret as evidence that their host galaxies are in a post-starburst phase, as explored in detail by \citet{Onoue25}.

Three targets (J1146$-$0005, J0844$-$0132, and J0217$-$0208) exhibit remarkably strong narrow Balmer emission lines on top of broad components.
In particular, the Balmer line luminosities of J1146$-$0005 and J0844$-$0132 lie well above the local relation between the 5100~\AA\ continuum and Balmer line luminosity, suggesting the presence of dense gas both around the nucleus and in the interstellar medium.
J0217$-$0208 shows the weakest broad Balmer lines in our sample and may be either an AGN hosting a relatively low-mass SMBH or a star-forming galaxy with outflowing gas.

\item
We detect broad \ion{Mg}{2} emission lines in all targets observed with MOIRCS $K_s$-band spectroscopy.
One target, J1350$-$0027, exhibits a double-peaked line profile, which could be interpreted either as narrow absorption at the peak of a broader emission line or as resolved \ion{Mg}{2} doublet emission.
For the other eight targets, the spectra are decomposed into an accretion-disk continuum, a pseudo-iron continuum, and \ion{Mg}{2} emission.

\item
We derive single-epoch virial black hole masses for a total of 26 SHELLQs quasars.
The combined SHELLQs sample with available $\mBH$ estimates spans $7.2 < \log (M_{\rm BH}/M_\odot) < 9.4$ and $-1.3 < \log (L_{\rm bol}/L_{\rm Edd}) < 0.4$, with a median of  $\log (M_{\rm BH}/M_\odot)=8.6$ and $\log (L_{\rm bol}/L_{\rm Edd})=-0.63$, respectively.

\item At fixed bolometric luminosity, the $z\sim6$ SHELLQs quasars have systematically smaller black hole masses and higher Eddington ratios than a luminosity-matched SDSS sample at $z\approx1$--2. 
We find median offsets of $\Delta \log (M_{\rm BH}/M_\odot) = -0.4 \pm 0.1$ and $\Delta \log (L_{\rm bol}/L_{\rm Edd}) = +0.2 \pm 0.1$. 
The high-redshift sample also shows an extended tail toward the super-Eddington regime and a deficit of objects at $\edd < 0.05$, supported by a statistical test. 
These results suggest that the anti-hierarchical growth of distant SMBHs extends to a representative quasar population in the epoch of cosmic reionization.

\item We test the impact of the \citet{Woo26} single-epoch $\mBH$ calibration for our high-redshift sample, which accounts for the Eddington-ratio dependence of the BLR size. 
This calibration shifts our $z>6$ quasars toward lower $\mBH$ and higher $\edd$, increasing the fraction of super-Eddington SMBHs from 11\% to 22\%. 
This result further supports the possible importance of frequent near- or super-Eddington accretion episodes in the early growth of SMBHs.

\end{enumerate}

Overall, this study demonstrates that the combination of wide-field Subaru surveys and JWST spectroscopy provides a powerful route to establishing the demographic properties of SMBHs in the reionization era.
Our observations reveal that a non-negligible fraction of the SHELLQs quasars are caught in a phase of rapid black hole growth, although their accretion efficiencies are less extreme than those of the most luminous quasars at similar redshifts.
We will investigate the intrinsic BH mass function and Eddington ratio distribution function with selection function taken into account in our future paper (Onoue et al. in prep.).
In addition, we will  use the rich JWST spectral data set to investigate physical properties of the BLR and NLR gas at $z\sim6$ in our future papers.

\begin{acknowledgments}
We thank Jong-Hak Woo and his group members for valuable inputs on the black hole mass calibrations.

We thank the Subaru support astronomers for their assistance in carrying out our Subaru/MOIRCS observing programs, and Akatoki Noboriguchi and Rikako Ishimoto for their help with the  observations.

This work is based on observations made with the NASA/ESA/CSA James Webb Space Telescope. 
The data were obtained from the Mikulski Archive for Space Telescopes at the Space Telescope Science Institute, which is operated by the Association of Universities for Research in Astronomy, Inc., under NASA contract NAS 5-03127 for JWST. 
These observations are associated with programs GO \#1967. 
Support for these programs was provided by NASA through a grant from the Space Telescope Science Institute (JWST-GO-01967.002-A), which is operated by the Association of Universities for Research in Astronomy, Inc., under NASA contract NAS 5-03127.
This work was supported by World Premier International Research Center Initiative (WPI), MEXT, Japan. 
This work used computing resources at Kavli IPMU.

M.O., Y.M., M.I., K.K., T.K., N.K., T.N., and T.I. acknowledge support from JSPS KAKENHI Grant Numbers 24K22894, 26K07155, 21H04494, 26K00745, JP25K07359, JP23K20035, JP24H00004, JP25K07370, JP25H00663, JP25K01038, JP25K01044, JP23K25911, and JP25H00671.
K.Inayoshi acknowledges support from the National Natural Science Foundation of China (12573015, W2532003), the Beijing Natural Science Foundation (IS25003), and the China Manned Space Program (CMS-CSST-2025-A09).
K.Iwasawa acknowledges support under the grant PID2022-136828NB-C44 provided by MCIN/AEI/10.13039/501100011033 / FEDER, UE.
M.Sawamura is supported by JST SPRING, Grant Number JPMJSP2108.
This paper is a part of the outcome of research performed under a Waseda University Grant for Special Research Projects (Project number: 2025R-062).

\end{acknowledgments}

\vspace{5mm}
\facilities{JWST (NIRSpec), Subaru (MOIRCS)}

\software{astropy \citep{Astropy},  matplotlib \citep{matplotlib}, scipy \citep{scipy}, IRAF \citep{Tody86}, QSOFitMORE \citep{Fu21_QSOfitmore}
          }

\appendix

\section{Black hole masses based on Tsuzuki et al. iron template} \label{sec:BHmass_T06}
Here, we present in Table~\ref{tab:MgII_T06} the \ion{Mg}{2} line profiles of the MOIRCS targets obtained using the \citet{Tsuzuki06} iron template, which better represents the iron emission underlying the \ion{Mg}{2} line than the \citet{Vestergaard01} template.
The format of the table is the same as in Table~\ref{tab:MgII_VO09}.

\begin{deluxetable*}{lCCCC}[htb]
\tabletypesize{\small}
\tablecaption{\ion{Mg}{2} profiles (\citealt{Tsuzuki06} iron template) \label{tab:MgII_T06}}
\tablehead{
\colhead{ID} &
\colhead{$z_{\rm MgII}$} &
\colhead{$L_{3000}$} &
\colhead{{\rm FWHM}(Mg\,\textsc{ii})} &
\colhead{$L_{\rm MgII}$} \\
\colhead{} &
\colhead{} &
\colhead{[10$^{45}$ erg s$^{-1}$]} &
\colhead{[km s$^{-1}$]} &
\colhead{[10$^{43}$ erg s$^{-1}$]}
}
\startdata
J2210+0304 & 6.861_{-0.005}^{+0.006} & 4.7 \pm 0.4 & 3900_{-1000}^{+900} & 4.4_{-0.8}^{+0.6} \\
J0921+0007 & 6.563_{-0.001}^{+0.001} & 14.3 \pm 0.6 & 1700_{-100}^{+100} & 11.3_{-0.7}^{+0.7} \\
J1545+4232 & 6.514_{-0.002}^{+0.002} & 4.5 \pm 0.3 & 2900_{-200}^{+200} & 7.5_{-0.4}^{+0.4} \\
J1004+0239 & 6.403_{-0.003}^{+0.003} & 3.7 \pm 0.2 & 2400_{-500}^{+900} & 4.2_{-0.5}^{+0.6} \\
J1137+0045 & 6.361_{-0.003}^{+0.003} & 2.7 \pm 0.2 & 4600_{-400}^{+200} & 5.7_{-0.4}^{+0.5} \\
J2304+0045 & 6.336_{-0.003}^{+0.004} & 2.3 \pm 0.1 & 2900_{-400}^{+400} & 3.1_{-0.3}^{+0.5} \\
J1406$-$0116 & 6.292_{-0.002}^{+0.003} & 3.5 \pm 0.2 & 2500_{-200}^{+200} & 2.7_{-0.2}^{+0.2} \\
J1217+0131 & 6.205_{-0.004}^{+0.004} & 11.5 \pm 0.3 & 2000_{-300}^{+100} & 5.4_{-0.8}^{+0.9} \\
\enddata
\end{deluxetable*}

\clearpage

\section{Full table for the JWST line measurements} \label{sec:table_full}

Here, we provide the full set of parameters derived from the continuum+line fitting of the JWST/NIRSpec G395M spectra of the 12 targets in machine-readable form. 
Because of the large number of columns, Table~\ref{tab:line_full} provides a description of each column rather than displaying the individual measurements. 
The complete table is provided as ancillary material with this manuscript.

\startlongtable
\begin{deluxetable*}{clll}
\tabletypesize{\small}
\tablecaption{Description of the Machine-readable Table of the JWST Spectral Measurements
\label{tab:line_full}}
\tablehead{
\colhead{Column} &
\colhead{Label} &
\colhead{Unit} &
\colhead{Description}
}
\startdata
1 & \texttt{ID} & \nodata & Object identifier \\
2 & \texttt{redshift} & \nodata & Best-fit [\ion{O}{3}] redshift \\
3 & \texttt{redshift\_err} & \nodata & Uncertainty in the [\ion{O}{3}] redshift \\
4 & \texttt{ra} & deg & Right ascension  \\
5 & \texttt{dec} & deg & Declination  \\
6 & \texttt{PL\_norm} & $10^{-17} {\rm erg~cm^{-2}~s^{-1}~\AA}^{-1}$ & Normalization of the power-law continuum at rest-frame $3000~{\rm \AA}$ \\
7 & \texttt{PL\_norm\_err} & $10^{-17} {\rm erg~cm^{-2}~s^{-1}~\AA}^{-1}$ & Uncertainty in the power-law normalization \\
8 & \texttt{PL\_slope} & \nodata & Slope of the power-law continuum \\
9 & \texttt{PL\_slope\_err} & \nodata & Uncertainty in the power-law slope \\
10 & \texttt{L\_Fe\_w3980\_w4040} & $10^{42}$ erg s$^{-1}$ & Integrated Fe II luminosity over 3980--4040 ${\rm \AA}$ \\
11 & \texttt{L\_Fe\_w3980\_w4040\_err} & $10^{42}$ erg s$^{-1}$ & Uncertainty in the integrated Fe II luminosity over 3980--4040 ${\rm \AA}$ \\
12 & \texttt{L\_Fe\_w4435\_w4685} & $10^{42}$ erg s$^{-1}$ & Integrated Fe II luminosity over 4435--4685 ${\rm \AA}$ \\
13 & \texttt{L\_Fe\_w4435\_w4685\_err} & $10^{42}$ erg s$^{-1}$ & Uncertainty in the integrated Fe II luminosity over 4435--4685 ${\rm \AA}$ \\
14 & \texttt{L\_Fe\_w5100\_w5500} & $10^{42}$ erg s$^{-1}$ & Integrated Fe II luminosity over 5100--5500 ${\rm \AA}$ \\
15 & \texttt{L\_Fe\_w5100\_w5500\_err} & $10^{42}$ erg s$^{-1}$ & Uncertainty in the integrated Fe II luminosity over 5100--5500 ${\rm \AA}$ \\
16 & \texttt{L\_Ha\_broad} & $10^{42}$ erg s$^{-1}$ & Luminosity of the broad H$\alpha$ component \\
17 & \texttt{L\_Ha\_broad\_err} & $10^{42}$ erg s$^{-1}$ & Uncertainty in the luminosity of the broad H$\alpha$ component \\
18 & \texttt{L\_Hb\_broad} & $10^{42}$ erg s$^{-1}$ & Luminosity of the broad H$\beta$ component \\
19 & \texttt{L\_Hb\_broad\_err} & $10^{42}$ erg s$^{-1}$ & Uncertainty in the luminosity of the broad H$\beta$ component \\
20 & \texttt{L\_Hg\_broad} & $10^{42}$ erg s$^{-1}$ & Luminosity of the broad H$\gamma$ component \\
21 & \texttt{L\_Hg\_broad\_err} & $10^{42}$ erg s$^{-1}$ & Uncertainty in the luminosity of the broad H$\gamma$ component \\
22 & \texttt{L\_Hd\_broad} & $10^{42}$ erg s$^{-1}$ & Luminosity of the broad H$\delta$ component \\
23 & \texttt{L\_Hd\_broad\_err} &$10^{42}$  erg s$^{-1}$ & Uncertainty in the luminosity of the broad H$\delta$ component \\
24 & \texttt{L\_Ha\_narrow} & $10^{42}$ erg s$^{-1}$ & Luminosity of the narrow H$\alpha$ component \\
25 & \texttt{L\_Ha\_narrow\_err} & $10^{42}$ erg s$^{-1}$ & Uncertainty in the luminosity of the narrow H$\alpha$ component \\
26 & \texttt{L\_Hb\_narrow} & $10^{42}$ erg s$^{-1}$ & Luminosity of the narrow H$\beta$ component \\
27 & \texttt{L\_Hb\_narrow\_err} & $10^{42}$ erg s$^{-1}$ & Uncertainty in the luminosity of the narrow H$\beta$ component \\
28 & \texttt{L\_Hg\_narrow} & $10^{42}$ erg s$^{-1}$ & Luminosity of the narrow H$\gamma$ component \\
29 & \texttt{L\_Hg\_narrow\_err} & $10^{42}$ erg s$^{-1}$ & Uncertainty in the luminosity of the narrow H$\gamma$ component \\
30 & \texttt{L\_Hd\_narrow} & $10^{42}$ erg s$^{-1}$ & Luminosity of the narrow H$\delta$ component \\
31 & \texttt{L\_Hd\_narrow\_err} & $10^{42}$ erg s$^{-1}$ & Uncertainty in the luminosity of the narrow H$\delta$ component \\
32 & \texttt{L\_O3\_5007\_c} & $10^{42}$ erg s$^{-1}$ & Luminosity of the core [\ion{O}{3}] $\lambda5007$ component \\
33 & \texttt{L\_O3\_5007\_c\_err} & $10^{42}$ erg s$^{-1}$ & Uncertainty in the luminosity of the core [\ion{O}{3}] $\lambda5007$ component \\
34 & \texttt{L\_O3\_5007\_w} & $10^{42}$ erg s$^{-1}$ & Luminosity of the wing [\ion{O}{3}] $\lambda5007$ component \\
35 & \texttt{L\_O3\_5007\_w\_err} & $10^{42}$ erg s$^{-1}$ & Uncertainty in the luminosity of the wing [\ion{O}{3}] $\lambda5007$ component \\
36 & \texttt{L\_O3\_4959\_c} & $10^{42}$ erg s$^{-1}$ & Luminosity of the core [\ion{O}{3}] $\lambda4959$ component \\
37 & \texttt{L\_O3\_4959\_c\_err} & $10^{42}$ erg s$^{-1}$ & Uncertainty in the luminosity of the core [\ion{O}{3}] $\lambda4959$ component \\
38 & \texttt{L\_O3\_4959\_w} & $10^{42}$ erg s$^{-1}$ & Luminosity of the wing [\ion{O}{3}] $\lambda4959$ component \\
39 & \texttt{L\_O3\_4959\_w\_err} & $10^{42}$ erg s$^{-1}$ & Uncertainty in the luminosity of the wing [\ion{O}{3}] $\lambda4959$ component \\
40 & \texttt{L\_N2\_6585} & $10^{42}$ erg s$^{-1}$ & Luminosity of [\ion{N}{2}] $\lambda6585$ \\
41 & \texttt{L\_N2\_6585\_err} & $10^{42}$ erg s$^{-1}$ & Uncertainty in the luminosity of [\ion{N}{2}] $\lambda6585$ \\
42 & \texttt{FWHM\_Ha\_broad} & km s$^{-1}$ & FWHM of the broad H$\alpha$ component \\
43 & \texttt{FWHM\_Ha\_broad\_err} & km s$^{-1}$ & Uncertainty in the FWHM of the broad H$\alpha$ component \\
44 & \texttt{FWHM\_Hb\_broad} & km s$^{-1}$ & FWHM of the broad H$\beta$ component \\
45 & \texttt{FWHM\_Hb\_broad\_err} & km s$^{-1}$ & Uncertainty in the FWHM of the broad H$\beta$ component \\
46 & \texttt{FWHM\_Hg\_broad} & km s$^{-1}$ & FWHM of the broad H$\gamma$ component \\
47 & \texttt{FWHM\_Hg\_broad\_err} & km s$^{-1}$ & Uncertainty in the FWHM of the broad H$\gamma$ component \\
48 & \texttt{FWHM\_Hd\_broad} & km s$^{-1}$ & FWHM of the broad H$\delta$ component \\
49 & \texttt{FWHM\_Hd\_broad\_err} & km s$^{-1}$ & Uncertainty in the FWHM of the broad H$\delta$ component \\
50 & \texttt{FWHM\_Ha\_narrow} & km s$^{-1}$ & FWHM of the narrow H$\alpha$ component \\
51 & \texttt{FWHM\_Ha\_narrow\_err} & km s$^{-1}$ & Uncertainty in the FWHM of the narrow H$\alpha$ component \\
52 & \texttt{FWHM\_Hb\_narrow} & km s$^{-1}$ & FWHM of the narrow H$\beta$ component \\
53 & \texttt{FWHM\_Hb\_narrow\_err} & km s$^{-1}$ & Uncertainty in the FWHM of the narrow H$\beta$ component \\
54 & \texttt{FWHM\_Hg\_narrow} & km s$^{-1}$ & FWHM of the narrow H$\gamma$ component \\
55 & \texttt{FWHM\_Hg\_narrow\_err} & km s$^{-1}$ & Uncertainty in the FWHM of the narrow H$\gamma$ component \\
56 & \texttt{FWHM\_Hd\_narrow} & km s$^{-1}$ & FWHM of the narrow H$\delta$ component \\
57 & \texttt{FWHM\_Hd\_narrow\_err} & km s$^{-1}$ & Uncertainty in the FWHM of the narrow H$\delta$ component \\
58 & \texttt{FWHM\_O3\_5007\_c} & km s$^{-1}$ & FWHM of the core [\ion{O}{3}] $\lambda5007$ component \\
59 & \texttt{FWHM\_O3\_5007\_c\_err} & km s$^{-1}$ & Uncertainty in the FWHM of the core [\ion{O}{3}] $\lambda5007$ component \\
60 & \texttt{FWHM\_O3\_5007\_w} & km s$^{-1}$ & FWHM of the wing [\ion{O}{3}] $\lambda5007$ component \\
61 & \texttt{FWHM\_O3\_5007\_w\_err} & km s$^{-1}$ & Uncertainty in the FWHM of the wing [\ion{O}{3}] $\lambda5007$ component \\
62 & \texttt{FWHM\_O3\_4959\_c} & km s$^{-1}$ & FWHM of the core [\ion{O}{3}] $\lambda4959$ component \\
63 & \texttt{FWHM\_O3\_4959\_c\_err} & km s$^{-1}$ & Uncertainty in the FWHM of the core [\ion{O}{3}] $\lambda4959$ component \\
64 & \texttt{FWHM\_O3\_4959\_w} & km s$^{-1}$ & FWHM of the wing [\ion{O}{3}] $\lambda4959$ component \\
65 & \texttt{FWHM\_O3\_4959\_w\_err} & km s$^{-1}$ & Uncertainty in the FWHM of the wing [\ion{O}{3}] $\lambda4959$ component \\
\enddata

\tablecomments{
The complete table is provided in machine-readable form as ancillary material with this manuscript.
For [\ion{O}{3}], the suffixes ``c'' and ``w'' denote the core and wing components, respectively.
}
\end{deluxetable*}


\bibliographystyle{aasjournalv7}

\bibliography{ref_SHELLQsNIR}

@ARTICLE{Oke83,
       author = {{Oke}, J.~B. and {Gunn}, J.~E.},
        title = "{Secondary standard stars for absolute spectrophotometry.}",
      journal = {\apj},
         year = 1983,
        month = mar,
       volume = {266},
        pages = {713-717},
          doi = {10.1086/160817},
       adsurl = {https://ui.adsabs.harvard.edu/abs/1983ApJ...266..713O}
}

@INPROCEEDINGS{Tody86,
       author = {{Tody}, Doug},
        title = "{The IRAF Data Reduction and Analysis System}",
    booktitle = {Instrumentation in astronomy VI},
         year = 1986,
       editor = {{Crawford}, David L.},
       series = {Society of Photo-Optical Instrumentation Engineers (SPIE) Conference Series},
       volume = {627},
        month = jan,
        pages = {733},
          doi = {10.1117/12.968154},
       adsurl = {https://ui.adsabs.harvard.edu/abs/1986SPIE..627..733T}
}

@Article{matplotlib,
  Author  = {Hunter, J. D.},
  Title   = {Matplotlib: A 2D Graphics Environment},
  Journal = {Computing in Science \& Engineering},
  Volume  = {9},
  Number  = {3},
  Pages   = {90--95},
  Year    = {2007},
  doi     = {10.1109/MCSE.2007.55},
}

@Article{scipy,
  Author  = {Virtanen, Pauli and Gommers, Ralf and Oliphant, Travis E. and others},
  Title   = {{SciPy} 1.0: Fundamental Algorithms for Scientific Computing in Python},
  Journal = {Nature Methods},
  Year    = {2020},
  Volume  = {17},
  Pages   = {261--272},
  doi     = {10.1038/s41592-019-0686-2},
}

@ARTICLE{Rigby23,
       author = {{Rigby}, Jane and {Perrin}, Marshall and {McElwain}, Michael and {Kimble}, Randy and {Friedman}, Scott and {Lallo}, Matt and {Doyon}, Ren{\'e} and {Feinberg}, Lee and {Ferruit}, Pierre and {Glasse}, Alistair and {Rieke}, Marcia and {Rieke}, George and {Wright}, Gillian and {Willott}, Chris and {Colon}, Knicole and {Milam}, Stefanie and {Neff}, Susan and {Stark}, Christopher and {Valenti}, Jeff and {Abell}, Jim and {Abney}, Faith and {Abul-Huda}, Yasin and {Acton}, D. Scott and {Adams}, Evan and {Adler}, David and {Aguilar}, Jonathan and {Ahmed}, Nasif and {Albert}, Lo{\"\i}c and {Alberts}, Stacey and {Aldridge}, David and {Allen}, Marsha and {Altenburg}, Martin and {{\'A}lvarez-M{\'a}rquez}, Javier and {Alves de Oliveira}, Catarina and {Andersen}, Greg and {Anderson}, Harry and {Anderson}, Sara and {Argyriou}, Ioannis and {Armstrong}, Amber and {Arribas}, Santiago and {Artigau}, Etienne and {Arvai}, Amanda and {Atkinson}, Charles and {Bacon}, Gregory and {Bair}, Thomas and {Banks}, Kimberly and {Barrientes}, Jaclyn and {Barringer}, Bruce and {Bartosik}, Peter and {Bast}, William and {Baudoz}, Pierre and {Beatty}, Thomas and {Bechtold}, Katie and {Beck}, Tracy and {Bergeron}, Eddie and {Bergkoetter}, Matthew and {Bhatawdekar}, Rachana and {Birkmann}, Stephan and {Blazek}, Ronald and {Blome}, Claire and {Boccaletti}, Anthony and {B{\"o}ker}, Torsten and {Boia}, John and {Bonaventura}, Nina and {Bond}, Nicholas and {Bosley}, Kari and {Boucarut}, Ray and {Bourque}, Matthew and {Bouwman}, Jeroen and {Bower}, Gary and {Bowers}, Charles and {Boyer}, Martha and {Bradley}, Larry and {Brady}, Greg and {Braun}, Hannah and {Breda}, David and {Bresnahan}, Pamela and {Bright}, Stacey and {Britt}, Christopher and {Bromenschenkel}, Asa and {Brooks}, Brian and {Brooks}, Keira and {Brown}, Bob and {Brown}, Matthew and {Brown}, Patricia and {Bunker}, Andy and {Burger}, Matthew and {Bushouse}, Howard and {Cale}, Steven and {Cameron}, Alex and {Cameron}, Peter and {Canipe}, Alicia and {Caplinger}, James and {Caputo}, Francis and {Cara}, Mihai and {Carey}, Larkin and {Carniani}, Stefano and {Carrasquilla}, Maria and {Carruthers}, Margaret and {Case}, Michael and {Catherine}, Riggs and {Chance}, Don and {Chapman}, George and {Charlot}, St{\'e}phane and {Charlow}, Brian and {Chayer}, Pierre and {Chen}, Bin and {Cherinka}, Brian and {Chichester}, Sarah and {Chilton}, Zack and {Chonis}, Taylor and {Clampin}, Mark and {Clark}, Charles and {Clark}, Kerry and {Coe}, Dan and {Coleman}, Benee and {Comber}, Brian and {Comeau}, Tom and {Connolly}, Dennis and {Cooper}, James and {Cooper}, Rachel and {Coppock}, Eric and {Correnti}, Matteo and {Cossou}, Christophe and {Coulais}, Alain and {Coyle}, Laura and {Cracraft}, Misty and {Curti}, Mirko and {Cuturic}, Steven and {Davis}, Katherine and {Davis}, Michael and {Dean}, Bruce and {DeLisa}, Amy and {deMeester}, Wim and {Dencheva}, Nadia and {Dencheva}, Nadezhda and {DePasquale}, Joseph and {Deschenes}, Jeremy and {Hunor Detre}, {\"O}rs and {Diaz}, Rosa and {Dicken}, Dan and {DiFelice}, Audrey and {Dillman}, Matthew and {Dixon}, William and {Doggett}, Jesse and {Donaldson}, Tom and {Douglas}, Rob and {DuPrie}, Kimberly and {Dupuis}, Jean and {Durning}, John and {Easmin}, Nilufar and {Eck}, Weston and {Edeani}, Chinwe and {Egami}, Eiichi and {Ehrenwinkler}, Ralf and {Eisenhamer}, Jonathan and {Eisenhower}, Michael and {Elie}, Michelle and {Elliott}, James and {Elliott}, Kyle and {Ellis}, Tracy and {Engesser}, Michael and {Espinoza}, Nestor and {Etienne}, Odessa and {Etxaluze}, Mireya and {Falini}, Patrick and {Feeney}, Matthew and {Ferry}, Malcolm and {Filippazzo}, Joseph and {Fincham}, Brian and {Fix}, Mees and {Flagey}, Nicolas and {Florian}, Michael and {Flynn}, Jim and {Fontanella}, Erin and {Ford}, Terrance and {Forshay}, Peter and {Fox}, Ori and {Franz}, David and {Fu}, Henry and {Fullerton}, Alexander and {Galkin}, Sergey and {Galyer}, Anthony and {Garc{\'\i}a Mar{\'\i}n}, Macarena and {Gardner}, Jonathan P. and {Gardner}, Lisa and {Garland}, Dennis and {Garrett}, Bruce and {Gasman}, Danny and {Gaspar}, Andras and {Gaudreau}, Daniel and {Gauthier}, Peter and {Geers}, Vincent and {Geithner}, Paul and {Gennaro}, Mario and {Giardino}, Giovanna and {Girard}, Julien and {Giuliano}, Mark and {Glassmire}, Kirk and {Glauser}, Adrian},
        title = "{The Science Performance of JWST as Characterized in Commissioning}",
      journal = {\pasp},
         year = 2023,
        month = apr,
       volume = {135},
       number = {1046},
          eid = {048001},
        pages = {048001},
          doi = {10.1088/1538-3873/acb293},
archivePrefix = {arXiv},
       eprint = {2207.05632},
 primaryClass = {astro-ph.IM},
       adsurl = {https://ui.adsabs.harvard.edu/abs/2023PASP..135d8001R}
}

@misc{PyQSOFit,
       author = {{Guo}, Hengxiao and {Shen}, Yue and {Wang}, Shu},
        title = "{PyQSOFit: Python code to fit the spectrum of quasars}",
 howpublished = {Astrophysics Source Code Library, record ascl:1809.008},
         year = 2018,
        month = sep,
          eid = {ascl:1809.008},
       adsurl = {https://ui.adsabs.harvard.edu/abs/2018ascl.soft09008G}
}

@misc{Fu21_QSOfitmore,
       author = {{Fu}, Yuming},
        title = "{QSOFITMORE: a python package for fitting UV-optical spectra of quasars}",
         year = 2021,
        month = dec,
          eid = {10.5281/zenodo.5810042},
          doi = {10.5281/zenodo.5810042},
      version = {v1.1.0},
    publisher = {Zenodo},
       adsurl = {https://ui.adsabs.harvard.edu/abs/2021zndo...5810042F}
}

@ARTICLE{Ghez08,
       author = {{Ghez}, A.~M. and {Salim}, S. and {Weinberg}, N.~N. and {Lu}, J.~R. and {Do}, T. and {Dunn}, J.~K. and {Matthews}, K. and {Morris}, M.~R. and {Yelda}, S. and {Becklin}, E.~E. and {Kremenek}, T. and {Milosavljevic}, M. and {Naiman}, J.},
        title = "{Measuring Distance and Properties of the Milky Way's Central Supermassive Black Hole with Stellar Orbits}",
      journal = {\apj},
         year = 2008,
        month = dec,
       volume = {689},
       number = {2},
        pages = {1044-1062},
          doi = {10.1086/592738},
archivePrefix = {arXiv},
       eprint = {0808.2870},
 primaryClass = {astro-ph},
       adsurl = {https://ui.adsabs.harvard.edu/abs/2008ApJ...689.1044G}
}

@ARTICLE{Genzel10,
       author = {{Genzel}, Reinhard and {Eisenhauer}, Frank and {Gillessen}, Stefan},
        title = "{The Galactic Center massive black hole and nuclear star cluster}",
      journal = {Reviews of Modern Physics},
         year = 2010,
        month = oct,
       volume = {82},
       number = {4},
        pages = {3121-3195},
          doi = {10.1103/RevModPhys.82.3121},
archivePrefix = {arXiv},
       eprint = {1006.0064},
 primaryClass = {astro-ph.GA},
       adsurl = {https://ui.adsabs.harvard.edu/abs/2010RvMP...82.3121G}
}

@ARTICLE{Choi22,
       author = {{Choi}, Hyunseop and {Leighly}, Karen M. and {Terndrup}, Donald M. and {Dabbieri}, Collin and {Gallagher}, Sarah C. and {Richards}, Gordon T.},
        title = "{The Physical Properties of Low-redshift FeLoBAL Quasars. I. Spectral-synthesis Analysis of the Broad Absorption-line (BAL) Outflows Using SimBAL}",
      journal = {\apj},
         year = 2022,
        month = oct,
       volume = {937},
       number = {2},
          eid = {74},
        pages = {74},
          doi = {10.3847/1538-4357/ac61d9},
archivePrefix = {arXiv},
       eprint = {2203.11964},
 primaryClass = {astro-ph.GA},
       adsurl = {https://ui.adsabs.harvard.edu/abs/2022ApJ...937...74C}
}

@ARTICLE{Eracleous09,
       author = {{Eracleous}, Michael and {Lewis}, Karen T. and {Flohic}, H{\'e}l{\`e}ne M.~L.~G.},
        title = "{Double-peaked emission lines as a probe of the broad-line regions of active galactic nuclei}",
      journal = {\nar},
         year = 2009,
        month = jul,
       volume = {53},
       number = {7-10},
        pages = {133-139},
          doi = {10.1016/j.newar.2009.07.005},
       adsurl = {https://ui.adsabs.harvard.edu/abs/2009NewAR..53..133E}
}

@ARTICLE{Ward24,
       author = {{Ward}, Charlotte and {Gezari}, Suvi and {Nugent}, Peter and {Kerr}, Matthew and {Eracleous}, Michael and {Frederick}, Sara and {Hammerstein}, Erica and {Graham}, Matthew J. and {van Velzen}, Sjoert and {Kasliwal}, Mansi M. and {Laher}, Russ R. and {Masci}, Frank J. and {Purdum}, Josiah and {Racine}, Benjamin and {Smith}, Roger},
        title = "{Panic at the ISCO: Time-varying Double-peaked Broad Lines from Evolving Accretion Disks Are Common among Optically Variable AGNs}",
      journal = {\apj},
         year = 2024,
        month = feb,
       volume = {961},
       number = {2},
          eid = {172},
        pages = {172},
          doi = {10.3847/1538-4357/ad147d},
archivePrefix = {arXiv},
       eprint = {2309.02516},
 primaryClass = {astro-ph.GA},
       adsurl = {https://ui.adsabs.harvard.edu/abs/2024ApJ...961..172W}
}

@ARTICLE{Vito18,
       author = {{Vito}, F. and {Brandt}, W.~N. and {Yang}, G. and {Gilli}, R. and {Luo}, B. and {Vignali}, C. and {Xue}, Y.~Q. and {Comastri}, A. and {Koekemoer}, A.~M. and {Lehmer}, B.~D. and {Liu}, T. and {Paolillo}, M. and {Ranalli}, P. and {Schneider}, D.~P. and {Shemmer}, O. and {Volonteri}, M. and {Wang}, J.},
        title = "{High-redshift AGN in the Chandra Deep Fields: the obscured fraction and space density of the sub-L$_{*}$ population}",
      journal = {\mnras},
         year = 2018,
        month = jan,
       volume = {473},
       number = {2},
        pages = {2378-2406},
          doi = {10.1093/mnras/stx2486},
archivePrefix = {arXiv},
       eprint = {1709.07892},
 primaryClass = {astro-ph.GA},
       adsurl = {https://ui.adsabs.harvard.edu/abs/2018MNRAS.473.2378V}
}

@ARTICLE{Ni20,
       author = {{Ni}, Yueying and {Di Matteo}, Tiziana and {Gilli}, Roberto and {Croft}, Rupert A.~C. and {Feng}, Yu and {Norman}, Colin},
        title = "{QSO obscuration at high redshift (z {\ensuremath{\gtrsim}} 7): predictions from the BLUETIDES simulation}",
      journal = {\mnras},
         year = 2020,
        month = jun,
       volume = {495},
       number = {2},
        pages = {2135-2151},
          doi = {10.1093/mnras/staa1313},
archivePrefix = {arXiv},
       eprint = {1912.03780},
 primaryClass = {astro-ph.GA},
       adsurl = {https://ui.adsabs.harvard.edu/abs/2020MNRAS.495.2135N}
}

@ARTICLE{Matsuoka25,
       author = {{Matsuoka}, Yoshiki and {Onoue}, Masafusa and {Iwasawa}, Kazushi and {Aoki}, Kentaro and {Strauss}, Michael A. and {Silverman}, John D. and {Ding}, Xuheng and {Phillips}, Camryn L. and {Akiyama}, Masayuki and {Arita}, Junya and {Imanishi}, Masatoshi and {Izumi}, Takuma and {Kashikawa}, Nobunari and {Kawaguchi}, Toshihiro and {Kikuta}, Satoshi and {Kohno}, Kotaro and {Lee}, Chien-Hsiu and {Nagao}, Tohru and {Takahashi}, Ayumi and {Toba}, Yoshiki},
        title = "{SHELLQs. Bridging the Gap: JWST Unveils Obscured Quasars in the Most Luminous Galaxies at z > 6}",
      journal = {\apj},
         year = 2025,
        month = jul,
       volume = {988},
       number = {1},
          eid = {57},
        pages = {57},
          doi = {10.3847/1538-4357/addf4e},
archivePrefix = {arXiv},
       eprint = {2505.04825},
 primaryClass = {astro-ph.GA},
       adsurl = {https://ui.adsabs.harvard.edu/abs/2025ApJ...988...57M}
}

@ARTICLE{Matthee24,
       author = {{Matthee}, Jorryt and {Naidu}, Rohan P. and {Brammer}, Gabriel and {Chisholm}, John and {Eilers}, Anna-Christina and {Goulding}, Andy and {Greene}, Jenny and {Kashino}, Daichi and {Labbe}, Ivo and {Lilly}, Simon J. and {Mackenzie}, Ruari and {Oesch}, Pascal A. and {Weibel}, Andrea and {Wuyts}, Stijn and {Xiao}, Mengyuan and {Bordoloi}, Rongmon and {Bouwens}, Rychard and {van Dokkum}, Pieter and {Illingworth}, Garth and {Kramarenko}, Ivan and {Maseda}, Michael V. and {Mason}, Charlotte and {Meyer}, Romain A. and {Nelson}, Erica J. and {Reddy}, Naveen A. and {Shivaei}, Irene and {Simcoe}, Robert A. and {Yue}, Minghao},
        title = "{Little Red Dots: An Abundant Population of Faint Active Galactic Nuclei at z {\ensuremath{\sim}} 5 Revealed by the EIGER and FRESCO JWST Surveys}",
      journal = {\apj},
         year = 2024,
        month = mar,
       volume = {963},
       number = {2},
          eid = {129},
        pages = {129},
          doi = {10.3847/1538-4357/ad2345},
archivePrefix = {arXiv},
       eprint = {2306.05448},
 primaryClass = {astro-ph.GA},
       adsurl = {https://ui.adsabs.harvard.edu/abs/2024ApJ...963..129M}
}

@ARTICLE{Yang23,
       author = {{Yang}, Jinyi and {Wang}, Feige and {Fan}, Xiaohui and {Hennawi}, Joseph F. and {Barth}, Aaron J. and {Ba{\~n}ados}, Eduardo and {Sun}, Fengwu and {Liu}, Weizhe and {Cai}, Zheng and {Jiang}, Linhua and {Li}, Zihao and {Onoue}, Masafusa and {Schindler}, Jan-Torge and {Shen}, Yue and {Wu}, Yunjing and {Bhowmick}, Aklant K. and {Bieri}, Rebekka and {Blecha}, Laura and {Bosman}, Sarah and {Champagne}, Jaclyn B. and {Colina}, Luis and {Connor}, Thomas and {Costa}, Tiago and {Davies}, Frederick B. and {Decarli}, Roberto and {De Rosa}, Gisella and {Drake}, Alyssa B. and {Egami}, Eiichi and {Eilers}, Anna-Christina and {Evans}, Analis E. and {Farina}, Emanuele Paolo and {Habouzit}, Melanie and {Haiman}, Zoltan and {Jin}, Xiangyu and {Jun}, Hyunsung D. and {Kakiichi}, Koki and {Khusanova}, Yana and {Kulkarni}, Girish and {Loiacono}, Federica and {Lupi}, Alessandro and {Mazzucchelli}, Chiara and {Pan}, Zhiwei and {Rojas-Ruiz}, Sof{\'\i}a and {Strauss}, Michael A. and {Tee}, Wei Leong and {Trakhtenbrot}, Benny and {Trebitsch}, Maxime and {Venemans}, Bram and {Vestergaard}, Marianne and {Volonteri}, Marta and {Walter}, Fabian and {Xie}, Zhang-Liang and {Yue}, Minghao and {Zhang}, Haowen and {Zhang}, Huanian and {Zou}, Siwei},
        title = "{A SPectroscopic Survey of Biased Halos in the Reionization Era (ASPIRE): A First Look at the Rest-frame Optical Spectra of z > 6.5 Quasars Using JWST}",
      journal = {\apjl},
         year = 2023,
        month = jul,
       volume = {951},
       number = {1},
          eid = {L5},
        pages = {L5},
          doi = {10.3847/2041-8213/acc9c8},
archivePrefix = {arXiv},
       eprint = {2304.09888},
 primaryClass = {astro-ph.GA},
       adsurl = {https://ui.adsabs.harvard.edu/abs/2023ApJ...951L...5Y}
}

@ARTICLE{Farina22,
       author = {{Farina}, Emanuele Paolo and {Schindler}, Jan-Torge and {Walter}, Fabian and {Ba{\~n}ados}, Eduardo and {Davies}, Frederick B. and {Decarli}, Roberto and {Eilers}, Anna-Christina and {Fan}, Xiaohui and {Hennawi}, Joseph F. and {Mazzucchelli}, Chiara and {Meyer}, Romain A. and {Trakhtenbrot}, Benny and {Volonteri}, Marta and {Wang}, Feige and {Worseck}, G{\'a}bor and {Yang}, Jinyi and {Gutcke}, Thales A. and {Venemans}, Bram P. and {Bosman}, Sarah E.~I. and {Costa}, Tiago and {De Rosa}, Gisella and {Drake}, Alyssa B. and {Onoue}, Masafusa},
        title = "{The X-shooter/ALMA Sample of Quasars in the Epoch of Reionization. II. Black Hole Masses, Eddington Ratios, and the Formation of the First Quasars}",
      journal = {\apj},
         year = 2022,
        month = dec,
       volume = {941},
       number = {2},
          eid = {106},
        pages = {106},
          doi = {10.3847/1538-4357/ac9626},
archivePrefix = {arXiv},
       eprint = {2207.05113},
 primaryClass = {astro-ph.GA},
       adsurl = {https://ui.adsabs.harvard.edu/abs/2022ApJ...941..106F}
}

@ARTICLE{Mazzucchelli23,
       author = {{Mazzucchelli}, C. and {Bischetti}, M. and {D'Odorico}, V. and {Feruglio}, C. and {Schindler}, J.-T. and {Onoue}, M. and {Ba{\~n}ados}, E. and {Becker}, G.~D. and {Bian}, F. and {Carniani}, S. and {Decarli}, R. and {Eilers}, A.-C. and {Farina}, E.~P. and {Gallerani}, S. and {Lai}, S. and {Meyer}, R.~A. and {Rojas-Ruiz}, S. and {Satyavolu}, S. and {Venemans}, B.~P. and {Wang}, F. and {Yang}, J. and {Zhu}, Y.},
        title = "{XQR-30: Black hole masses and accretion rates of 42 z {\ensuremath{\gtrsim}} 6 quasars}",
      journal = {\aap},
         year = 2023,
        month = aug,
       volume = {676},
          eid = {A71},
        pages = {A71},
          doi = {10.1051/0004-6361/202346317},
archivePrefix = {arXiv},
       eprint = {2306.16474},
 primaryClass = {astro-ph.GA},
       adsurl = {https://ui.adsabs.harvard.edu/abs/2023A&A...676A..71M}
}

@ARTICLE{Ishimoto20,
       author = {{Ishimoto}, Rikako and {Kashikawa}, Nobunari and {Onoue}, Masafusa and {Matsuoka}, Yoshiki and {Izumi}, Takuma and {Strauss}, Michael A. and {Fujimoto}, Seiji and {Imanishi}, Masatoshi and {Ito}, Kei and {Iwasawa}, Kazushi and {Kawaguchi}, Toshihiro and {Lee}, Chien-Hsiu and {Liang}, Yongming and {Lu}, Ting-Yi and {Momose}, Rieko and {Toba}, Yoshiki and {Uchiyama}, Hisakazu},
        title = "{Subaru High-z Exploration of Low-luminosity Quasars (SHELLQs). XI. Proximity Zone Analysis for Faint Quasar Spectra at z {\ensuremath{\sim}} 6}",
      journal = {\apj},
         year = 2020,
        month = nov,
       volume = {903},
       number = {1},
          eid = {60},
        pages = {60},
          doi = {10.3847/1538-4357/abb80b},
archivePrefix = {arXiv},
       eprint = {2009.06648},
 primaryClass = {astro-ph.GA},
       adsurl = {https://ui.adsabs.harvard.edu/abs/2020ApJ...903...60I}
}

@ARTICLE{Takahashi24,
       author = {{Takahashi}, Ayumi and {Matsuoka}, Yoshiki and {Onoue}, Masafusa and {Strauss}, Michael A. and {Kashikawa}, Nobunari and {Toba}, Yoshiki and {Iwasawa}, Kazushi and {Imanishi}, Masatoshi and {Akiyama}, Masayuki and {Kawaguchi}, Toshihiro and {Noboriguchi}, Akatoki and {Lee}, Chien-Hsiu},
        title = "{Subaru High-z Exploration of Low-luminosity Quasars (SHELLQs). XVII. Black Hole Mass Distribution at z 6 Estimated via Spectral Comparison with Low-z Quasars}",
      journal = {\apj},
         year = 2024,
        month = jan,
       volume = {960},
       number = {2},
          eid = {112},
        pages = {112},
          doi = {10.3847/1538-4357/ad045e},
archivePrefix = {arXiv},
       eprint = {2310.12222},
 primaryClass = {astro-ph.GA},
       adsurl = {https://ui.adsabs.harvard.edu/abs/2024ApJ...960..112T}
}

@ARTICLE{Ding23,
       author = {{Ding}, Xuheng and {Onoue}, Masafusa and {Silverman}, John D. and {Matsuoka}, Yoshiki and {Izumi}, Takuma and {Strauss}, Michael A. and {Jahnke}, Knud and {Phillips}, Camryn L. and {Li}, Junyao and {Volonteri}, Marta and {Haiman}, Zoltan and {Andika}, Irham Taufik and {Aoki}, Kentaro and {Baba}, Shunsuke and {Bieri}, Rebekka and {Bosman}, Sarah E.~I. and {Bottrell}, Connor and {Eilers}, Anna-Christina and {Fujimoto}, Seiji and {Habouzit}, Melanie and {Imanishi}, Masatoshi and {Inayoshi}, Kohei and {Iwasawa}, Kazushi and {Kashikawa}, Nobunari and {Kawaguchi}, Toshihiro and {Kohno}, Kotaro and {Lee}, Chien-Hsiu and {Lupi}, Alessandro and {Lyu}, Jianwei and {Nagao}, Tohru and {Overzier}, Roderik and {Schindler}, Jan-Torge and {Schramm}, Malte and {Shimasaku}, Kazuhiro and {Toba}, Yoshiki and {Trakhtenbrot}, Benny and {Trebitsch}, Maxime and {Treu}, Tommaso and {Umehata}, Hideki and {Venemans}, Bram P. and {Vestergaard}, Marianne and {Walter}, Fabian and {Wang}, Feige and {Yang}, Jinyi},
        title = "{Detection of stellar light from quasar host galaxies at redshifts above 6}",
      journal = {\nat},
         year = 2023,
        month = sep,
       volume = {621},
       number = {7977},
        pages = {51-55},
          doi = {10.1038/s41586-023-06345-5},
archivePrefix = {arXiv},
       eprint = {2211.14329},
 primaryClass = {astro-ph.GA},
       adsurl = {https://ui.adsabs.harvard.edu/abs/2023Natur.621...51D}
}

@ARTICLE{Wolf23,
       author = {{Wolf}, J. and {Nandra}, K. and {Salvato}, M. and {Buchner}, J. and {Onoue}, M. and {Liu}, T. and {Arcodia}, R. and {Merloni}, A. and {Ciroi}, S. and {Di Mille}, F. and {Burwitz}, V. and {Brusa}, M. and {Ishimoto}, R. and {Kashikawa}, N. and {Matsuoka}, Y. and {Urrutia}, T. and {Waddell}, S.~G.~H.},
        title = "{X-ray emission from a rapidly accreting narrow-line Seyfert 1 galaxy at z = 6.56}",
      journal = {\aap},
         year = 2023,
        month = jan,
       volume = {669},
          eid = {A127},
        pages = {A127},
          doi = {10.1051/0004-6361/202244688},
archivePrefix = {arXiv},
       eprint = {2211.13820},
 primaryClass = {astro-ph.HE},
       adsurl = {https://ui.adsabs.harvard.edu/abs/2023A&A...669A.127W}
}

@ARTICLE{Lyu25,
       author = {{Lyu}, Jianwei and {Rieke}, George H. and {Stone}, Meredith and {Morrison}, Jane and {Alberts}, Stacey and {Jin}, Xiangyu and {Zhu}, Yongda and {Liu}, Weizhe and {Yang}, Jinyi},
        title = "{Fading Light, Fierce Winds: JWST Snapshot of a Sub-Eddington Quasar at Cosmic Dawn}",
      journal = {\apjl},
         year = 2025,
        month = mar,
       volume = {981},
       number = {1},
          eid = {L20},
        pages = {L20},
          doi = {10.3847/2041-8213/adb613},
archivePrefix = {arXiv},
       eprint = {2412.04548},
 primaryClass = {astro-ph.GA},
       adsurl = {https://ui.adsabs.harvard.edu/abs/2025ApJ...981L..20L}
}

@ARTICLE{Matsuoka24,
       author = {{Matsuoka}, Yoshiki and {Izumi}, Takuma and {Onoue}, Masafusa and {Strauss}, Michael A. and {Iwasawa}, Kazushi and {Kashikawa}, Nobunari and {Akiyama}, Masayuki and {Aoki}, Kentaro and {Arita}, Junya and {Imanishi}, Masatoshi and {Ishimoto}, Rikako and {Kawaguchi}, Toshihiro and {Kohno}, Kotaro and {Lee}, Chien-Hsiu and {Nagao}, Tohru and {Silverman}, John D. and {Toba}, Yoshiki},
        title = "{Discovery of Merging Twin Quasars at z = 6.05}",
      journal = {\apjl},
         year = 2024,
        month = apr,
       volume = {965},
       number = {1},
          eid = {L4},
        pages = {L4},
          doi = {10.3847/2041-8213/ad35c7},
archivePrefix = {arXiv},
       eprint = {2405.02465},
 primaryClass = {astro-ph.GA},
       adsurl = {https://ui.adsabs.harvard.edu/abs/2024ApJ...965L...4M}
}

@ARTICLE{Matsuoka22,
       author = {{Matsuoka}, Yoshiki and {Iwasawa}, Kazushi and {Onoue}, Masafusa and {Izumi}, Takuma and {Kashikawa}, Nobunari and {Strauss}, Michael A. and {Imanishi}, Masatoshi and {Nagao}, Tohru and {Akiyama}, Masayuki and {Silverman}, John D. and {Asami}, Naoko and {Bosch}, James and {Furusawa}, Hisanori and {Goto}, Tomotsugu and {Gunn}, James E. and {Harikane}, Yuichi and {Ikeda}, Hiroyuki and {Ishimoto}, Rikako and {Kawaguchi}, Toshihiro and {Kato}, Nanako and {Kikuta}, Satoshi and {Kohno}, Kotaro and {Komiyama}, Yutaka and {Lee}, Chien-Hsiu and {Lupton}, Robert H. and {Minezaki}, Takeo and {Miyazaki}, Satoshi and {Murayama}, Hitoshi and {Nishizawa}, Atsushi J. and {Oguri}, Masamune and {Ono}, Yoshiaki and {Ouchi}, Masami and {Price}, Paul A. and {Sameshima}, Hiroaki and {Sugiyama}, Naoshi and {Tait}, Philip J. and {Takada}, Masahiro and {Takahashi}, Ayumi and {Takata}, Tadafumi and {Tanaka}, Masayuki and {Toba}, Yoshiki and {Utsumi}, Yousuke and {Wang}, Shiang-Yu and {Yamashita}, Takuji},
        title = "{Subaru High-z Exploration of Low-luminosity Quasars (SHELLQs). XVI. 69 New Quasars at 5.8 < z < 7.0}",
      journal = {\apjs},
         year = 2022,
        month = mar,
       volume = {259},
       number = {1},
          eid = {18},
        pages = {18},
          doi = {10.3847/1538-4365/ac3d31},
archivePrefix = {arXiv},
       eprint = {2111.12766},
 primaryClass = {astro-ph.GA},
       adsurl = {https://ui.adsabs.harvard.edu/abs/2022ApJS..259...18M}
}

@ARTICLE{Barnett19,
       author = {{Euclid Collaboration} and {Barnett}, R. and {Warren}, S.~J. and {Mortlock}, D.~J. and {Cuby}, J. -G. and {Conselice}, C. and {Hewett}, P.~C. and {Willott}, C.~J. and {Auricchio}, N. and {Balaguera-Antol{\'\i}nez}, A. and {Baldi}, M. and {Bardelli}, S. and {Bellagamba}, F. and {Bender}, R. and {Biviano}, A. and {Bonino}, D. and {Bozzo}, E. and {Branchini}, E. and {Brescia}, M. and {Brinchmann}, J. and {Burigana}, C. and {Camera}, S. and {Capobianco}, V. and {Carbone}, C. and {Carretero}, J. and {Carvalho}, C.~S. and {Castander}, F.~J. and {Castellano}, M. and {Cavuoti}, S. and {Cimatti}, A. and {Cl{\'e}dassou}, R. and {Congedo}, G. and {Conversi}, L. and {Copin}, Y. and {Corcione}, L. and {Coupon}, J. and {Courtois}, H.~M. and {Cropper}, M. and {Da Silva}, A. and {Duncan}, C.~A.~J. and {Dusini}, S. and {Ealet}, A. and {Farrens}, S. and {Fosalba}, P. and {Fotopoulou}, S. and {Fourmanoit}, N. and {Frailis}, M. and {Fumana}, M. and {Galeotta}, S. and {Garilli}, B. and {Gillard}, W. and {Gillis}, B.~R. and {Graci{\'a}-Carpio}, J. and {Grupp}, F. and {Hoekstra}, H. and {Hormuth}, F. and {Israel}, H. and {Jahnke}, K. and {Kermiche}, S. and {Kilbinger}, M. and {Kirkpatrick}, C.~C. and {Kitching}, T. and {Kohley}, R. and {Kubik}, B. and {Kunz}, M. and {Kurki-Suonio}, H. and {Laureijs}, R. and {Ligori}, S. and {Lilje}, P.~B. and {Lloro}, I. and {Maiorano}, E. and {Mansutti}, O. and {Marggraf}, O. and {Martinet}, N. and {Marulli}, F. and {Massey}, R. and {Mauri}, N. and {Medinaceli}, E. and {Mei}, S. and {Mellier}, Y. and {Metcalf}, R.~B. and {Metge}, J.~J. and {Meylan}, G. and {Moresco}, M. and {Moscardini}, L. and {Munari}, E. and {Neissner}, C. and {Niemi}, S.~M. and {Nutma}, T. and {Padilla}, C. and {Paltani}, S. and {Pasian}, F. and {Paykari}, P. and {Percival}, W.~J. and {Pettorino}, V. and {Polenta}, G. and {Poncet}, M. and {Pozzetti}, L. and {Raison}, F. and {Renzi}, A. and {Rhodes}, J. and {Rix}, H. -W. and {Romelli}, E. and {Roncarelli}, M. and {Rossetti}, E. and {Saglia}, R. and {Sapone}, D. and {Scaramella}, R. and {Schneider}, P. and {Scottez}, V. and {Secroun}, A. and {Serrano}, S. and {Sirri}, G. and {Stanco}, L. and {Sureau}, F. and {Tallada-Cresp{\'\i}}, P. and {Tavagnacco}, D. and {Taylor}, A.~N. and {Tenti}, M. and {Tereno}, I. and {Toledo-Moreo}, R. and {Torradeflot}, F. and {Valenziano}, L. and {Vassallo}, T. and {Wang}, Y. and {Zacchei}, A. and {Zamorani}, G. and {Zoubian}, J. and {Zucca}, E.},
        title = "{Euclid preparation. V. Predicted yield of redshift 7 < z < 9 quasars from the wide survey}",
      journal = {\aap},
         year = 2019,
        month = nov,
       volume = {631},
          eid = {A85},
        pages = {A85},
          doi = {10.1051/0004-6361/201936427},
archivePrefix = {arXiv},
       eprint = {1908.04310},
 primaryClass = {astro-ph.GA},
       adsurl = {https://ui.adsabs.harvard.edu/abs/2019A&A...631A..85E}
}

@ARTICLE{Banados25,
       author = {{Ba{\~n}ados}, E. and {Le Brun}, V. and {Belladitta}, S. and {Momcheva}, I. and {Stern}, D. and {Wolf}, J. and {Ezziati}, M. and {Mortlock}, D.~J. and {Humphrey}, A. and {Smart}, R.~L. and {Casewell}, S.~L. and {P{\'e}rez-Garrido}, A. and {Goldman}, B. and {Mart{\'\i}n}, E.~L. and {Mohandasan}, A. and {Reyl{\'e}}, C. and {Dominguez-Tagle}, C. and {Copin}, Y. and {Lusso}, E. and {Matsuoka}, Y. and {McCarthy}, K. and {Ricci}, F. and {Rix}, H.-W. and {Rottgering}, H.~J.~A. and {Schindler}, J.-T. and {Weaver}, J.~R. and {Allaoui}, A. and {Bedrine}, T. and {Castellano}, M. and {Chabaud}, P.-Y. and {Daste}, G. and {Dufresne}, F. and {Gracia-Carpio}, J. and {K{\"u}mmel}, M. and {Moresco}, M. and {Scodeggio}, M. and {Surace}, C. and {Vibert}, D. and {Balestra}, A. and {Bonnefoi}, A. and {Caillat}, A. and {Cogato}, F. and {Costille}, A. and {Dusini}, S. and {Ferriol}, S. and {Franceschi}, E. and {Gillard}, W. and {Jahnke}, K. and {Le Mignant}, D. and {Ligori}, S. and {Medinaceli}, E. and {Morgante}, G. and {Passalacqua}, F. and {Paterson}, K. and {Pires}, S. and {Sirignano}, C. and {Andika}, I.~T. and {Atek}, H. and {Barrado}, D. and {Bisogni}, S. and {Conselice}, C.~J. and {Dannerbauer}, H. and {Decarli}, R. and {Dole}, H. and {Dupuy}, T. and {Feltre}, A. and {Fotopoulou}, S. and {Gillis}, B. and {Lopez}, X. Lopez and {Onoue}, M. and {Rodighiero}, G. and {Sedighi}, N. and {Shankar}, F. and {Siudek}, M. and {Spinoglio}, L. and {Vergani}, D. and {Vietri}, G. and {Walter}, F. and {Zamorani}, G. and {Zapatero Osorio}, M.~R. and {Zhang}, J.-Y. and {Bethermin}, M. and {Aghanim}, N. and {Altieri}, B. and {Amara}, A. and {Andreon}, S. and {Baccigalupi}, C. and {Baldi}, M. and {Bardelli}, S. and {Basset}, A. and {Battaglia}, P. and {Biviano}, A. and {Bonchi}, A. and {Bonino}, D. and {Branchini}, E. and {Brescia}, M. and {Brinchmann}, J. and {Camera}, S. and {Capobianco}, V. and {Carbone}, C. and {Carretero}, J. and {Casas}, S. and {Castignani}, G. and {Cavuoti}, S. and {Cimatti}, A. and {Colodro-Conde}, C. and {Congedo}, G. and {Conversi}, L. and {Courbin}, F. and {Courtois}, H.~M. and {Cropper}, M. and {Cuby}, J.-G. and {Da Silva}, A. and {Degaudenzi}, H. and {De Lucia}, G. and {Giorgio}, A.~M. Di and {Dolding}, C. and {Dubath}, F. and {Duncan}, C.~A.~J. and {Dupac}, X. and {Ealet}, A. and {Farina}, M. and {Faustini}, F. and {Fourmanoit}, N. and {Frailis}, M. and {Galeotta}, S. and {George}, K. and {Giocoli}, C. and {Granett}, B.~R. and {Grazian}, A. and {Grupp}, F. and {Guzzo}, L. and {Haugan}, S.~V.~H. and {Hoar}, J. and {Hoekstra}, H. and {Holmes}, W. and {Hook}, I. and {Hormuth}, F. and {Hornstrup}, A. and {Hudelot}, P. and {Jhabvala}, M. and {Joachimi}, B. and {Keih{\"a}nen}, E. and {Kermiche}, S. and {Kubik}, B. and {Kuijken}, K. and {Kunz}, M. and {Kurki-Suonio}, H. and {Lilje}, P.~B. and {Lindholm}, V. and {Lloro}, I. and {Mainetti}, G. and {Maino}, D. and {Maiorano}, E. and {Mansutti}, O. and {Marggraf}, O. and {Markovic}, K. and {Martinelli}, M. and {Martinet}, N. and {Marulli}, F. and {Massey}, R. and {Mei}, S. and {Mellier}, Y. and {Meneghetti}, M. and {Merlin}, E. and {Meylan}, G. and {Mora}, A. and {Moscardini}, L. and {Neissner}, C. and {Niemi}, S.-M. and {Nightingale}, J.~W. and {Padilla}, C. and {Paltani}, S. and {Pasian}, F. and {Pedersen}, K. and {Percival}, W.~J. and {Pettorino}, V. and {Polenta}, G. and {Poncet}, M. and {Popa}, L.~A. and {Pozzetti}, L. and {Raison}, F. and {Rebolo}, R. and {Renzi}, A. and {Rhodes}, J. and {Riccio}, G. and {Romelli}, E. and {Roncarelli}, M. and {Rossetti}, E. and {Saglia}, R. and {Sakr}, Z. and {Sapone}, D. and {Sartoris}, B. and {Schewtschenko}, J.~A. and {Schirmer}, M. and {Schneider}, P. and {Schrabback}, T. and {Secroun}, A. and {Sefusatti}, E. and {Seidel}, G.},
        title = "{Euclid: the potential of slitless infrared spectroscopy: a z = 5.4 quasar and new ultracool dwarfs}",
      journal = {\mnras},
         year = 2025,
        month = sep,
       volume = {542},
       number = {2},
        pages = {1088-1102},
          doi = {10.1093/mnras/staf1274},
archivePrefix = {arXiv},
       eprint = {2506.13945},
 primaryClass = {astro-ph.GA},
       adsurl = {https://ui.adsabs.harvard.edu/abs/2025MNRAS.542.1088B}
}

@ARTICLE{Fan23ARAA,
       author = {{Fan}, Xiaohui and {Ba{\~n}ados}, Eduardo and {Simcoe}, Robert A.},
        title = "{Quasars and the Intergalactic Medium at Cosmic Dawn}",
      journal = {\araa},
         year = 2023,
        month = aug,
       volume = {61},
        pages = {373-426},
          doi = {10.1146/annurev-astro-052920-102455},
archivePrefix = {arXiv},
       eprint = {2212.06907},
 primaryClass = {astro-ph.GA},
       adsurl = {https://ui.adsabs.harvard.edu/abs/2023ARA&A..61..373F}
}

@ARTICLE{Wang21,
       author = {{Wang}, Feige and {Yang}, Jinyi and {Fan}, Xiaohui and {Hennawi}, Joseph F. and {Barth}, Aaron J. and {Banados}, Eduardo and {Bian}, Fuyan and {Boutsia}, Konstantina and {Connor}, Thomas and {Davies}, Frederick B. and {Decarli}, Roberto and {Eilers}, Anna-Christina and {Farina}, Emanuele Paolo and {Green}, Richard and {Jiang}, Linhua and {Li}, Jiang-Tao and {Mazzucchelli}, Chiara and {Nanni}, Riccardo and {Schindler}, Jan-Torge and {Venemans}, Bram and {Walter}, Fabian and {Wu}, Xue-Bing and {Yue}, Minghao},
        title = "{A Luminous Quasar at Redshift 7.642}",
      journal = {\apjl},
         year = 2021,
        month = jan,
       volume = {907},
       number = {1},
          eid = {L1},
        pages = {L1},
          doi = {10.3847/2041-8213/abd8c6},
archivePrefix = {arXiv},
       eprint = {2101.03179},
 primaryClass = {astro-ph.GA},
       adsurl = {https://ui.adsabs.harvard.edu/abs/2021ApJ...907L...1W}
}

@ARTICLE{Yang20,
       author = {{Yang}, Jinyi and {Wang}, Feige and {Fan}, Xiaohui and {Hennawi}, Joseph F. and {Davies}, Frederick B. and {Yue}, Minghao and {Banados}, Eduardo and {Wu}, Xue-Bing and {Venemans}, Bram and {Barth}, Aaron J. and {Bian}, Fuyan and {Boutsia}, Konstantina and {Decarli}, Roberto and {Farina}, Emanuele Paolo and {Green}, Richard and {Jiang}, Linhua and {Li}, Jiang-Tao and {Mazzucchelli}, Chiara and {Walter}, Fabian},
        title = "{P{\={o}}niu{\={a}}'ena: A Luminous z = 7.5 Quasar Hosting a 1.5 Billion Solar Mass Black Hole}",
      journal = {\apjl},
         year = 2020,
        month = jul,
       volume = {897},
       number = {1},
          eid = {L14},
        pages = {L14},
          doi = {10.3847/2041-8213/ab9c26},
archivePrefix = {arXiv},
       eprint = {2006.13452},
 primaryClass = {astro-ph.GA},
       adsurl = {https://ui.adsabs.harvard.edu/abs/2020ApJ...897L..14Y}
}

@ARTICLE{Tee23,
       author = {{Tee}, Wei Leong and {Fan}, Xiaohui and {Wang}, Feige and {Yang}, Jinyi and {Malhotra}, Sangeeta and {Rhoads}, James E.},
        title = "{Predicting the Yields of z > 6.5 Quasar Surveys in the Era of Roman and Rubin}",
      journal = {\apj},
         year = 2023,
        month = oct,
       volume = {956},
       number = {1},
          eid = {52},
        pages = {52},
          doi = {10.3847/1538-4357/acf12d},
archivePrefix = {arXiv},
       eprint = {2308.12278},
 primaryClass = {astro-ph.GA},
       adsurl = {https://ui.adsabs.harvard.edu/abs/2023ApJ...956...52T}
}

@ARTICLE{Nanni17,
       author = {{Nanni}, R. and {Vignali}, C. and {Gilli}, R. and {Moretti}, A. and {Brandt}, W.~N.},
        title = "{The X-ray properties of z   6 luminous quasars}",
      journal = {\aap},
         year = 2017,
        month = jul,
       volume = {603},
          eid = {A128},
        pages = {A128},
          doi = {10.1051/0004-6361/201730484},
archivePrefix = {arXiv},
       eprint = {1704.08693},
 primaryClass = {astro-ph.GA},
       adsurl = {https://ui.adsabs.harvard.edu/abs/2017A&A...603A.128N}
}

@ARTICLE{Vestergaard02,
       author = {{Vestergaard}, M.},
        title = "{Determining Central Black Hole Masses in Distant Active Galaxies}",
      journal = {\apj},
         year = 2002,
        month = jun,
       volume = {571},
       number = {2},
        pages = {733-752},
          doi = {10.1086/340045},
archivePrefix = {arXiv},
       eprint = {astro-ph/0204106},
 primaryClass = {astro-ph},
       adsurl = {https://ui.adsabs.harvard.edu/abs/2002ApJ...571..733V}
}

@ARTICLE{Wu22,
       author = {{Wu}, Qiaoya and {Shen}, Yue},
        title = "{A Catalog of Quasar Properties from Sloan Digital Sky Survey Data Release 16}",
      journal = {\apjs},
         year = 2022,
        month = dec,
       volume = {263},
       number = {2},
          eid = {42},
        pages = {42},
          doi = {10.3847/1538-4365/ac9ead},
archivePrefix = {arXiv},
       eprint = {2209.03987},
 primaryClass = {astro-ph.GA},
       adsurl = {https://ui.adsabs.harvard.edu/abs/2022ApJS..263...42W}
}

@ARTICLE{GH05,
       author = {{Greene}, Jenny E. and {Ho}, Luis C.},
        title = "{Estimating Black Hole Masses in Active Galaxies Using the H{\ensuremath{\alpha}} Emission Line}",
      journal = {\apj},
         year = 2005,
        month = sep,
       volume = {630},
       number = {1},
        pages = {122-129},
          doi = {10.1086/431897},
archivePrefix = {arXiv},
       eprint = {astro-ph/0508335},
 primaryClass = {astro-ph},
       adsurl = {https://ui.adsabs.harvard.edu/abs/2005ApJ...630..122G}
}

@ARTICLE{Onoue25,
       author = {{Onoue}, Masafusa and {Ding}, Xuheng and {Silverman}, John D. and {Matsuoka}, Yoshiki and {Izumi}, Takuma and {Strauss}, Michael A. and {Ward}, Charlotte and {Phillips}, Camryn L. and {Ito}, Kei and {Andika}, Irham T. and {Aoki}, Kentaro and {Arita}, Junya and {Baba}, Shunsuke and {Bieri}, Rebekka and {Bosman}, Sarah E.~I. and {Eilers}, Anna-Christina and {Fujimoto}, Seiji and {Habouzit}, Melanie and {Haiman}, Zoltan and {Imanishi}, Masatoshi and {Inayoshi}, Kohei and {Iwasawa}, Kazushi and {Jahnke}, Knud and {Kashikawa}, Nobunari and {Kawaguchi}, Toshihiro and {Kohno}, Kotaro and {Lee}, Chien-Hsiu and {Li}, Junyao and {Lupi}, Alessandro and {Lyu}, Jianwei and {Nagao}, Tohru and {Overzier}, Roderik and {Schindler}, Jan-Torge and {Schramm}, Malte and {Scoggins}, Matthew T. and {Shimasaku}, Kazuhiro and {Toba}, Yoshiki and {Trakhtenbrot}, Benny and {Trebitsch}, Maxime and {Treu}, Tommaso and {Umehata}, Hideki and {Venemans}, Bram and {Vestergaard}, Marianne and {Volonteri}, Marta and {Walter}, Fabian and {Wang}, Feige and {Yang}, Jinyi and {Zhang}, Haowen},
        title = "{A post-starburst pathway for the formation of massive galaxies and black holes at z > 6}",
      journal = {Nature Astronomy},
         year = 2025,
        month = aug,
       volume = {9},
        pages = {1541-1552},
          doi = {10.1038/s41550-025-02628-1},
archivePrefix = {arXiv},
       eprint = {2409.07113},
 primaryClass = {astro-ph.GA},
       adsurl = {https://ui.adsabs.harvard.edu/abs/2025NatAs...9.1541O}
}

@ARTICLE{Kaspi00,
       author = {{Kaspi}, Shai and {Smith}, Paul S. and {Netzer}, Hagai and {Maoz}, Dan and {Jannuzi}, Buell T. and {Giveon}, Uriel},
        title = "{Reverberation Measurements for 17 Quasars and the Size-Mass-Luminosity Relations in Active Galactic Nuclei}",
      journal = {\apj},
         year = 2000,
        month = apr,
       volume = {533},
       number = {2},
        pages = {631-649},
          doi = {10.1086/308704},
archivePrefix = {arXiv},
       eprint = {astro-ph/9911476},
 primaryClass = {astro-ph},
       adsurl = {https://ui.adsabs.harvard.edu/abs/2000ApJ...533..631K}
}

@ARTICLE{Yue24,
       author = {{Yue}, Minghao and {Eilers}, Anna-Christina and {Simcoe}, Robert A. and {Mackenzie}, Ruari and {Matthee}, Jorryt and {Kashino}, Daichi and {Bordoloi}, Rongmon and {Lilly}, Simon J. and {Naidu}, Rohan P.},
        title = "{EIGER. V. Characterizing the Host Galaxies of Luminous Quasars at z {\ensuremath{\gtrsim}} 6}",
      journal = {\apj},
         year = 2024,
        month = may,
       volume = {966},
       number = {2},
          eid = {176},
        pages = {176},
          doi = {10.3847/1538-4357/ad3914},
archivePrefix = {arXiv},
       eprint = {2309.04614},
 primaryClass = {astro-ph.GA},
       adsurl = {https://ui.adsabs.harvard.edu/abs/2024ApJ...966..176Y}
}

@ARTICLE{Wang23,
       author = {{Wang}, Feige and {Yang}, Jinyi and {Hennawi}, Joseph F. and {Fan}, Xiaohui and {Sun}, Fengwu and {Champagne}, Jaclyn B. and {Costa}, Tiago and {Habouzit}, Melanie and {Endsley}, Ryan and {Li}, Zihao and {Lin}, Xiaojing and {Meyer}, Romain A. and {Schindler}, Jan-Torge and {Wu}, Yunjing and {Ba{\~n}ados}, Eduardo and {Barth}, Aaron J. and {Bhowmick}, Aklant K. and {Bieri}, Rebekka and {Blecha}, Laura and {Bosman}, Sarah and {Cai}, Zheng and {Colina}, Luis and {Connor}, Thomas and {Davies}, Frederick B. and {Decarli}, Roberto and {De Rosa}, Gisella and {Drake}, Alyssa B. and {Egami}, Eiichi and {Eilers}, Anna-Christina and {Evans}, Analis E. and {Farina}, Emanuele Paolo and {Haiman}, Zoltan and {Jiang}, Linhua and {Jin}, Xiangyu and {Jun}, Hyunsung D. and {Kakiichi}, Koki and {Khusanova}, Yana and {Kulkarni}, Girish and {Li}, Mingyu and {Liu}, Weizhe and {Loiacono}, Federica and {Lupi}, Alessandro and {Mazzucchelli}, Chiara and {Onoue}, Masafusa and {Pudoka}, Maria A. and {Rojas-Ruiz}, Sof{\'\i}a and {Shen}, Yue and {Strauss}, Michael A. and {Tee}, Wei Leong and {Trakhtenbrot}, Benny and {Trebitsch}, Maxime and {Venemans}, Bram and {Volonteri}, Marta and {Walter}, Fabian and {Xie}, Zhang-Liang and {Yue}, Minghao and {Zhang}, Haowen and {Zhang}, Huanian and {Zou}, Siwei},
        title = "{A SPectroscopic Survey of Biased Halos in the Reionization Era (ASPIRE): JWST Reveals a Filamentary Structure around a z = 6.61 Quasar}",
      journal = {\apjl},
         year = 2023,
        month = jul,
       volume = {951},
       number = {1},
          eid = {L4},
        pages = {L4},
          doi = {10.3847/2041-8213/accd6f},
archivePrefix = {arXiv},
       eprint = {2304.09894},
 primaryClass = {astro-ph.GA},
       adsurl = {https://ui.adsabs.harvard.edu/abs/2023ApJ...951L...4W}
}

@ARTICLE{Kashikawa15,
   author = {{Kashikawa}, N. and {Ishizaki}, Y. and {Willott}, C.~J. and 
	{Onoue}, M. and {Im}, M. and {Furusawa}, H. and {Toshikawa}, J. and 
	{Ishikawa}, S. and {Niino}, Y. and {Shimasaku}, K. and {Ouchi}, M. and 
	{Hibon}, P.},
    title = "{The Subaru High-z Quasar Survey: Discovery of Faint z \~{} 6 Quasars}",
  journal = {\apj},
archivePrefix = "arXiv",
   eprint = {1410.7401},
     year = 2015,
    month = jan,
   volume = 798,
      eid = {28},
    pages = {28},
      doi = {10.1088/0004-637X/798/1/28},
   adsurl = {http://ads.nao.ac.jp/abs/2015ApJ...798...28K}
}

@ARTICLE{Mortlock11,
   author = {{Mortlock}, D.~J. and {Warren}, S.~J. and {Venemans}, B.~P. and 
	{Patel}, M. and {Hewett}, P.~C. and {McMahon}, R.~G. and {Simpson}, C. and 
	{Theuns}, T. and {Gonz{\'a}les-Solares}, E.~A. and {Adamson}, A. and 
	{Dye}, S. and {Hambly}, N.~C. and {Hirst}, P. and {Irwin}, M.~J. and 
	{Kuiper}, E. and {Lawrence}, A. and {R{\"o}ttgering}, H.~J.~A.
	},
    title = "{A luminous quasar at a redshift of z = 7.085}",
  journal = {\nat},
archivePrefix = "arXiv",
   eprint = {1106.6088},
 primaryClass = "astro-ph.CO",
     year = 2011,
    month = jun,
   volume = 474,
    pages = {616-619},
      doi = {10.1038/nature10159},
   adsurl = {http://ads.nao.ac.jp/abs/2011Natur.474..616M}
}

@ARTICLE{Matsuoka18a,
       author = {{Matsuoka}, Yoshiki and {Onoue}, Masafusa and {Kashikawa}, Nobunari and
        {Iwasawa}, Kazushi and {Strauss}, Michael A. and {Nagao}, Tohru
        and {Imanishi}, Masatoshi and {Lee}, Chien-Hsiu and {Akiyama},
        Masayuki and {Asami}, Naoko and {Bosch}, James and {Foucaud},
        S{\'e}bastien and {Furusawa}, Hisanori and {Goto}, Tomotsugu and
        {Gunn}, James E. and {Harikane}, Yuichi and {Ikeda}, Hiroyuki
        and {Izumi}, Takuma and {Kawaguchi}, Toshihiro and {Kikuta},
        Satoshi and {Kohno}, Kotaro and {Komiyama}, Yutaka and {Lupton},
        Robert H. and {Minezaki}, Takeo and {Miyazaki}, Satoshi and
        {Morokuma}, Tomoki and {Murayama}, Hitoshi and {Niida}, Mana and
        {Nishizawa}, Atsushi J. and {Oguri}, Masamune and {Ono},
        Yoshiaki and {Ouchi}, Masami and {Price}, Paul A. and
        {Sameshima}, Hiroaki and {Schulze}, Andreas and {Shirakata},
        Hikari and {Silverman}, John D. and {Sugiyama}, Naoshi and
        {Tait}, Philip J. and {Takada}, Masahiro and {Takata}, Tadafumi
        and {Tanaka}, Masayuki and {Tang}, Ji-Jia and {Toba}, Yoshiki
        and {Utsumi}, Yousuke and {Wang}, Shiang-Yu},
        title = "{Subaru High-z Exploration of Low-Luminosity Quasars (SHELLQs). II.
        Discovery of 32 quasars and luminous galaxies at 5.7 \&lt; z
        {\ensuremath{\leq}} 6.8}",
      journal = {Publications of the Astronomical Society of Japan},
         year = 2018,
        month = Jan,
       volume = {70},
          eid = {S35},
        pages = {S35},
          doi = {10.1093/pasj/psx046},
archivePrefix = {arXiv},
       eprint = {1704.05854},
 primaryClass = {astro-ph.GA},
       adsurl = {https://ui.adsabs.harvard.edu/\#abs/2018PASJ...70S..35M}
}

@ARTICLE{Banados16,
   author = {{Ba{\~n}ados}, E. and {Venemans}, B.~P. and {Decarli}, R. and 
	{Farina}, E.~P. and {Mazzucchelli}, C. and {Walter}, F. and 
	{Fan}, X. and {Stern}, D. and {Schlafly}, E. and {Chambers}, K.~C. and 
	{Rix}, H.-W. and {Jiang}, L. and {McGreer}, I. and {Simcoe}, R. and 
	{Wang}, F. and {Yang}, J. and {Morganson}, E. and {De Rosa}, G. and 
	{Greiner}, J. and {Balokovi{\'c}}, M. and {Burgett}, W.~S. and 
	{Cooper}, T. and {Draper}, P.~W. and {Flewelling}, H. and {Hodapp}, K.~W. and 
	{Jun}, H.~D. and {Kaiser}, N. and {Kudritzki}, R.-P. and {Magnier}, E.~A. and 
	{Metcalfe}, N. and {Miller}, D. and {Schindler}, J.-T. and {Tonry}, J.~L. and 
	{Wainscoat}, R.~J. and {Waters}, C. and {Yang}, Q.},
    title = "{The Pan-STARRS1 Distant $z > 5.6$ Quasar Survey: More than 100 Quasars within the First Gyr of the Universe}",
  journal = {\apjs},
archivePrefix = "arXiv",
   eprint = {1608.03279},
     year = 2016,
    month = nov,
   volume = 227,
      eid = {11},
    pages = {11},
      doi = {10.3847/0067-0049/227/1/11},
   adsurl = {http://ads.nao.ac.jp/abs/2016ApJS..227...11B}
}

@ARTICLE{Matsuoka16,
   author = {{Matsuoka}, Y. and {Onoue}, M. and {Kashikawa}, N. and {Iwasawa}, K. and 
	{Strauss}, M.~A. and {Nagao}, T. and {Imanishi}, M. and {Niida}, M. and 
	{Toba}, Y. and {Akiyama}, M. and {Asami}, N. and {Bosch}, J. and 
	{Foucaud}, S. and {Furusawa}, H. and {Goto}, T. and {Gunn}, J.~E. and 
	{Harikane}, Y. and {Ikeda}, H. and {Kawaguchi}, T. and {Kikuta}, S. and 
	{Komiyama}, Y. and {Lupton}, R.~H. and {Minezaki}, T. and {Miyazaki}, S. and 
	{Morokuma}, T. and {Murayama}, H. and {Nishizawa}, A.~J. and 
	{Ono}, Y. and {Ouchi}, M. and {Price}, P.~A. and {Sameshima}, H. and 
	{Silverman}, J.~D. and {Sugiyama}, N. and {Tait}, P.~J. and 
	{Takada}, M. and {Takata}, T. and {Tanaka}, M. and {Tang}, J.-J. and 
	{Utsumi}, Y.},
    title = "{Subaru High-z Exploration of Low-luminosity Quasars (SHELLQs). I. Discovery of 15 Quasars and Bright Galaxies at 5.7 $\lt$ z $\lt$ 6.9}",
  journal = {\apj},
archivePrefix = "arXiv",
   eprint = {1603.02281},
     year = 2016,
    month = sep,
   volume = 828,
      eid = {26},
    pages = {26},
      doi = {10.3847/0004-637X/828/1/26},
   adsurl = {http://ads.nao.ac.jp/abs/2016ApJ...828...26M}
}

@ARTICLE{Wu15,
   author = {{Wu}, X.-B. and {Wang}, F. and {Fan}, X. and {Yi}, W. and {Zuo}, W. and 
	{Bian}, F. and {Jiang}, L. and {McGreer}, I.~D. and {Wang}, R. and 
	{Yang}, J. and {Yang}, Q. and {Thompson}, D. and {Beletsky}, Y.
	},
    title = "{An ultraluminous quasar with a twelve-billion-solar-mass black hole at redshift 6.30}",
  journal = {\nat},
archivePrefix = "arXiv",
   eprint = {1502.07418},
     year = 2015,
    month = feb,
   volume = 518,
    pages = {512-515},
      doi = {10.1038/nature14241},
   adsurl = {http://ads.nao.ac.jp/abs/2015Natur.518..512W}
}

@ARTICLE{SSP_PDR3,
       author = {{Aihara}, Hiroaki and {AlSayyad}, Yusra and {Ando}, Makoto and {Armstrong}, Robert and {Bosch}, James and {Egami}, Eiichi and {Furusawa}, Hisanori and {Furusawa}, Junko and {Harasawa}, Sumiko and {Harikane}, Yuichi and {Hsieh}, Bau-Ching and {Ikeda}, Hiroyuki and {Ito}, Kei and {Iwata}, Ikuru and {Kodama}, Tadayuki and {Koike}, Michitaro and {Kokubo}, Mitsuru and {Komiyama}, Yutaka and {Li}, Xiangchong and {Liang}, Yongming and {Lin}, Yen-Ting and {Lupton}, Robert H. and {Lust}, Nate B. and {MacArthur}, Lauren A. and {Mawatari}, Ken and {Mineo}, Sogo and {Miyatake}, Hironao and {Miyazaki}, Satoshi and {More}, Surhud and {Morishima}, Takahiro and {Murayama}, Hitoshi and {Nakajima}, Kimihiko and {Nakata}, Fumiaki and {Nishizawa}, Atsushi J. and {Oguri}, Masamune and {Okabe}, Nobuhiro and {Okura}, Yuki and {Ono}, Yoshiaki and {Osato}, Ken and {Ouchi}, Masami and {Pan}, Yen-Chen and {Plazas Malag{\'o}n}, Andr{\'e}s A. and {Price}, Paul A. and {Reed}, Sophie L. and {Rykoff}, Eli S. and {Shibuya}, Takatoshi and {Simunovic}, Mirko and {Strauss}, Michael A. and {Sugimori}, Kanako and {Suto}, Yasushi and {Suzuki}, Nao and {Takada}, Masahiro and {Takagi}, Yuhei and {Takata}, Tadafumi and {Takita}, Satoshi and {Tanaka}, Masayuki and {Tang}, Shenli and {Taranu}, Dan S. and {Terai}, Tsuyoshi and {Toba}, Yoshiki and {Turner}, Edwin L. and {Uchiyama}, Hisakazu and {Vijarnwannaluk}, Bovornpratch and {Waters}, Christopher Z. and {Yamada}, Yoshihiko and {Yamamoto}, Naoaki and {Yamashita}, Takuji},
        title = "{Third data release of the Hyper Suprime-Cam Subaru Strategic Program}",
      journal = {\pasj},
         year = 2022,
        month = apr,
       volume = {74},
       number = {2},
        pages = {247-272},
          doi = {10.1093/pasj/psab122},
archivePrefix = {arXiv},
       eprint = {2108.13045},
 primaryClass = {astro-ph.IM},
       adsurl = {https://ui.adsabs.harvard.edu/abs/2022PASJ...74..247A}
}

@ARTICLE{Jiang07,
   author = {{Jiang}, L. and {Fan}, X. and {Vestergaard}, M. and {Kurk}, J.~D. and 
	{Walter}, F. and {Kelly}, B.~C. and {Strauss}, M.~A.},
    title = "{Gemini Near-Infrared Spectroscopy of Luminous z \~{} 6 Quasars: Chemical Abundances, Black Hole Masses, and Mg II Absorption}",
  journal = {\aj},
archivePrefix = "arXiv",
   eprint = {0707.1663},
     year = 2007,
    month = sep,
   volume = 134,
    pages = {1150},
      doi = {10.1086/520811},
   adsurl = {http://ads.nao.ac.jp/abs/2007AJ....134.1150J}
}

@ARTICLE{Matsuoka18b,
       author = {{Matsuoka}, Yoshiki and {Iwasawa}, Kazushi and {Onoue}, Masafusa and
        {Kashikawa}, Nobunari and {Strauss}, Michael A. and {Lee},
        Chien-Hsiu and {Imanishi}, Masatoshi and {Nagao}, Tohru and
        {Akiyama}, Masayuki and {Asami}, Naoko and {Bosch}, James and
        {Furusawa}, Hisanori and {Goto}, Tomotsugu and {Gunn}, James E.
        and {Harikane}, Yuichi and {Ikeda}, Hiroyuki and {Izumi}, Takuma
        and {Kawaguchi}, Toshihiro and {Kato}, Nanako and {Kikuta},
        Satoshi and {Kohno}, Kotaro and {Komiyama}, Yutaka and {Lupton},
        Robert H. and {Minezaki}, Takeo and {Miyazaki}, Satoshi and
        {Morokuma}, Tomoki and {Murayama}, Hitoshi and {Niida}, Mana and
        {Nishizawa}, Atsushi J. and {Oguri}, Masamune and {Ono},
        Yoshiaki and {Ouchi}, Masami and {Price}, Paul A. and
        {Sameshima}, Hiroaki and {Schulze}, Andreas and {Shirakata},
        Hikari and {Silverman}, John D. and {Sugiyama}, Naoshi and
        {Tait}, Philip J. and {Takada}, Masahiro and {Takata}, Tadafumi
        and {Tanaka}, Masayuki and {Tang}, Ji-Jia and {Toba}, Yoshiki
        and {Utsumi}, Yousuke and {Wang}, Shiang-Yu and {Yamashita},
        Takuji},
        title = "{Subaru High-z Exploration of Low-luminosity Quasars (SHELLQs). IV. Discovery of 41 Quasars and Luminous Galaxies at 5.7 {\ensuremath{\leq}} z {\ensuremath{\leq}} 6.9}",
      journal = {The Astrophysical Journal Supplement Series},
         year = 2018,
        month = Jul,
       volume = {237},
          eid = {5},
        pages = {5},
          doi = {10.3847/1538-4365/aac724},
archivePrefix = {arXiv},
       eprint = {1803.01861},
 primaryClass = {astro-ph.GA},
       adsurl = {https://ui.adsabs.harvard.edu/\#abs/2018ApJS..237....5M}
}

@ARTICLE{Matsuoka18c,
       author = {{Matsuoka}, Yoshiki and {Strauss}, Michael A. and {Kashikawa}, Nobunari and
         {Onoue}, Masafusa and {Iwasawa}, Kazushi and {Tang}, Ji-Jia and
         {Lee}, Chien-Hsiu and {Imanishi}, Masatoshi and {Nagao}, Tohru and
         {Akiyama}, Masayuki and {Asami}, Naoko and {Bosch}, James and
         {Furusawa}, Hisanori and {Goto}, Tomotsugu and {Gunn}, James E. and
         {Harikane}, Yuichi and {Ikeda}, Hiroyuki and {Izumi}, Takuma and
         {Kawaguchi}, Toshihiro and {Kato}, Nanako and {Kikuta}, Satoshi and
         {Kohno}, Kotaro and {Komiyama}, Yutaka and {Lupton}, Robert H. and
         {Minezaki}, Takeo and {Miyazaki}, Satoshi and {Murayama}, Hitoshi and
         {Niida}, Mana and {Nishizawa}, Atsushi J. and {Noboriguchi}, Akatoki and
         {Oguri}, Masamune and {Ono}, Yoshiaki and {Ouchi}, Masami and
         {Price}, Paul A. and {Sameshima}, Hiroaki and {Schulze}, Andreas and
         {Shirakata}, Hikari and {Silverman}, John D. and {Sugiyama}, Naoshi and
         {Tait}, Philip J. and {Takada}, Masahiro and {Takata}, Tadafumi and
         {Tanaka}, Masayuki and {Toba}, Yoshiki and {Utsumi}, Yousuke and
         {Wang}, Shiang-Yu and {Yamashita}, Takuji},
        title = "{Subaru High-z  Exploration of Low-luminosity Quasars (SHELLQs). V. Quasar Luminosity Function and Contribution to Cosmic Reionization at z = 6}",
      journal = {\apj},
         year = "2018",
        month = "Dec",
       volume = {869},
       number = {2},
          eid = {150},
        pages = {150},
          doi = {10.3847/1538-4357/aaee7a},
archivePrefix = {arXiv},
       eprint = {1811.01963},
 primaryClass = {astro-ph.GA},
       adsurl = {https://ui.adsabs.harvard.edu/abs/2018ApJ...869..150M}
}

@ARTICLE{Onoue19,
       author = {{Onoue}, Masafusa and {Kashikawa}, Nobunari and {Matsuoka}, Yoshiki and {Kato}, Nanako and {Izumi}, Takuma and {Nagao}, Tohru and {Strauss}, Michael A. and {Harikane}, Yuichi and {Imanishi}, Masatoshi and {Ito}, Kei and {Iwasawa}, Kazushi and {Kawaguchi}, Toshihiro and {Lee}, Chien-Hsiu and {Noboriguchi}, Akatoki and {Suh}, Hyewon and {Tanaka}, Masayuki and {Toba}, Yoshiki},
        title = "{Subaru High-z Exploration of Low-luminosity Quasars (SHELLQs). VI. Black Hole Mass Measurements of Six Quasars at 6.1 {\ensuremath{\leq}} z {\ensuremath{\leq}} 6.7}",
      journal = {\apj},
         year = 2019,
        month = aug,
       volume = {880},
       number = {2},
          eid = {77},
        pages = {77},
          doi = {10.3847/1538-4357/ab29e9},
archivePrefix = {arXiv},
       eprint = {1904.07278},
 primaryClass = {astro-ph.GA},
       adsurl = {https://ui.adsabs.harvard.edu/abs/2019ApJ...880...77O}
}

@ARTICLE{Onoue17,
       author = {{Onoue}, Masafusa and {Kashikawa}, Nobunari and {Willott}, Chris J. and
         {Hibon}, Pascale and {Im}, Myungshin and {Furusawa}, Hisanori and
         {Harikane}, Yuichi and {Imanishi}, Masatoshi and {Ishikawa}, Shogo and
         {Kikuta}, Satoshi and {Matsuoka}, Yoshiki and {Nagao}, Tohru and
         {Niino}, Yuu and {Ono}, Yoshiaki and {Ouchi}, Masami and
         {Tanaka}, Masayuki and {Tang}, Ji-Jia and {Toshikawa}, Jun and
         {Uchiyama}, Hisakazu},
        title = "{Minor Contribution of Quasars to Ionizing Photon Budget at z\CID{3} 6: Update on Quasar Luminosity Function at the Faint End with Subaru/Suprime-Cam}",
      journal = {\apj},
         year = "2017",
        month = "Oct",
       volume = {847},
          eid = {L15},
        pages = {L15},
          doi = {10.3847/2041-8213/aa8cc6},
archivePrefix = {arXiv},
       eprint = {1709.04413},
 primaryClass = {astro-ph.GA},
       adsurl = {https://ui.adsabs.harvard.edu/\#abs/2017ApJ...847L..15O}
}

@ARTICLE{Wang17,
       author = {{Wang}, Feige and {Fan}, Xiaohui and {Yang}, Jinyi and {Wu}, Xue-Bing
        and {Yang}, Qian and {Bian}, Fuyan and {McGreer}, Ian D. and
        {Li}, Jiang-Tao and {Li}, Zefeng and {Ding}, Jiani and {Dey},
        Arjun and {Dye}, Simon and {Findlay}, Joseph R. and {Green},
        Richard and {James}, David and {Jiang}, Linhua and {Lang},
        Dustin and {Lawrence}, Andy and {Myers}, Adam D. and {Ross},
        Nicholas P. and {Schlegel}, David J. and {Shanks}, Tom},
        title = "{First Discoveries of z \&gt; 6 Quasars with the DECam Legacy Survey and
        UKIRT Hemisphere Survey}",
      journal = {\apj},
         year = 2017,
        month = Apr,
       volume = {839},
          eid = {27},
        pages = {27},
          doi = {10.3847/1538-4357/aa689f},
archivePrefix = {arXiv},
       eprint = {1703.07490},
 primaryClass = {astro-ph.GA},
       adsurl = {https://ui.adsabs.harvard.edu/\#abs/2017ApJ...839...27W}
}

@ARTICLE{Mazzucchelli17,
       author = {{Mazzucchelli}, C. and {Ba{\~n}ados}, E. and {Venemans}, B.~P. and
        {Decarli}, R. and {Farina}, E.~P. and {Walter}, F. and {Eilers},
        A. -C. and {Rix}, H. -W. and {Simcoe}, R. and {Stern}, D. and
        {Fan}, X. and {Schlafly}, E. and {De Rosa}, G. and {Hennawi}, J.
        and {Chambers}, K.~C. and {Greiner}, J. and {Burgett}, W. and
        {Draper}, P.~W. and {Kaiser}, N. and {Kudritzki}, R. -P. and
        {Magnier}, E. and {Metcalfe}, N. and {Waters}, C. and
        {Wainscoat}, R.~J.},
        title = "{Physical Properties of 15 Quasars at z {\ensuremath{\gtrsim}} 6.5}",
      journal = {\apj},
         year = 2017,
        month = Nov,
       volume = {849},
          eid = {91},
        pages = {91},
          doi = {10.3847/1538-4357/aa9185},
archivePrefix = {arXiv},
       eprint = {1710.01251},
 primaryClass = {astro-ph.GA},
       adsurl = {https://ui.adsabs.harvard.edu/\#abs/2017ApJ...849...91M}
}

@ARTICLE{Banados18,
       author = {{Ba{\~n}ados}, Eduardo and {Venemans}, Bram P. and {Mazzucchelli},
        Chiara and {Farina}, Emanuele P. and {Walter}, Fabian and
        {Wang}, Feige and {Decarli}, Roberto and {Stern}, Daniel and
        {Fan}, Xiaohui and {Davies}, Frederick B. and {Hennawi}, Joseph
        F. and {Simcoe}, Robert A. and {Turner}, Monica L. and {Rix},
        Hans-Walter and {Yang}, Jinyi and {Kelson}, Daniel D. and
        {Rudie}, Gwen C. and {Winters}, Jan Martin},
        title = "{An 800-million-solar-mass black hole in a significantly neutral Universe
        at a redshift of 7.5}",
      journal = {\nat},
         year = 2018,
        month = Jan,
       volume = {553},
        pages = {473-476},
          doi = {10.1038/nature25180},
archivePrefix = {arXiv},
       eprint = {1712.01860},
 primaryClass = {astro-ph.GA},
       adsurl = {https://ui.adsabs.harvard.edu/\#abs/2018Natur.553..473B}
}

@ARTICLE{Kurk07,
       author = {{Kurk}, Jaron D. and {Walter}, Fabian and {Fan}, Xiaohui and {Jiang},
        Linhua and {Riechers}, Dominik A. and {Rix}, Hans-Walter and
        {Pentericci}, Laura and {Strauss}, Michael A. and {Carilli},
        Chris and {Wagner}, Stefan},
        title = "{Black Hole Masses and Enrichment of z \textasciitilde 6 SDSS Quasars}",
      journal = {\apj},
         year = 2007,
        month = Nov,
       volume = {669},
        pages = {32-44},
          doi = {10.1086/521596},
archivePrefix = {arXiv},
       eprint = {0707.1662},
 primaryClass = {astro-ph},
       adsurl = {https://ui.adsabs.harvard.edu/\#abs/2007ApJ...669...32K}
}

@ARTICLE{Willott10a,
       author = {{Willott}, Chris J. and {Albert}, Loic and {Arzoumanian}, Doris and
        {Bergeron}, Jacqueline and {Crampton}, David and {Delorme},
        Philippe and {Hutchings}, John B. and {Omont}, Alain and
        {Reyl{\'e}}, C{\'e}line and {Schade}, David},
        title = "{Eddington-limited Accretion and the Black Hole Mass Function at Redshift
        6}",
      journal = {\aj},
         year = 2010,
        month = Aug,
       volume = {140},
        pages = {546-560},
          doi = {10.1088/0004-6256/140/2/546},
archivePrefix = {arXiv},
       eprint = {1006.1342},
 primaryClass = {astro-ph.CO},
       adsurl = {https://ui.adsabs.harvard.edu/\#abs/2010AJ....140..546W}
}

@ARTICLE{Selsing16,
       author = {{Selsing}, J. and {Fynbo}, J.~P.~U. and {Christensen}, L. and
        {Krogager}, J. -K.},
        title = "{An X-Shooter composite of bright 1 \&lt; z \&lt; 2 quasars from UV to infrared}",
      journal = {\aap},
         year = 2016,
        month = Jan,
       volume = {585},
          eid = {A87},
        pages = {A87},
          doi = {10.1051/0004-6361/201527096},
archivePrefix = {arXiv},
       eprint = {1510.08058},
 primaryClass = {astro-ph.GA},
       adsurl = {https://ui.adsabs.harvard.edu/\#abs/2016A&A...585A..87S}
}

@ARTICLE{VB01,
       author = {{Vanden Berk}, Daniel E. and {Richards}, Gordon T. and {Bauer}, Amanda
        and {Strauss}, Michael A. and {Schneider}, Donald P. and
        {Heckman}, Timothy M. and {York}, Donald G. and {Hall}, Patrick
        B. and {Fan}, Xiaohui and {Knapp}, G.~R. and {Anderson}, Scott
        F. and {Annis}, James and {Bahcall}, Neta A. and {Bernardi},
        Mariangela and {Briggs}, John W. and {Brinkmann}, J. and
        {Brunner}, Robert and {Burles}, Scott and {Carey}, Larry and
        {Castander}, Francisco J. and {Connolly}, A.~J. and {Crocker},
        J.~H. and {Csabai}, Istv{\'a}n and {Doi}, Mamoru and
        {Finkbeiner}, Douglas and {Friedman}, Scott and {Frieman},
        Joshua A. and {Fukugita}, Masataka and {Gunn}, James E. and
        {Hennessy}, G.~S. and {Ivezi{\'c}}, {\v{Z}}eljko and {Kent},
        Stephen and {Kunszt}, Peter Z. and {Lamb}, D.~Q. and {Leger}, R.
        French and {Long}, Daniel C. and {Loveday}, Jon and {Lupton},
        Robert H. and {Meiksin}, Avery and {Merelli}, Aronne and {Munn},
        Jeffrey A. and {Newberg}, Heidi Jo and {Newcomb}, Matt and
        {Nichol}, R.~C. and {Owen}, Russell and {Pier}, Jeffrey R. and
        {Pope}, Adrian and {Rockosi}, Constance M. and {Schlegel}, David
        J. and {Siegmund}, Walter A. and {Smee}, Stephen and {Snir},
        Yehuda and {Stoughton}, Chris and {Stubbs}, Christopher and
        {SubbaRao}, Mark and {Szalay}, Alexander S. and {Szokoly}, Gyula
        P. and {Tremonti}, Christy and {Uomoto}, Alan and {Waddell},
        Patrick and {Yanny}, Brian and {Zheng}, Wei},
        title = "{Composite Quasar Spectra from the Sloan Digital Sky Survey}",
      journal = {\aj},
         year = 2001,
        month = Aug,
       volume = {122},
        pages = {549-564},
          doi = {10.1086/321167},
archivePrefix = {arXiv},
       eprint = {astro-ph/0105231},
 primaryClass = {astro-ph},
       adsurl = {https://ui.adsabs.harvard.edu/\#abs/2001AJ....122..549V}
}

@ARTICLE{Vestergaard01,
       author = {{Vestergaard}, M. and {Wilkes}, B.~J.},
        title = "{An Empirical Ultraviolet Template for Iron Emission in Quasars as Derived from I Zwicky 1}",
      journal = {The Astrophysical Journal Supplement Series},
         year = 2001,
        month = May,
       volume = {134},
        pages = {1-33},
          doi = {10.1086/320357},
archivePrefix = {arXiv},
       eprint = {astro-ph/0104320},
 primaryClass = {astro-ph},
       adsurl = {https://ui.adsabs.harvard.edu/\#abs/2001ApJS..134....1V}
}

@ARTICLE{Vestergaard09,
       author = {{Vestergaard}, M. and {Osmer}, Patrick S.},
        title = "{Mass Functions of the Active Black Holes in Distant Quasars from the Large Bright Quasar Survey, the Bright Quasar Survey, and the Color-selected Sample of the SDSS Fall Equatorial Stripe}",
      journal = {\apj},
         year = 2009,
        month = Jul,
       volume = {699},
        pages = {800-816},
          doi = {10.1088/0004-637X/699/1/800},
archivePrefix = {arXiv},
       eprint = {0904.3348},
 primaryClass = {astro-ph.CO},
       adsurl = {https://ui.adsabs.harvard.edu/\#abs/2009ApJ...699..800V}
}

@ARTICLE{Vestergaard06,
       author = {{Vestergaard}, Marianne and {Peterson}, Bradley M.},
        title = "{Determining Central Black Hole Masses in Distant Active Galaxies and Quasars. II. Improved Optical and UV Scaling Relationships}",
      journal = {\apj},
         year = 2006,
        month = Apr,
       volume = {641},
        pages = {689-709},
          doi = {10.1086/500572},
archivePrefix = {arXiv},
       eprint = {astro-ph/0601303},
 primaryClass = {astro-ph},
       adsurl = {https://ui.adsabs.harvard.edu/\#abs/2006ApJ...641..689V}
}

@ARTICLE{Shen11,
       author = {{Shen}, Yue and {Richards}, Gordon T. and {Strauss}, Michael A. and
        {Hall}, Patrick B. and {Schneider}, Donald P. and {Snedden},
        Stephanie and {Bizyaev}, Dmitry and {Brewington}, Howard and
        {Malanushenko}, Viktor and {Malanushenko}, Elena and {Oravetz},
        Dan and {Pan}, Kaike and {Simmons}, Audrey},
        title = "{A Catalog of Quasar Properties from Sloan Digital Sky Survey Data Release 7}",
      journal = {The Astrophysical Journal Supplement Series},
         year = 2011,
        month = Jun,
       volume = {194},
          eid = {45},
        pages = {45},
          doi = {10.1088/0067-0049/194/2/45},
archivePrefix = {arXiv},
       eprint = {1006.5178},
 primaryClass = {astro-ph.CO},
       adsurl = {https://ui.adsabs.harvard.edu/\#abs/2011ApJS..194...45S}
}

@ARTICLE{Inayoshi20,
       author = {{Inayoshi}, Kohei and {Visbal}, Eli and {Haiman}, Zolt{\'a}n},
        title = "{The Assembly of the First Massive Black Holes}",
      journal = {\araa},
         year = 2020,
        month = aug,
       volume = {58},
        pages = {27-97},
          doi = {10.1146/annurev-astro-120419-014455},
archivePrefix = {arXiv},
       eprint = {1911.05791},
 primaryClass = {astro-ph.GA},
       adsurl = {https://ui.adsabs.harvard.edu/abs/2020ARA&A..58...27I}
}

@ARTICLE{Madau14,
       author = {{Madau}, Piero and {Haardt}, Francesco and {Dotti}, Massimo},
        title = "{Super-critical Growth of Massive Black Holes from Stellar-mass Seeds}",
      journal = {\apjl},
         year = 2014,
        month = apr,
       volume = {784},
       number = {2},
          eid = {L38},
        pages = {L38},
          doi = {10.1088/2041-8205/784/2/L38},
archivePrefix = {arXiv},
       eprint = {1402.6995},
 primaryClass = {astro-ph.CO},
       adsurl = {https://ui.adsabs.harvard.edu/abs/2014ApJ...784L..38M}
}

@ARTICLE{Ueda03,
       author = {{Ueda}, Yoshihiro and {Akiyama}, Masayuki and {Ohta}, Kouji and
        {Miyaji}, Takamitsu},
        title = "{Cosmological Evolution of the Hard X-Ray Active Galactic Nucleus Luminosity Function and the Origin of the Hard X-Ray Background}",
      journal = {\apj},
         year = 2003,
        month = Dec,
       volume = {598},
        pages = {886-908},
          doi = {10.1086/378940},
archivePrefix = {arXiv},
       eprint = {astro-ph/0308140},
 primaryClass = {astro-ph},
       adsurl = {https://ui.adsabs.harvard.edu/\#abs/2003ApJ...598..886U}
}

@ARTICLE{Volonteri15,
       author = {{Volonteri}, Marta and {Silk}, Joseph and {Dubus}, Guillaume},
        title = "{The Case for Supercritical Accretion onto Massive Black Holes at High Redshift}",
      journal = {\apj},
         year = 2015,
        month = May,
       volume = {804},
          eid = {148},
        pages = {148},
          doi = {10.1088/0004-637X/804/2/148},
archivePrefix = {arXiv},
       eprint = {1401.3513},
 primaryClass = {astro-ph.GA},
       adsurl = {https://ui.adsabs.harvard.edu/\#abs/2015ApJ...804..148V}
}

@ARTICLE{Hall02,
   author = {{Hall}, P.~B. and {Anderson}, S.~F. and {Strauss}, M.~A. and 
	{York}, D.~G. and {Richards}, G.~T. and {Fan}, X. and {Knapp}, G.~R. and 
	{Schneider}, D.~P. and {Vanden Berk}, D.~E. and {Geballe}, T.~R. and 
	{Bauer}, A.~E. and {Becker}, R.~H. and {Davis}, M. and {Rix}, H.-W. and 
	{Nichol}, R.~C. and {Bahcall}, N.~A. and {Brinkmann}, J. and 
	{Brunner}, R. and {Connolly}, A.~J. and {Csabai}, I. and {Doi}, M. and 
	{Fukugita}, M. and {Gunn}, J.~E. and {Haiman}, Z. and {Harvanek}, M. and 
	{Heckman}, T.~M. and {Hennessy}, G.~S. and {Inada}, N. and {Ivezi{\'c}}, {\v Z}. and 
	{Johnston}, D. and {Kleinman}, S. and {Krolik}, J.~H. and {Krzesinski}, J. and 
	{Kunszt}, P.~Z. and {Lamb}, D.~Q. and {Long}, D.~C. and {Lupton}, R.~H. and 
	{Miknaitis}, G. and {Munn}, J.~A. and {Narayanan}, V.~K. and 
	{Neilsen}, E. and {Newman}, P.~R. and {Nitta}, A. and {Okamura}, S. and 
	{Pentericci}, L. and {Pier}, J.~R. and {Schlegel}, D.~J. and 
	{Snedden}, S. and {Szalay}, A.~S. and {Thakar}, A.~R. and {Tsvetanov}, Z. and 
	{White}, R.~L. and {Zheng}, W.},
    title = "{Unusual Broad Absorption Line Quasars from the Sloan Digital Sky Survey}",
  journal = {\apjs},
   eprint = {astro-ph/0203252},
     year = 2002,
    month = aug,
   volume = 141,
    pages = {267-309},
      doi = {10.1086/340546},
   adsurl = {http://adsabs.harvard.edu/abs/2002ApJS..141..267H}
}

@ARTICLE{Astropy,
       author = {{Astropy Collaboration} and {Robitaille}, Thomas P. and {Tollerud}, Erik
        J. and {Greenfield}, Perry and {Droettboom}, Michael and {Bray},
        Erik and {Aldcroft}, Tom and {Davis}, Matt and {Ginsburg}, Adam
        and {Price-Whelan}, Adrian M. and {Kerzendorf}, Wolfgang E. and
        {Conley}, Alexander and {Crighton}, Neil and {Barbary}, Kyle and
        {Muna}, Demitri and {Ferguson}, Henry and {Grollier},
        Fr{\'e}d{\'e}ric and {Parikh}, Madhura M. and {Nair}, Prasanth
        H. and {Unther}, Hans M. and {Deil}, Christoph and {Woillez},
        Julien and {Conseil}, Simon and {Kramer}, Roban and {Turner},
        James E.~H. and {Singer}, Leo and {Fox}, Ryan and {Weaver},
        Benjamin A. and {Zabalza}, Victor and {Edwards}, Zachary I. and
        {Azalee Bostroem}, K. and {Burke}, D.~J. and {Casey}, Andrew R.
        and {Crawford}, Steven M. and {Dencheva}, Nadia and {Ely},
        Justin and {Jenness}, Tim and {Labrie}, Kathleen and {Lim}, Pey
        Lian and {Pierfederici}, Francesco and {Pontzen}, Andrew and
        {Ptak}, Andy and {Refsdal}, Brian and {Servillat}, Mathieu and
        {Streicher}, Ole},
        title = "{Astropy: A community Python package for astronomy}",
      journal = {\aap},
         year = 2013,
        month = Oct,
       volume = {558},
          eid = {A33},
        pages = {A33},
          doi = {10.1051/0004-6361/201322068},
archivePrefix = {arXiv},
       eprint = {1307.6212},
 primaryClass = {astro-ph.IM},
       adsurl = {https://ui.adsabs.harvard.edu/\#abs/2013A&A...558A..33A}
}

@ARTICLE{Matsuoka19,
       author = {{Matsuoka}, Yoshiki and {Onoue}, Masafusa and {Kashikawa}, Nobunari and
         {Strauss}, Michael A. and {Iwasawa}, Kazushi and {Lee}, Chien-Hsiu and
         {Imanishi}, Masatoshi and {Nagao}, Tohru and {Akiyama}, Masayuki and
         {Asami}, Naoko and {Bosch}, James and {Furusawa}, Hisanori and
         {Goto}, Tomotsugu and {Gunn}, James E. and {Harikane}, Yuichi and
         {Ikeda}, Hiroyuki and {Izumi}, Takuma and {Kawaguchi}, Toshihiro and
         {Kato}, Nanako and {Kikuta}, Satoshi and {Kohno}, Kotaro and
         {Komiyama}, Yutaka and {Koyama}, Shuhei and {Lupton}, Robert H. and
         {Minezaki}, Takeo and {Miyazaki}, Satoshi and {Murayama}, Hitoshi and
         {Niida}, Mana and {Nishizawa}, Atsushi J. and {Noboriguchi}, Akatoki and
         {Oguri}, Masamune and {Ono}, Yoshiaki and {Ouchi}, Masami and
         {Price}, Paul A. and {Sameshima}, Hiroaki and {Schulze}, Andreas and
         {Shirakata}, Hikari and {Silverman}, John D. and {Sugiyama}, Naoshi and
         {Tait}, Philip J. and {Takada}, Masahiro and {Takata}, Tadafumi and
         {Tanaka}, Masayuki and {Tang}, Ji-Jia and {Toba}, Yoshiki and
         {Utsumi}, Yousuke and {Wang}, Shiang-Yu and {Yamashita}, Takuji},
        title = "{Discovery of the First Low-luminosity Quasar at z \&gt; 7}",
      journal = {\apj},
         year = "2019",
        month = "Feb",
       volume = {872},
       number = {1},
          eid = {L2},
        pages = {L2},
          doi = {10.3847/2041-8213/ab0216},
archivePrefix = {arXiv},
       eprint = {1901.10487},
 primaryClass = {astro-ph.GA},
       adsurl = {https://ui.adsabs.harvard.edu/abs/2019ApJ...872L...2M}
}

@ARTICLE{Matsuoka19b,
       author = {{Matsuoka}, Yoshiki and {Iwasawa}, Kazushi and {Onoue}, Masafusa and {Kashikawa}, Nobunari and {Strauss}, Michael A. and {Lee}, Chien-Hsiu and {Imanishi}, Masatoshi and {Nagao}, Tohru and {Akiyama}, Masayuki and {Asami}, Naoko and {Bosch}, James and {Furusawa}, Hisanori and {Goto}, Tomotsugu and {Gunn}, James E. and {Harikane}, Yuichi and {Ikeda}, Hiroyuki and {Izumi}, Takuma and {Kawaguchi}, Toshihiro and {Kato}, Nanako and {Kikuta}, Satoshi and {Kohno}, Kotaro and {Komiyama}, Yutaka and {Koyama}, Shuhei and {Lupton}, Robert H. and {Minezaki}, Takeo and {Miyazaki}, Satoshi and {Murayama}, Hitoshi and {Niida}, Mana and {Nishizawa}, Atsushi J. and {Noboriguchi}, Akatoki and {Oguri}, Masamune and {Ono}, Yoshiaki and {Ouchi}, Masami and {Price}, Paul A. and {Sameshima}, Hiroaki and {Schulze}, Andreas and {Silverman}, John D. and {Sugiyama}, Naoshi and {Tait}, Philip J. and {Takada}, Masahiro and {Takata}, Tadafumi and {Tanaka}, Masayuki and {Tang}, Ji-Jia and {Toba}, Yoshiki and {Utsumi}, Yousuke and {Wang}, Shiang-Yu and {Yamashita}, Takuji},
        title = "{Subaru High-z Exploration of Low-luminosity Quasars (SHELLQs). X. Discovery of 35 Quasars and Luminous Galaxies at 5.7 {\ensuremath{\leq}} z {\ensuremath{\leq}} 7.0}",
      journal = {\apj},
         year = 2019,
        month = oct,
       volume = {883},
       number = {2},
          eid = {183},
        pages = {183},
          doi = {10.3847/1538-4357/ab3c60},
archivePrefix = {arXiv},
       eprint = {1908.07910},
 primaryClass = {astro-ph.GA},
       adsurl = {https://ui.adsabs.harvard.edu/abs/2019ApJ...883..183M}
}

@ARTICLE{Kawaguchi04,
       author = {{Kawaguchi}, T. and {Aoki}, K. and {Ohta}, K. and {Collin}, S.},
        title = "{Growth of massive black holes by super-Eddington accretion}",
      journal = {\aap},
         year = 2004,
        month = Jun,
       volume = {420},
        pages = {L23-L26},
          doi = {10.1051/0004-6361:20040157},
archivePrefix = {arXiv},
       eprint = {astro-ph/0405024},
 primaryClass = {astro-ph},
       adsurl = {https://ui.adsabs.harvard.edu/\#abs/2004A&A...420L..23K}
}

@ARTICLE{HSCSSP,
       author = {{Aihara}, Hiroaki and {Arimoto}, Nobuo and {Armstrong}, Robert and
         {Arnouts}, St{\'e}phane and {Bahcall}, Neta A. and {Bickerton}, Steven and
         {Bosch}, James and {Bundy}, Kevin and {Capak}, Peter L. and
         {Chan}, James H.~H. and {Chiba}, Masashi and {Coupon}, Jean and
         {Egami}, Eiichi and {Enoki}, Motohiro and {Finet}, Francois and
         {Fujimori}, Hiroki and {Fujimoto}, Seiji and {Furusawa}, Hisanori and
         {Furusawa}, Junko and {Goto}, Tomotsugu and {Goulding}, Andy and
         {Greco}, Johnny P. and {Greene}, Jenny E. and {Gunn}, James E. and
         {Hamana}, Takashi and {Harikane}, Yuichi and {Hashimoto}, Yasuhiro and
         {Hattori}, Takashi and {Hayashi}, Masao and {Hayashi}, Yusuke and
         {He{\l}miniak}, Krzysztof G. and {Higuchi}, Ryo and {Hikage}, Chiaki and
         {Ho}, Paul T.~P. and {Hsieh}, Bau-Ching and {Huang}, Kuiyun and
         {Huang}, Song and {Ikeda}, Hiroyuki and {Imanishi}, Masatoshi and
         {Inoue}, Akio K. and {Iwasawa}, Kazushi and {Iwata}, Ikuru and
         {Jaelani}, Anton T. and {Jian}, Hung-Yu and {Kamata}, Yukiko and
         {Karoji}, Hiroshi and {Kashikawa}, Nobunari and {Katayama}, Nobuhiko and
         {Kawanomoto}, Satoshi and {Kayo}, Issha and {Koda}, Jin and
         {Koike}, Michitaro and {Kojima}, Takashi and {Komiyama}, Yutaka and
         {Konno}, Akira and {Koshida}, Shintaro and {Koyama}, Yusei and
         {Kusakabe}, Haruka and {Leauthaud}, Alexie and {Lee}, Chien-Hsiu and
         {Lin}, Lihwai and {Lin}, Yen-Ting and {Lupton}, Robert H. and {Mand
        elbaum}, Rachel and {Matsuoka}, Yoshiki and {Medezinski}, Elinor and
         {Mineo}, Sogo and {Miyama}, Shoken and {Miyatake}, Hironao and
         {Miyazaki}, Satoshi and {Momose}, Rieko and {More}, Anupreeta and
         {More}, Surhud and {Moritani}, Yuki and {Moriya}, Takashi J. and
         {Morokuma}, Tomoki and {Mukae}, Shiro and {Murata}, Ryoma and
         {Murayama}, Hitoshi and {Nagao}, Tohru and {Nakata}, Fumiaki and
         {Niida}, Mana and {Niikura}, Hiroko and {Nishizawa}, Atsushi J. and
         {Obuchi}, Yoshiyuki and {Oguri}, Masamune and {Oishi}, Yukie and
         {Okabe}, Nobuhiro and {Okamoto}, Sakurako and {Okura}, Yuki and
         {Ono}, Yoshiaki and {Onodera}, Masato and {Onoue}, Masafusa and
         {Osato}, Ken and {Ouchi}, Masami and {Price}, Paul A. and
         {Pyo}, Tae-Soo and {Sako}, Masao and {Sawicki}, Marcin and
         {Shibuya}, Takatoshi and {Shimasaku}, Kazuhiro and {Shimono}, Atsushi and
         {Shirasaki}, Masato and {Silverman}, John D. and {Simet}, Melanie and
         {Speagle}, Joshua and {Spergel}, David N. and {Strauss}, Michael A. and
         {Sugahara}, Yuma and {Sugiyama}, Naoshi and {Suto}, Yasushi and
         {Suyu}, Sherry H. and {Suzuki}, Nao and {Tait}, Philip J. and
         {Takada}, Masahiro and {Takata}, Tadafumi and {Tamura}, Naoyuki and
         {Tanaka}, Manobu M. and {Tanaka}, Masaomi and {Tanaka}, Masayuki and
         {Tanaka}, Yoko and {Terai}, Tsuyoshi and {Terashima}, Yuichi and
         {Toba}, Yoshiki and {Tominaga}, Nozomu and {Toshikawa}, Jun and
         {Turner}, Edwin L. and {Uchida}, Tomohisa and {Uchiyama}, Hisakazu and
         {Umetsu}, Keiichi and {Uraguchi}, Fumihiro and {Urata}, Yuji and
         {Usuda}, Tomonori and {Utsumi}, Yousuke and {Wang}, Shiang-Yu and
         {Wang}, Wei-Hao and {Wong}, Kenneth C. and {Yabe}, Kiyoto and
         {Yamada}, Yoshihiko and {Yamanoi}, Hitomi and {Yasuda}, Naoki and
         {Yeh}, Sherry and {Yonehara}, Atsunori and {Yuma}, Suraphong},
        title = "{The Hyper Suprime-Cam SSP Survey: Overview and survey design}",
      journal = {Publications of the Astronomical Society of Japan},
         year = "2018",
        month = "Jan",
       volume = {70},
          eid = {S4},
        pages = {S4},
          doi = {10.1093/pasj/psx066},
archivePrefix = {arXiv},
       eprint = {1704.05858},
 primaryClass = {astro-ph.IM},
       adsurl = {https://ui.adsabs.harvard.edu/\#abs/2018PASJ...70S...4A}
}

@ARTICLE{Miyazaki18,
       author = {{Miyazaki}, Satoshi and {Komiyama}, Yutaka and {Kawanomoto}, Satoshi and
         {Doi}, Yoshiyuki and {Furusawa}, Hisanori and {Hamana}, Takashi and
         {Hayashi}, Yusuke and {Ikeda}, Hiroyuki and {Kamata}, Yukiko and
         {Karoji}, Hiroshi and {Koike}, Michitaro and {Kurakami}, Tomio and
         {Miyama}, Shoken and {Morokuma}, Tomoki and {Nakata}, Fumiaki and
         {Namikawa}, Kazuhito and {Nakaya}, Hidehiko and {Nariai}, Kyoji and
         {Obuchi}, Yoshiyuki and {Oishi}, Yukie and {Okada}, Norio and
         {Okura}, Yuki and {Tait}, Philip and {Takata}, Tadafumi and
         {Tanaka}, Yoko and {Tanaka}, Masayuki and {Terai}, Tsuyoshi and
         {Tomono}, Daigo and {Uraguchi}, Fumihiro and {Usuda}, Tomonori and
         {Utsumi}, Yousuke and {Yamada}, Yoshihiko and {Yamanoi}, Hitomi and
         {Aihara}, Hiroaki and {Fujimori}, Hiroki and {Mineo}, Sogo and
         {Miyatake}, Hironao and {Oguri}, Masamune and {Uchida}, Tomohisa and
         {Tanaka}, Manobu M. and {Yasuda}, Naoki and {Takada}, Masahiro and
         {Murayama}, Hitoshi and {Nishizawa}, Atsushi J. and {Sugiyama}, Naoshi and
         {Chiba}, Masashi and {Futamase}, Toshifumi and {Wang}, Shiang-Yu and
         {Chen}, Hsin-Yo and {Ho}, Paul T.~P. and {Liaw}, Eric J.~Y. and
         {Chiu}, Chi-Fang and {Ho}, Cheng-Lin and {Lai}, Tsang-Chih and
         {Lee}, Yao-Cheng and {Jeng}, Dun-Zen and {Iwamura}, Satoru and
         {Armstrong}, Robert and {Bickerton}, Steve and {Bosch}, James and
         {Gunn}, James E. and {Lupton}, Robert H. and {Loomis}, Craig and
         {Price}, Paul and {Smith}, Steward and {Strauss}, Michael A. and
         {Turner}, Edwin L. and {Suzuki}, Hisanori and {Miyazaki}, Yasuhito and
         {Muramatsu}, Masaharu and {Yamamoto}, Koei and {Endo}, Makoto and
         {Ezaki}, Yutaka and {Ito}, Noboru and {Kawaguchi}, Noboru and
         {Sofuku}, Satoshi and {Taniike}, Tomoaki and {Akutsu}, Kotaro and
         {Dojo}, Naoto and {Kasumi}, Kazuyuki and {Matsuda}, Toru and
         {Imoto}, Kohei and {Miwa}, Yoshinori and {Suzuki}, Masayuki and
         {Takeshi}, Kunio and {Yokota}, Hideo},
        title = "{Hyper Suprime-Cam: System design and verification of image quality}",
      journal = {Publications of the Astronomical Society of Japan},
         year = "2018",
        month = "Jan",
       volume = {70},
          eid = {S1},
        pages = {S1},
          doi = {10.1093/pasj/psx063},
       adsurl = {https://ui.adsabs.harvard.edu/\#abs/2018PASJ...70S...1M}
}

@ARTICLE{Boroson92,
       author = {{Boroson}, Todd A. and {Green}, Richard F.},
        title = "{The Emission-Line Properties of Low-Redshift Quasi-stellar Objects}",
      journal = {\apjs},
         year = 1992,
        month = may,
       volume = {80},
        pages = {109},
          doi = {10.1086/191661},
       adsurl = {https://ui.adsabs.harvard.edu/abs/1992ApJS...80..109B}
}

@ARTICLE{Maiolino24,
       author = {{Maiolino}, Roberto and {Scholtz}, Jan and {Curtis-Lake}, Emma and {Carniani}, Stefano and {Baker}, William and {de Graaff}, Anna and {Tacchella}, Sandro and {{\"U}bler}, Hannah and {D'Eugenio}, Francesco and {Witstok}, Joris and {Curti}, Mirko and {Arribas}, Santiago and {Bunker}, Andrew J. and {Charlot}, St{\'e}phane and {Chevallard}, Jacopo and {Eisenstein}, Daniel J. and {Egami}, Eiichi and {Ji}, Zhiyuan and {Jones}, Gareth C. and {Lyu}, Jianwei and {Rawle}, Tim and {Robertson}, Brant and {Rujopakarn}, Wiphu and {Perna}, Michele and {Sun}, Fengwu and {Venturi}, Giacomo and {Williams}, Christina C. and {Willott}, Chris},
        title = "{JADES: The diverse population of infant black holes at 4 < z < 11: Merging, tiny, poor, but mighty}",
      journal = {\aap},
         year = 2024,
        month = nov,
       volume = {691},
          eid = {A145},
        pages = {A145},
          doi = {10.1051/0004-6361/202347640},
archivePrefix = {arXiv},
       eprint = {2308.01230},
 primaryClass = {astro-ph.GA},
       adsurl = {https://ui.adsabs.harvard.edu/abs/2024A&A...691A.145M}
}

@ARTICLE{Collin04,
       author = {{Collin}, S. and {Kawaguchi}, T.},
        title = "{Super-Eddington accretion rates in Narrow Line Seyfert 1 galaxies}",
      journal = {\aap},
         year = "2004",
        month = "Nov",
       volume = {426},
        pages = {797-808},
          doi = {10.1051/0004-6361:20040528},
archivePrefix = {arXiv},
       eprint = {astro-ph/0407181},
 primaryClass = {astro-ph},
       adsurl = {https://ui.adsabs.harvard.edu/\#abs/2004A&A...426..797C}
}

@ARTICLE{Richards03,
       author = {{Richards}, Gordon T. and {Hall}, Patrick B. and {Vand
        en Berk}, Daniel E. and {Strauss}, Michael A. and
         {Schneider}, Donald P. and {Weinstein}, Michael A. and
         {Reichard}, Timothy A. and {York}, Donald G. and {Knapp}, G.~R. and
         {Fan}, Xiaohui and {Ivezi{\'c}}, {\v{Z}}eljko and {Brinkmann}, J. and
         {Budav{\'a}ri}, Tam{\'a}s and {Csabai}, Istv{\'a}n and {Nichol}, R.~C.},
        title = "{Red and Reddened Quasars in the Sloan Digital Sky Survey}",
      journal = {\aj},
         year = "2003",
        month = "Sep",
       volume = {126},
        pages = {1131-1147},
          doi = {10.1086/377014},
archivePrefix = {arXiv},
       eprint = {astro-ph/0305305},
 primaryClass = {astro-ph},
       adsurl = {https://ui.adsabs.harvard.edu/\#abs/2003AJ....126.1131R}
}

@ARTICLE{Shen13_review,
       author = {{Shen}, Yue},
        title = "{The mass of quasars}",
      journal = {Bulletin of the Astronomical Society of India},
         year = "2013",
        month = "Mar",
       volume = {41},
        pages = {61-115},
archivePrefix = {arXiv},
       eprint = {1302.2643},
 primaryClass = {astro-ph.CO},
       adsurl = {https://ui.adsabs.harvard.edu/\#abs/2013BASI...41...61S}
}

@ARTICLE{Richards06a,
       author = {{Richards}, Gordon T. and {Lacy}, Mark and {Storrie-Lombardi}, Lisa J. and
         {Hall}, Patrick B. and {Gallagher}, S.~C. and {Hines}, Dean C. and
         {Fan}, Xiaohui and {Papovich}, Casey and {Vanden Berk}, Daniel E. and
         {Trammell}, George B. and {Schneider}, Donald P. and
         {Vestergaard}, Marianne and {York}, Donald G. and {Jester}, Sebastian and
         {Anderson}, Scott F. and {Budav{\'a}ri}, Tam{\'a}s and {Szalay}, Alexand
        er S.},
        title = "{Spectral Energy Distributions and Multiwavelength Selection of Type 1 Quasars}",
      journal = {The Astrophysical Journal Supplement Series},
         year = "2006",
        month = "Oct",
       volume = {166},
        pages = {470-497},
          doi = {10.1086/506525},
archivePrefix = {arXiv},
       eprint = {astro-ph/0601558},
 primaryClass = {astro-ph},
       adsurl = {https://ui.adsabs.harvard.edu/\#abs/2006ApJS..166..470R}
}

@ARTICLE{Shen19,
       author = {{Shen}, Yue and {Wu}, Jin and {Jiang}, Linhua and
         {Ba{\~n}ados}, Eduardo and {Fan}, Xiaohui and {Ho}, Luis C. and
         {Riechers}, Dominik A. and {Strauss}, Michael A. and {Venemans}, Bram and
         {Vestergaard}, Marianne and {Walter}, Fabian and {Wang}, Feige and
         {Willott}, Chris and {Wu}, Xue-Bing and {Yang}, Jinyi},
        title = "{Gemini GNIRS Near-infrared Spectroscopy of 50 Quasars at z {\ensuremath{\gtrsim}} 5.7}",
      journal = {\apj},
         year = "2019",
        month = "Mar",
       volume = {873},
          eid = {35},
        pages = {35},
          doi = {10.3847/1538-4357/ab03d9},
archivePrefix = {arXiv},
       eprint = {1809.05584},
 primaryClass = {astro-ph.GA},
       adsurl = {https://ui.adsabs.harvard.edu/\#abs/2019ApJ...873...35S}
}

@ARTICLE{Kato20,
       author = {{Kato}, Nanako and {Matsuoka}, Yoshiki and {Onoue}, Masafusa and {Koyama}, Shuhei and {Toba}, Yoshiki and {Akiyama}, Masayuki and {Fujimoto}, Seiji and {Imanishi}, Masatoshi and {Iwasawa}, Kazushi and {Izumi}, Takuma and {Kashikawa}, Nobunari and {Kawaguchi}, Toshihiro and {Lee}, Chien-Hsiu and {Minezaki}, Takeo and {Nagao}, Tohru and {Noboriguchi}, Akatoki and {Strauss}, Michael A.},
        title = "{Subaru High-z Exploration of Low-Luminosity Quasars (SHELLQs). IX. Identification of two red quasars at z > 5.6}",
      journal = {\pasj},
         year = 2020,
        month = oct,
       volume = {72},
       number = {5},
          eid = {84},
        pages = {84},
          doi = {10.1093/pasj/psaa074},
archivePrefix = {arXiv},
       eprint = {2007.08685},
 primaryClass = {astro-ph.GA},
       adsurl = {https://ui.adsabs.harvard.edu/abs/2020PASJ...72...84K}
}

@ARTICLE{Schlegel98,
       author = {{Schlegel}, David J. and {Finkbeiner}, Douglas P. and {Davis}, Marc},
        title = "{Maps of Dust Infrared Emission for Use in Estimation of Reddening and Cosmic Microwave Background Radiation Foregrounds}",
      journal = {\apj},
         year = "1998",
        month = "Jun",
       volume = {500},
       number = {2},
        pages = {525-553},
          doi = {10.1086/305772},
archivePrefix = {arXiv},
       eprint = {astro-ph/9710327},
 primaryClass = {astro-ph},
       adsurl = {https://ui.adsabs.harvard.edu/abs/1998ApJ...500..525S}
}

@INPROCEEDINGS{Fabricius16,
       author = {{Fabricius}, Maximilian and {Walawender}, Joshua and {Arimoto}, Nobuo and {Cook}, David and {Elms}, Brian and {Hashiba}, Yasuhito and {Hattori}, Takashi and {Hu}, Yen-Sang and {Iwata}, Ikuru and {Nishimura}, Tetsuo and {Omata}, Koji and {Tait}, Philip and {Takato}, Naruhisa and {Tanaka}, Ichi and {Wang}, Shiang-Yu and {Weber}, Mark and {Wung}, Matthew},
        title = "{Detector upgrade of Subaru's Multi-object Infrared Camera and Spectrograph (MOIRCS)}",
    booktitle = {Ground-based and Airborne Instrumentation for Astronomy VI},
         year = 2016,
       editor = {{Evans}, Christopher J. and {Simard}, Luc and {Takami}, Hideki},
       series = {Society of Photo-Optical Instrumentation Engineers (SPIE) Conference Series},
       volume = {9908},
        month = aug,
          eid = {990828},
        pages = {990828},
          doi = {10.1117/12.2231417},
       adsurl = {https://ui.adsabs.harvard.edu/abs/2016SPIE.9908E..28F}
}

@ARTICLE{Ebizuka11,
       author = {{Ebizuka}, Noboru and {Ichiyama}, Kotaro and {Yamada}, Toru and {Tokoku}, Chihiro and {Onodera}, Masato and {Hanesaka}, Mai and {Kodate}, Kashiko and {Katsuno Uchimoto}, Yuka and {Maruyama}, Miyoko and {Shimasaku}, Kazuhiro and {Tanaka}, Ichi and {Yoshikawa}, Tomohiro and {Kashikawa}, Nobunari and {Iye}, Masanori and {Ichikawa}, Takashi},
        title = "{Cryogenic Volume-Phase Holographic Grisms for MOIRCS}",
      journal = {\pasj},
         year = 2011,
        month = mar,
       volume = {63},
        pages = {605},
          doi = {10.1093/pasj/63.sp2.S605},
archivePrefix = {arXiv},
       eprint = {1105.0996},
 primaryClass = {astro-ph.IM},
       adsurl = {https://ui.adsabs.harvard.edu/abs/2011PASJ...63S.605E}
}

@ARTICLE{Suzuki08,
       author = {{Suzuki}, Ryuji and {Tokoku}, Chihiro and {Ichikawa}, Takashi and {Uchimoto}, Yuka Katsuno and {Konishi}, Masahiro and {Yoshikawa}, Tomohiro and {Tanaka}, Ichi and {Yamada}, Toru and {Omata}, Koji and {Nishimura}, Tetsuo},
        title = "{Multi-Object Infrared Camera and Spectrograph (MOIRCS) for the Subaru Telescope I. Imaging}",
      journal = {\pasj},
         year = 2008,
        month = dec,
       volume = {60},
        pages = {1347},
          doi = {10.1093/pasj/60.6.1347},
       adsurl = {https://ui.adsabs.harvard.edu/abs/2008PASJ...60.1347S}
}

@INPROCEEDINGS{Ichikawa06,
       author = {{Ichikawa}, Takashi and {Suzuki}, Ryuji and {Tokoku}, Chihiro and {Uchimoto}, Yuka Katsuno and {Konishi}, Masahiro and {Yoshikawa}, Tomohiro and {Yamada}, Toru and {Tanaka}, Ichi and {Omata}, Koji and {Nishimura}, Tetsuo},
        title = "{MOIRCS: multi-object infrared camera and spectrograph for SUBARU}",
    booktitle = {Society of Photo-Optical Instrumentation Engineers (SPIE) Conference Series},
         year = 2006,
       editor = {{McLean}, Ian S. and {Iye}, Masanori},
       series = {Society of Photo-Optical Instrumentation Engineers (SPIE) Conference Series},
       volume = {6269},
        month = jun,
          eid = {626916},
        pages = {626916},
          doi = {10.1117/12.670078},
       adsurl = {https://ui.adsabs.harvard.edu/abs/2006SPIE.6269E..16I}
}

@INPROCEEDINGS{Walawender16,
       author = {{Walawender}, Josh and {Wung}, Matthew and {Fabricius}, Maximilian and {Tanaka}, Ichi and {Arimoto}, Nobuo and {Cook}, David and {Elms}, Brian and {Hashiba}, Yasuhito and {Hu}, Yen-Sang and {Iwata}, Ikuru and {Nishimura}, Tetsuo and {Omata}, Koji and {Takato}, Naruhisa and {Wang}, Shiang-Yu and {Weber}, Mark},
        title = "{The nuMOIRCS project: detector upgrade overview and early commissioning results}",
    booktitle = {Ground-based and Airborne Instrumentation for Astronomy VI},
         year = 2016,
       editor = {{Evans}, Christopher J. and {Simard}, Luc and {Takami}, Hideki},
       series = {Society of Photo-Optical Instrumentation Engineers (SPIE) Conference Series},
       volume = {9908},
        month = aug,
          eid = {99082G},
        pages = {99082G},
          doi = {10.1117/12.2231812},
       adsurl = {https://ui.adsabs.harvard.edu/abs/2016SPIE.9908E..2GW}
}

@ARTICLE{Leggett06,
       author = {{Leggett}, S.~K. and {Currie}, M.~J. and {Varricatt}, W.~P. and {Hawarden}, T.~G. and {Adamson}, A.~J. and {Buckle}, J. and {Carroll}, T. and {Davies}, J.~K. and {Davis}, C.~J. and {Kerr}, T.~H. and {Kuhn}, O.~P. and {Seigar}, M.~S. and {Wold}, T.},
        title = "{JHK observations of faint standard stars in the Mauna Kea Observatories near-infrared photometric system}",
      journal = {\mnras},
         year = 2006,
        month = dec,
       volume = {373},
       number = {2},
        pages = {781-792},
          doi = {10.1111/j.1365-2966.2006.11069.x},
archivePrefix = {arXiv},
       eprint = {astro-ph/0609461},
 primaryClass = {astro-ph},
       adsurl = {https://ui.adsabs.harvard.edu/abs/2006MNRAS.373..781L}
}

@ARTICLE{Tsuzuki06,
       author = {{Tsuzuki}, Yumihiko and {Kawara}, Kimiaki and {Yoshii}, Yuzuru and
         {Oyabu}, Shinki and {Tanab{\'e}}, Toshihiko and {Matsuoka}, Yoshiki},
        title = "{Fe II Emission in 14 Low-Redshift Quasars. I. Observations}",
      journal = {\apj},
         year = "2006",
        month = "Oct",
       volume = {650},
       number = {1},
        pages = {57-79},
          doi = {10.1086/506376},
archivePrefix = {arXiv},
       eprint = {astro-ph/0606040},
 primaryClass = {astro-ph},
       adsurl = {https://ui.adsabs.harvard.edu/abs/2006ApJ...650...57T}
}

@ARTICLE{Woo18,
       author = {{Woo}, Jong-Hak and {Le}, Huynh Anh N. and {Karouzos}, Marios and
         {Park}, Dawoo and {Park}, Daeseong and {Malkan}, Matthew A. and
         {Treu}, Tommaso and {Bennert}, Vardha N.},
        title = "{Calibration and Limitations of the Mg II Line-based Black Hole Masses}",
      journal = {\apj},
         year = "2018",
        month = "Jun",
       volume = {859},
       number = {2},
          eid = {138},
        pages = {138},
          doi = {10.3847/1538-4357/aabf3e},
archivePrefix = {arXiv},
       eprint = {1804.02798},
 primaryClass = {astro-ph.GA},
       adsurl = {https://ui.adsabs.harvard.edu/abs/2018ApJ...859..138W}
}

@ARTICLE{Sameshima17,
       author = {{Sameshima}, H. and {Yoshii}, Y. and {Kawara}, K.},
        title = "{Chemical Evolution of the Universe at 0.7 \&lt; z \&lt; 1.6 Derived from Abundance Diagnostics of the Broad-line Region of Quasars}",
      journal = {\apj},
         year = "2017",
        month = "Jan",
       volume = {834},
       number = {2},
          eid = {203},
        pages = {203},
          doi = {10.3847/1538-4357/834/2/203},
archivePrefix = {arXiv},
       eprint = {1611.06027},
 primaryClass = {astro-ph.GA},
       adsurl = {https://ui.adsabs.harvard.edu/abs/2017ApJ...834..203S}
}

@ARTICLE{Yang21,
       author = {{Yang}, Jinyi and {Wang}, Feige and {Fan}, Xiaohui and {Barth}, Aaron J. and {Hennawi}, Joseph F. and {Nanni}, Riccardo and {Bian}, Fuyan and {Davies}, Frederick B. and {Farina}, Emanuele P. and {Schindler}, Jan-Torge and {Ba{\~n}ados}, Eduardo and {Decarli}, Roberto and {Eilers}, Anna-Christina and {Green}, Richard and {Guo}, Hengxiao and {Jiang}, Linhua and {Li}, Jiang-Tao and {Venemans}, Bram and {Walter}, Fabian and {Wu}, Xue-Bing and {Yue}, Minghao},
        title = "{Probing Early Supermassive Black Hole Growth and Quasar Evolution with Near-infrared Spectroscopy of 37 Reionization-era Quasars at 6.3 < z {\ensuremath{\leq}} 7.64}",
      journal = {\apj},
         year = 2021,
        month = dec,
       volume = {923},
       number = {2},
          eid = {262},
        pages = {262},
          doi = {10.3847/1538-4357/ac2b32},
archivePrefix = {arXiv},
       eprint = {2109.13942},
 primaryClass = {astro-ph.GA},
       adsurl = {https://ui.adsabs.harvard.edu/abs/2021ApJ...923..262Y}
}

@ARTICLE{Schindler20,
       author = {{Schindler}, Jan-Torge and {Farina}, Emanuele Paolo and {Ba{\~n}ados}, Eduardo and {Eilers}, Anna-Christina and {Hennawi}, Joseph F. and {Onoue}, Masafusa and {Venemans}, Bram P. and {Walter}, Fabian and {Wang}, Feige and {Davies}, Frederick B. and {Decarli}, Roberto and {Rosa}, Gisella De and {Drake}, Alyssa and {Fan}, Xiaohui and {Mazzucchelli}, Chiara and {Rix}, Hans-Walter and {Worseck}, G{\'a}bor and {Yang}, Jinyi},
        title = "{The X-SHOOTER/ALMA Sample of Quasars in the Epoch of Reionization. I. NIR Spectral Modeling, Iron Enrichment, and Broad Emission Line Properties}",
      journal = {\apj},
         year = 2020,
        month = dec,
       volume = {905},
       number = {1},
          eid = {51},
        pages = {51},
          doi = {10.3847/1538-4357/abc2d7},
archivePrefix = {arXiv},
       eprint = {2010.06902},
 primaryClass = {astro-ph.GA},
       adsurl = {https://ui.adsabs.harvard.edu/abs/2020ApJ...905...51S}
}

@ARTICLE{Onoue20,
       author = {{Onoue}, Masafusa and {Ba{\~n}ados}, Eduardo and {Mazzucchelli}, Chiara and {Venemans}, Bram P. and {Schindler}, Jan-Torge and {Walter}, Fabian and {Hennawi}, Joseph F. and {Andika}, Irham Taufik and {Davies}, Frederick B. and {Decarli}, Roberto and {Farina}, Emanuele P. and {Jahnke}, Knud and {Nagao}, Tohru and {Tominaga}, Nozomu and {Wang}, Feige},
        title = "{No Redshift Evolution in the Broad-line-region Metallicity up to z = 7.54: Deep Near-infrared Spectroscopy of ULAS J1342+0928}",
      journal = {\apj},
         year = 2020,
        month = aug,
       volume = {898},
       number = {2},
          eid = {105},
        pages = {105},
          doi = {10.3847/1538-4357/aba193},
archivePrefix = {arXiv},
       eprint = {2006.16268},
 primaryClass = {astro-ph.GA},
       adsurl = {https://ui.adsabs.harvard.edu/abs/2020ApJ...898..105O}
}

@ARTICLE{Onoue23,
       author = {{Onoue}, Masafusa and {Inayoshi}, Kohei and {Ding}, Xuheng and {Li}, Wenxiu and {Li}, Zhengrong and {Molina}, Juan and {Inoue}, Akio K. and {Jiang}, Linhua and {Ho}, Luis C.},
        title = "{A Candidate for the Least-massive Black Hole in the First 1.1 Billion Years of the Universe}",
      journal = {\apjl},
         year = 2023,
        month = jan,
       volume = {942},
       number = {1},
          eid = {L17},
        pages = {L17},
          doi = {10.3847/2041-8213/aca9d3},
archivePrefix = {arXiv},
       eprint = {2209.07325},
 primaryClass = {astro-ph.GA},
       adsurl = {https://ui.adsabs.harvard.edu/abs/2023ApJ...942L..17O}
}

@ARTICLE{Kocevski23,
       author = {{Kocevski}, Dale D. and {Onoue}, Masafusa and {Inayoshi}, Kohei and {Trump}, Jonathan R. and {Arrabal Haro}, Pablo and {Grazian}, Andrea and {Dickinson}, Mark and {Finkelstein}, Steven L. and {Kartaltepe}, Jeyhan S. and {Hirschmann}, Michaela and {Aird}, James and {Holwerda}, Benne W. and {Fujimoto}, Seiji and {Juneau}, St{\'e}phanie and {Amor{\'\i}n}, Ricardo O. and {Backhaus}, Bren E. and {Bagley}, Micaela B. and {Barro}, Guillermo and {Bell}, Eric F. and {Bisigello}, Laura and {Calabr{\`o}}, Antonello and {Cleri}, Nikko J. and {Cooper}, M.~C. and {Ding}, Xuheng and {Grogin}, Norman A. and {Ho}, Luis C. and {Hutchison}, Taylor A. and {Inoue}, Akio K. and {Jiang}, Linhua and {Jones}, Brenda and {Koekemoer}, Anton M. and {Li}, Wenxiu and {Li}, Zhengrong and {McGrath}, Elizabeth J. and {Molina}, Juan and {Papovich}, Casey and {P{\'e}rez-Gonz{\'a}lez}, Pablo G. and {Pirzkal}, Nor and {Wilkins}, Stephen M. and {Yang}, Guang and {Yung}, L.~Y. Aaron},
        title = "{Hidden Little Monsters: Spectroscopic Identification of Low-mass, Broad-line AGNs at z > 5 with CEERS}",
      journal = {\apjl},
         year = 2023,
        month = sep,
       volume = {954},
       number = {1},
          eid = {L4},
        pages = {L4},
          doi = {10.3847/2041-8213/ace5a0},
archivePrefix = {arXiv},
       eprint = {2302.00012},
 primaryClass = {astro-ph.GA},
       adsurl = {https://ui.adsabs.harvard.edu/abs/2023ApJ...954L...4K}
}

@ARTICLE{Kocevski25,
       author = {{Kocevski}, Dale D. and {Finkelstein}, Steven L. and {Barro}, Guillermo and {Taylor}, Anthony J. and {Calabr{\`o}}, Antonello and {Laloux}, Brivael and {Buchner}, Johannes and {Trump}, Jonathan R. and {Leung}, Gene C.~K. and {Yang}, Guang and {Dickinson}, Mark and {P{\'e}rez-Gonz{\'a}lez}, Pablo G. and {Pacucci}, Fabio and {Inayoshi}, Kohei and {Somerville}, Rachel S. and {McGrath}, Elizabeth J. and {Akins}, Hollis B. and {Bagley}, Micaela B. and {Bowler}, Rebecca A.~A. and {Bisigello}, Laura and {Carnall}, Adam and {Casey}, Caitlin M. and {Cheng}, Yingjie and {Cleri}, Nikko J. and {Costantin}, Luca and {Cullen}, Fergus and {Davis}, Kelcey and {Donnan}, Callum T. and {Dunlop}, James S. and {Ellis}, Richard S. and {Ferguson}, Henry C. and {Fujimoto}, Seiji and {Fontana}, Adriano and {Giavalisco}, Mauro and {Grazian}, Andrea and {Grogin}, Norman A. and {Hathi}, Nimish P. and {Hirschmann}, Michaela and {Huertas-Company}, Marc and {Holwerda}, Benne W. and {Illingworth}, Garth and {Juneau}, St{\'e}phanie and {Kartaltepe}, Jeyhan S. and {Koekemoer}, Anton M. and {Li}, Wenxiu and {Lucas}, Ray A. and {Magee}, Dan and {Mason}, Charlotte and {McLeod}, Derek J. and {McLure}, Ross J. and {Napolitano}, Lorenzo and {Papovich}, Casey and {Pirzkal}, Nor and {Rodighiero}, Giulia and {Santini}, Paola and {Wilkins}, Stephen M. and {Yung}, L.~Y. Aaron},
        title = "{The Rise of Faint, Red Active Galactic Nuclei at z > 4: A Sample of Little Red Dots in the JWST Extragalactic Legacy Fields}",
      journal = {\apj},
         year = 2025,
        month = jun,
       volume = {986},
       number = {2},
          eid = {126},
        pages = {126},
          doi = {10.3847/1538-4357/adbc7d},
archivePrefix = {arXiv},
       eprint = {2404.03576},
 primaryClass = {astro-ph.GA},
       adsurl = {https://ui.adsabs.harvard.edu/abs/2025ApJ...986..126K}
}

@ARTICLE{Ding25,
       author = {{Ding}, Xuheng and {Onoue}, Masafusa and {Silverman}, John D. and {Matsuoka}, Yoshiki and {Izumi}, Takuma and {Strauss}, Michael A. and {Yang}, Lilan and {Jahnke}, Knud and {Phillips}, Camryn L. and {Treu}, Tommaso and {Andika}, Irham T. and {Aoki}, Kentaro and {Arita}, Junya and {Baba}, Shunsuke and {Bosman}, Sarah E.~I. and {Eilers}, Anna-Christina and {Fujimoto}, Seiji and {Haiman}, Zoltan and {Imanishi}, Masatoshi and {Inayoshi}, Kohei and {Iwasawa}, Kazushi and {Kartaltepe}, Jeyhan and {Kashikawa}, Nobunari and {Kawaguchi}, Toshihiro and {Li}, Junyao and {Lee}, Chien-Hsiu and {Lupi}, Alessandro and {Schindler}, Jan-Torge and {Schramm}, Malte and {Shimasaku}, Kazuhiro and {Shuntov}, Marko and {Tanaka}, Takumi S. and {Toba}, Yoshiki and {Trakhtenbrot}, Benny and {Umehata}, Hideki and {Vestergaard}, Marianne and {Wang}, Feige and {Yang}, Jinyi},
        title = "{SHELLQs-JWST Unveils the Host Galaxies of 12 Quasars at z > 6}",
      journal = {\apj},
         year = 2025,
        month = nov,
       volume = {993},
       number = {1},
          eid = {91},
        pages = {91},
          doi = {10.3847/1538-4357/ae045b},
archivePrefix = {arXiv},
       eprint = {2505.03876},
 primaryClass = {astro-ph.GA},
       adsurl = {https://ui.adsabs.harvard.edu/abs/2025ApJ...993...91D}
}

@ARTICLE{Naidu25,
       author = {{Naidu}, Rohan P. and {Matthee}, Jorryt and {Katz}, Harley and {de Graaff}, Anna and {Oesch}, Pascal and {Smith}, Aaron and {Greene}, Jenny E. and {Brammer}, Gabriel and {Weibel}, Andrea and {Hviding}, Raphael and {Chisholm}, John and {Labb\textbackslash'e}, Ivo and {Simcoe}, Robert A. and {Witten}, Callum and {Atek}, Hakim and {Baggen}, Josephine F.~W. and {Belli}, Sirio and {Bezanson}, Rachel and {Boogaard}, Leindert A. and {Bose}, Sownak and {Covelo-Paz}, Alba and {Dayal}, Pratika and {Fudamoto}, Yoshinobu and {Furtak}, Lukas J. and {Giovinazzo}, Emma and {Goulding}, Andy and {Gronke}, Max and {Heintz}, Kasper E. and {Hirschmann}, Michaela and {Illingworth}, Garth and {Inoue}, Akio K. and {Johnson}, Benjamin D. and {Leja}, Joel and {Leonova}, Ecaterina and {McConachie}, Ian and {Maseda}, Michael V. and {Natarajan}, Priyamvada and {Nelson}, Erica and {Setton}, David J. and {Shivaei}, Irene and {Sobral}, David and {Stefanon}, Mauro and {Tacchella}, Sandro and {Toft}, Sune and {Torralba}, Alberto and {van Dokkum}, Pieter and {van der Wel}, Arjen and {Volonteri}, Marta and {Walter}, Fabian and {Wang}, Bingjie and {Watson}, Darach},
        title = "{A ``Black Hole Star'' Reveals the Remarkable Gas-Enshrouded Hearts of the Little Red Dots}",
      journal = {arXiv e-prints},
         year = 2025,
        month = mar,
          eid = {arXiv:2503.16596},
        pages = {arXiv:2503.16596},
          doi = {10.48550/arXiv.2503.16596},
archivePrefix = {arXiv},
       eprint = {2503.16596},
 primaryClass = {astro-ph.GA},
       adsurl = {https://ui.adsabs.harvard.edu/abs/2025arXiv250316596N}
}

@ARTICLE{deGraaff25,
       author = {{de Graaff}, Anna and {Rix}, Hans-Walter and {Naidu}, Rohan P. and {Labb{\'e}}, Ivo and {Wang}, Bingjie and {Leja}, Joel and {Matthee}, Jorryt and {Katz}, Harley and {Greene}, Jenny E. and {Hviding}, Raphael E. and {Baggen}, Josephine and {Bezanson}, Rachel and {Boogaard}, Leindert A. and {Brammer}, Gabriel and {Dayal}, Pratika and {van Dokkum}, Pieter and {Goulding}, Andy D. and {Hirschmann}, Michaela and {Maseda}, Michael V. and {McConachie}, Ian and {Miller}, Tim B. and {Nelson}, Erica and {Oesch}, Pascal A. and {Setton}, David J. and {Shivaei}, Irene and {Weibel}, Andrea and {Whitaker}, Katherine E. and {Williams}, Christina C.},
        title = "{A remarkable ruby: Absorption in dense gas, rather than evolved stars, drives the extreme Balmer break of a little red dot at z = 3.5}",
      journal = {\aap},
         year = 2025,
        month = sep,
       volume = {701},
          eid = {A168},
        pages = {A168},
          doi = {10.1051/0004-6361/202554681},
archivePrefix = {arXiv},
       eprint = {2503.16600},
 primaryClass = {astro-ph.GA},
       adsurl = {https://ui.adsabs.harvard.edu/abs/2025A&A...701A.168D}
}

@ARTICLE{Greene24,
       author = {{Greene}, Jenny E. and {Labbe}, Ivo and {Goulding}, Andy D. and {Furtak}, Lukas J. and {Chemerynska}, Iryna and {Kokorev}, Vasily and {Dayal}, Pratika and {Volonteri}, Marta and {Williams}, Christina C. and {Wang}, Bingjie and {Setton}, David J. and {Burgasser}, Adam J. and {Bezanson}, Rachel and {Atek}, Hakim and {Brammer}, Gabriel and {Cutler}, Sam E. and {Feldmann}, Robert and {Fujimoto}, Seiji and {Glazebrook}, Karl and {de Graaff}, Anna and {Khullar}, Gourav and {Leja}, Joel and {Marchesini}, Danilo and {Maseda}, Michael V. and {Matthee}, Jorryt and {Miller}, Tim B. and {Naidu}, Rohan P. and {Nanayakkara}, Themiya and {Oesch}, Pascal A. and {Pan}, Richard and {Papovich}, Casey and {Price}, Sedona H. and {van Dokkum}, Pieter and {Weaver}, John R. and {Whitaker}, Katherine E. and {Zitrin}, Adi},
        title = "{UNCOVER Spectroscopy Confirms the Surprising Ubiquity of Active Galactic Nuclei in Red Sources at z > 5}",
      journal = {\apj},
         year = 2024,
        month = mar,
       volume = {964},
       number = {1},
          eid = {39},
        pages = {39},
          doi = {10.3847/1538-4357/ad1e5f},
archivePrefix = {arXiv},
       eprint = {2309.05714},
 primaryClass = {astro-ph.GA},
       adsurl = {https://ui.adsabs.harvard.edu/abs/2024ApJ...964...39G}
}

@ARTICLE{Greene26,
       author = {{Greene}, Jenny E. and {Setton}, David J. and {Furtak}, Lukas J. and {Naidu}, Rohan P. and {Volonteri}, Marta and {Dayal}, Pratika and {Labbe}, Ivo and {van Dokkum}, Pieter and {Bezanson}, Rachel and {Brammer}, Gabriel and {Cutler}, Sam E. and {Glazebrook}, Karl and {de Graaff}, Anna and {Hirschmann}, Michaela and {Hviding}, Raphael E. and {Kokorev}, Vasily and {Leja}, Joel and {Liu}, Hanpu and {Ma}, Yilun and {Matthee}, Jorryt and {Nanayakkara}, Themiya and {Oesch}, Pascal A. and {Pan}, Richard and {Price}, Sedona H. and {Spilker}, Justin S. and {Wang}, Bingjie and {Weaver}, John R. and {Whitaker}, Katherine E. and {Williams}, Christina C. and {Zitrin}, Adi},
        title = "{What You See Is What You Get: Empirically Measured Bolometric Luminosities of Little Red Dots}",
      journal = {\apj},
         year = 2026,
        month = jan,
       volume = {996},
       number = {2},
          eid = {129},
        pages = {129},
          doi = {10.3847/1538-4357/ae1836},
archivePrefix = {arXiv},
       eprint = {2509.05434},
 primaryClass = {astro-ph.GA},
       adsurl = {https://ui.adsabs.harvard.edu/abs/2026ApJ...996..129G}
}

@ARTICLE{Juodzbalis26,
       author = {{Juod{\v{z}}balis}, Ignas and {Maiolino}, Roberto and {Baker}, William M. and {Lake}, Emma Curtis and {Scholtz}, Jan and {D'Eugenio}, Francesco and {Trefoloni}, Bartolomeo and {Isobe}, Yuki and {Tacchella}, Sandro and {Bunker}, Andrew J. and {Carniani}, Stefano and {Charlot}, St{\'e}phane and {Jones}, Gareth C. and {Parlanti}, Eleonora and {Perna}, Michele and {Rinaldi}, Pierluigi and {Robertson}, Brant and {{\"U}bler}, Hannah and {Venturi}, Giacomo and {Willott}, Chris},
        title = "{JADES: comprehensive census of broad-line AGN from reionization to cosmic noon revealed by JWST}",
      journal = {\mnras},
         year = 2026,
        month = mar,
       volume = {546},
       number = {3},
          eid = {stag086},
        pages = {stag086},
          doi = {10.1093/mnras/stag086},
archivePrefix = {arXiv},
       eprint = {2504.03551},
 primaryClass = {astro-ph.GA},
       adsurl = {https://ui.adsabs.harvard.edu/abs/2026MNRAS.546ag086J}
}

@ARTICLE{Harikane23,
       author = {{Harikane}, Yuichi and {Zhang}, Yechi and {Nakajima}, Kimihiko and {Ouchi}, Masami and {Isobe}, Yuki and {Ono}, Yoshiaki and {Hatano}, Shun and {Xu}, Yi and {Umeda}, Hiroya},
        title = "{A JWST/NIRSpec First Census of Broad-line AGNs at z = 4-7: Detection of 10 Faint AGNs with M $_{BH}$ {}10$^{6}$-{}10$^{8}$ M $_{{\ensuremath{\odot}}}$ and Their Host Galaxy Properties}",
      journal = {\apj},
         year = 2023,
        month = dec,
       volume = {959},
       number = {1},
          eid = {39},
        pages = {39},
          doi = {10.3847/1538-4357/ad029e},
archivePrefix = {arXiv},
       eprint = {2303.11946},
 primaryClass = {astro-ph.GA},
       adsurl = {https://ui.adsabs.harvard.edu/abs/2023ApJ...959...39H}
}

@ARTICLE{Silverman25,
       author = {{Silverman}, John David and {Li}, Junyao and {Ding}, Xuheng and {Onoue}, Masafusa and {Strauss}, Michael A. and {Matsuoka}, Yoshiki and {Izumi}, Takuma and {Jahnke}, Knud and {Treu}, Tommaso and {Volonteri}, Marta and {Phillips}, Camryn L. and {Andika}, Irham T. and {Aoki}, Kentaro and {Arita}, Junya and {Baba}, Shunsuke and {Bosman}, Sarah E.~I. and {Eilers}, Anna-Christina and {Fan}, Xiaohui and {Fujimoto}, Seiji and {Habouzit}, Melanie and {Haiman}, Zoltan and {Imanishi}, Masatoshi and {Inayoshi}, Kohei and {Iwasawa}, Kazushi and {Kashikawa}, Nobunari and {Kawaguchi}, Toshihiro and {Lee}, Chien-Hsiu and {Lupi}, Alessandro and {Nagao}, Tohru and {Schindler}, Jan-Torge and {Schramm}, Malte and {Shimasaku}, Kazuhiro and {Toba}, Yoshiki and {Trakhtenbrot}, Benny and {Umehata}, Hideki and {Vestergaard}, Marianne and {Walter}, Fabian and {Wang}, Feige and {Yang}, Jinyi},
        title = "{SHELLQs─JWST Perspective on the Intrinsic Mass Relation between Supermassive Black Holes and Their Host Galaxies at z > 6}",
      journal = {\apjl},
         year = 2025,
        month = dec,
       volume = {995},
       number = {2},
          eid = {L67},
        pages = {L67},
          doi = {10.3847/2041-8213/ae279c},
archivePrefix = {arXiv},
       eprint = {2507.23066},
 primaryClass = {astro-ph.GA},
       adsurl = {https://ui.adsabs.harvard.edu/abs/2025ApJ...995L..67S}
}

@misc{msaexp,
       author = {{Brammer}, Gabriel},
        title = "{msaexp: NIRSpec analyis tools}",
         year = 2023,
        month = sep,
          eid = {10.5281/zenodo.8319596},
          doi = {10.5281/zenodo.8319596},
      version = {0.6.17},
    publisher = {Zenodo},
       adsurl = {https://ui.adsabs.harvard.edu/abs/2023zndo...8319596B}
}

@ARTICLE{Schulze18,
       author = {{Schulze}, Andreas and {Silverman}, John D. and {Kashino}, Daichi and {Akiyama}, Masayuki and {Schramm}, Malte and {Sanders}, Dave and {Kartaltepe}, Jeyhan and {Daddi}, Emanuele and {Rodighiero}, Giulia and {Renzini}, Alvio and {Arimoto}, Nobuo and {Nagao}, Tohru and {Puglisi}, Annagrazia and {Trakhtenbrot}, Benny and {Civano}, Francesca and {Suh}, Hyewon},
        title = "{An FMOS Survey of Moderate-luminosity, Broad-line AGNs in COSMOS, SXDS, and E-CDF-S}",
      journal = {\apjs},
         year = 2018,
        month = dec,
       volume = {239},
       number = {2},
          eid = {22},
        pages = {22},
          doi = {10.3847/1538-4365/aae82f},
archivePrefix = {arXiv},
       eprint = {1810.07445},
 primaryClass = {astro-ph.GA},
       adsurl = {https://ui.adsabs.harvard.edu/abs/2018ApJS..239...22S}
      
}

@ARTICLE{Halpern83,
       author = {{Halpern}, J.~P. and {Steiner}, J.~E.},
        title = "{Low ionization active galactic nuclei : X-ray or shock heated ?}",
      journal = {\apjl},
         year = 1983,
        month = jun,
       volume = {269},
        pages = {L37-L41},
          doi = {10.1086/184051},
       adsurl = {https://ui.adsabs.harvard.edu/abs/1983ApJ...269L..37H}
}

@BOOK{Osterbrock06,
       author = {{Osterbrock}, Donald E. and {Ferland}, Gary J.},
        title = "{Astrophysics of gaseous nebulae and active galactic nuclei}",
         year = 2006,
       adsurl = {https://ui.adsabs.harvard.edu/abs/2006agna.book.....O}
}

@ARTICLE{Onoue21,
       author = {{Onoue}, Masafusa and {Matsuoka}, Yoshiki and {Kashikawa}, Nobunari and {Strauss}, Michael A. and {Iwasawa}, Kazushi and {Izumi}, Takuma and {Nagao}, Tohru and {Asami}, Naoko and {Fujimoto}, Seiji and {Harikane}, Yuichi and {Hashimoto}, Takuya and {Imanishi}, Masatoshi and {Lee}, Chien-Hsiu and {Shibuya}, Takatoshi and {Toba}, Yoshiki},
        title = "{Subaru High-z Exploration of Low-luminosity Quasars (SHELLQs). XIV. A Candidate Type II Quasar at z = 6.1292}",
      journal = {\apj},
         year = 2021,
        month = sep,
       volume = {919},
       number = {1},
          eid = {61},
        pages = {61},
          doi = {10.3847/1538-4357/ac0f07},
archivePrefix = {arXiv},
       eprint = {2106.13807},
 primaryClass = {astro-ph.GA},
       adsurl = {https://ui.adsabs.harvard.edu/abs/2021ApJ...919...61O}
}

@ARTICLE{Eilers23,
       author = {{Eilers}, Anna-Christina and {Simcoe}, Robert A. and {Yue}, Minghao and {Mackenzie}, Ruari and {Matthee}, Jorryt and {{\v{D}}urov{\v{c}}{\'\i}kov{\'a}}, Dominika and {Kashino}, Daichi and {Bordoloi}, Rongmon and {Lilly}, Simon J.},
        title = "{EIGER. III. JWST/NIRCam Observations of the Ultraluminous High-redshift Quasar J0100+2802}",
      journal = {\apj},
         year = 2023,
        month = jun,
       volume = {950},
       number = {1},
          eid = {68},
        pages = {68},
          doi = {10.3847/1538-4357/acd776},
archivePrefix = {arXiv},
       eprint = {2211.16261},
 primaryClass = {astro-ph.GA},
       adsurl = {https://ui.adsabs.harvard.edu/abs/2023ApJ...950...68E}
}

@ARTICLE{Pei_SMC,
       author = {{Pei}, Yichuan C.},
        title = "{Interstellar Dust from the Milky Way to the Magellanic Clouds}",
      journal = {\apj},
         year = 1992,
        month = aug,
       volume = {395},
        pages = {130},
          doi = {10.1086/171637},
       adsurl = {https://ui.adsabs.harvard.edu/abs/1992ApJ...395..130P}
}

@ARTICLE{YanZ25,
       author = {{Yan}, Zu and {Inayoshi}, Kohei and {Chen}, Kejian and {Guo}, Jingsong},
        title = "{Balmer Transition Signatures from Gas-Enshrouded, Dust-Poor Active Galactic Nuclei}",
      journal = {arXiv e-prints},
         year = 2025,
        month = dec,
          eid = {arXiv:2512.11050},
        pages = {arXiv:2512.11050},
          doi = {10.48550/arXiv.2512.11050},
archivePrefix = {arXiv},
       eprint = {2512.11050},
 primaryClass = {astro-ph.GA},
       adsurl = {https://ui.adsabs.harvard.edu/abs/2025arXiv251211050Y}
}

@ARTICLE{Inayoshi_Maiolino25,
       author = {{Inayoshi}, Kohei and {Maiolino}, Roberto},
        title = "{Extremely Dense Gas around Little Red Dots and High-redshift Active Galactic Nuclei: A Nonstellar Origin of the Balmer Break and Absorption Features}",
      journal = {\apjl},
         year = 2025,
        month = feb,
       volume = {980},
       number = {2},
          eid = {L27},
        pages = {L27},
          doi = {10.3847/2041-8213/adaebd},
archivePrefix = {arXiv},
       eprint = {2409.07805},
 primaryClass = {astro-ph.GA},
       adsurl = {https://ui.adsabs.harvard.edu/abs/2025ApJ...980L..27I}
}

@ARTICLE{DeGraaff25b,
       author = {{de Graaff}, Anna and {Hviding}, Raphael E. and {Naidu}, Rohan P. and {Greene}, Jenny E. and {Miller}, Tim B. and {Leja}, Joel and {Matthee}, Jorryt and {Brammer}, Gabriel and {Katz}, Harley and {Bezanson}, Rachel and {Boogaard}, Leindert A. and {Bose}, Sownak and {Chisholm}, John and {Cleri}, Nikko J. and {Dayal}, Pratika and {Feldmann}, Robert and {Fudamoto}, Yoshinobu and {Fujimoto}, Seiji and {Furtak}, Lukas J. and {Glazebrook}, Karl and {Gottumukkala}, Rashmi and {Heintz}, Kasper E. and {Kokorev}, Vasily and {Labbe}, Ivo and {Maseda}, Michael V. and {McConachie}, Ian and {Nanayakkara}, Themiya and {Nelson}, Erica and {Nowaczyk}, Przemys{\l}aw and {Oesch}, Pascal A. and {Rix}, Hans-Walter and {Setton}, David J. and {Torralba}, Alberto and {Walter}, Fabian and {Wang}, Bingjie and {Weibel}, Andrea and {van der Wel}, Arjen},
        title = "{Little Red Dots host Black Hole Stars: A unified family of gas-reddened AGN revealed by JWST/NIRSpec spectroscopy}",
      journal = {\mnras},
         year = 2026,
        month = aug,
          doi = {10.1093/mnras/stag1567},
archivePrefix = {arXiv},
       eprint = {2511.21820},
 primaryClass = {astro-ph.GA},
       adsurl = {https://ui.adsabs.harvard.edu/abs/2026MNRAS.tmp.1477D}
}

@ARTICLE{Umeda25,
       author = {{Umeda}, Hiroya and {Inayoshi}, Kohei and {Harikane}, Yuichi and {Murase}, Kohta},
        title = "{A Black-Hole Envelope Interpretation for Cosmological Demographics of Little Red Dots}",
      journal = {arXiv e-prints},
         year = 2025,
        month = dec,
          eid = {arXiv:2512.04208},
        pages = {arXiv:2512.04208},
          doi = {10.48550/arXiv.2512.04208},
archivePrefix = {arXiv},
       eprint = {2512.04208},
 primaryClass = {astro-ph.GA},
       adsurl = {https://ui.adsabs.harvard.edu/abs/2025arXiv251204208U}
}

@ARTICLE{LiW23,
       author = {{Li}, Wenxiu and {Inayoshi}, Kohei and {Onoue}, Masafusa and {Toyouchi}, Daisuke},
        title = "{The Assembly of Black Hole Mass and Luminosity Functions of High-redshift Quasars via Multiple Accretion Episodes}",
      journal = {\apj},
         year = 2023,
        month = jun,
       volume = {950},
       number = {2},
          eid = {85},
        pages = {85},
          doi = {10.3847/1538-4357/accbbe},
archivePrefix = {arXiv},
       eprint = {2210.02308},
 primaryClass = {astro-ph.GA},
       adsurl = {https://ui.adsabs.harvard.edu/abs/2023ApJ...950...85L}
}

@ARTICLE{Matsuoka25b,
       author = {{Matsuoka}, Yoshiki and {Iwasawa}, Kazushi and {Onoue}, Masafusa and {Izumi}, Takuma and {Strauss}, Michael A. and {Akiyama}, Masayuki and {Aoki}, Kentaro and {Arita}, Junya and {Ding}, Xuheng and {Imanishi}, Masatoshi and {Kashikawa}, Nobunari and {Kawaguchi}, Toshihiro and {Kikuta}, Satoshi and {Kohno}, Kotaro and {Lee}, Chien-Hsiu and {Nagao}, Tohru and {Phillips}, Camryn L. and {Sawamura}, Mahoshi and {Silverman}, John D. and {Takahashi}, Ayumi and {Toba}, Yoshiki},
        title = "{Subaru High-z Exploration of Low-luminosity Quasars (SHELLQs). XXIV. 54 New Quasars and Candidate Obscured Quasars at 5.71 {\ensuremath{\leq}} z {\ensuremath{\leq}} 7.02}",
      journal = {\apjs},
         year = 2025,
        month = oct,
       volume = {280},
       number = {2},
          eid = {68},
        pages = {68},
          doi = {10.3847/1538-4365/ae0035},
archivePrefix = {arXiv},
       eprint = {2508.21229},
 primaryClass = {astro-ph.GA},
       adsurl = {https://ui.adsabs.harvard.edu/abs/2025ApJS..280...68M}
}

@ARTICLE{SDSS_DR14Q,
       author = {{P{\^a}ris}, Isabelle and {Petitjean}, Patrick and {Aubourg}, {\'E}ric and {Myers}, Adam D. and {Streblyanska}, Alina and {Lyke}, Brad W. and {Anderson}, Scott F. and {Armengaud}, {\'E}ric and {Bautista}, Julian and {Blanton}, Michael R. and {Blomqvist}, Michael and {Brinkmann}, Jonathan and {Brownstein}, Joel R. and {Brandt}, William Nielsen and {Burtin}, {\'E}tienne and {Dawson}, Kyle and {de la Torre}, Sylvain and {Georgakakis}, Antonis and {Gil-Mar{\'\i}n}, H{\'e}ctor and {Green}, Paul J. and {Hall}, Patrick B. and {Kneib}, Jean-Paul and {LaMassa}, Stephanie M. and {Le Goff}, Jean-Marc and {MacLeod}, Chelsea and {Mariappan}, Vivek and {McGreer}, Ian D. and {Merloni}, Andrea and {Noterdaeme}, Pasquier and {Palanque-Delabrouille}, Nathalie and {Percival}, Will J. and {Ross}, Ashley J. and {Rossi}, Graziano and {Schneider}, Donald P. and {Seo}, Hee-Jong and {Tojeiro}, Rita and {Weaver}, Benjamin A. and {Weijmans}, Anne-Marie and {Y{\`e}che}, Christophe and {Zarrouk}, Pauline and {Zhao}, Gong-Bo},
        title = "{The Sloan Digital Sky Survey Quasar Catalog: Fourteenth data release}",
      journal = {\aap},
         year = 2018,
        month = may,
       volume = {613},
          eid = {A51},
        pages = {A51},
          doi = {10.1051/0004-6361/201732445},
archivePrefix = {arXiv},
       eprint = {1712.05029},
 primaryClass = {astro-ph.GA},
       adsurl = {https://ui.adsabs.harvard.edu/abs/2018A&A...613A..51P}
}

@ARTICLE{LSST_2019,
       author = {{Ivezi{\'c}}, {\v{Z}}eljko and {Kahn}, Steven M. and {Tyson}, J. Anthony and {Abel}, Bob and {Acosta}, Emily and {Allsman}, Robyn and {Alonso}, David and {AlSayyad}, Yusra and {Anderson}, Scott F. and {Andrew}, John and {Angel}, James Roger P. and {Angeli}, George Z. and {Ansari}, Reza and {Antilogus}, Pierre and {Araujo}, Constanza and {Armstrong}, Robert and {Arndt}, Kirk T. and {Astier}, Pierre and {Aubourg}, {\'E}ric and {Auza}, Nicole and {Axelrod}, Tim S. and {Bard}, Deborah J. and {Barr}, Jeff D. and {Barrau}, Aurelian and {Bartlett}, James G. and {Bauer}, Amanda E. and {Bauman}, Brian J. and {Baumont}, Sylvain and {Bechtol}, Ellen and {Bechtol}, Keith and {Becker}, Andrew C. and {Becla}, Jacek and {Beldica}, Cristina and {Bellavia}, Steve and {Bianco}, Federica B. and {Biswas}, Rahul and {Blanc}, Guillaume and {Blazek}, Jonathan and {Blandford}, Roger D. and {Bloom}, Josh S. and {Bogart}, Joanne and {Bond}, Tim W. and {Booth}, Michael T. and {Borgland}, Anders W. and {Borne}, Kirk and {Bosch}, James F. and {Boutigny}, Dominique and {Brackett}, Craig A. and {Bradshaw}, Andrew and {Brandt}, William Nielsen and {Brown}, Michael E. and {Bullock}, James S. and {Burchat}, Patricia and {Burke}, David L. and {Cagnoli}, Gianpietro and {Calabrese}, Daniel and {Callahan}, Shawn and {Callen}, Alice L. and {Carlin}, Jeffrey L. and {Carlson}, Erin L. and {Chandrasekharan}, Srinivasan and {Charles-Emerson}, Glenaver and {Chesley}, Steve and {Cheu}, Elliott C. and {Chiang}, Hsin-Fang and {Chiang}, James and {Chirino}, Carol and {Chow}, Derek and {Ciardi}, David R. and {Claver}, Charles F. and {Cohen-Tanugi}, Johann and {Cockrum}, Joseph J. and {Coles}, Rebecca and {Connolly}, Andrew J. and {Cook}, Kem H. and {Cooray}, Asantha and {Covey}, Kevin R. and {Cribbs}, Chris and {Cui}, Wei and {Cutri}, Roc and {Daly}, Philip N. and {Daniel}, Scott F. and {Daruich}, Felipe and {Daubard}, Guillaume and {Daues}, Greg and {Dawson}, William and {Delgado}, Francisco and {Dellapenna}, Alfred and {de Peyster}, Robert and {de Val-Borro}, Miguel and {Digel}, Seth W. and {Doherty}, Peter and {Dubois}, Richard and {Dubois-Felsmann}, Gregory P. and {Durech}, Josef and {Economou}, Frossie and {Eifler}, Tim and {Eracleous}, Michael and {Emmons}, Benjamin L. and {Fausti Neto}, Angelo and {Ferguson}, Henry and {Figueroa}, Enrique and {Fisher-Levine}, Merlin and {Focke}, Warren and {Foss}, Michael D. and {Frank}, James and {Freemon}, Michael D. and {Gangler}, Emmanuel and {Gawiser}, Eric and {Geary}, John C. and {Gee}, Perry and {Geha}, Marla and {Gessner}, Charles J.~B. and {Gibson}, Robert R. and {Gilmore}, D. Kirk and {Glanzman}, Thomas and {Glick}, William and {Goldina}, Tatiana and {Goldstein}, Daniel A. and {Goodenow}, Iain and {Graham}, Melissa L. and {Gressler}, William J. and {Gris}, Philippe and {Guy}, Leanne P. and {Guyonnet}, Augustin and {Haller}, Gunther and {Harris}, Ron and {Hascall}, Patrick A. and {Haupt}, Justine and {Hernandez}, Fabio and {Herrmann}, Sven and {Hileman}, Edward and {Hoblitt}, Joshua and {Hodgson}, John A. and {Hogan}, Craig and {Howard}, James D. and {Huang}, Dajun and {Huffer}, Michael E. and {Ingraham}, Patrick and {Innes}, Walter R. and {Jacoby}, Suzanne H. and {Jain}, Bhuvnesh and {Jammes}, Fabrice and {Jee}, M. James and {Jenness}, Tim and {Jernigan}, Garrett and {Jevremovi{\'c}}, Darko and {Johns}, Kenneth and {Johnson}, Anthony S. and {Johnson}, Margaret W.~G. and {Jones}, R. Lynne and {Juramy-Gilles}, Claire and {Juri{\'c}}, Mario and {Kalirai}, Jason S. and {Kallivayalil}, Nitya J. and {Kalmbach}, Bryce and {Kantor}, Jeffrey P. and {Karst}, Pierre and {Kasliwal}, Mansi M. and {Kelly}, Heather and {Kessler}, Richard and {Kinnison}, Veronica and {Kirkby}, David and {Knox}, Lloyd and {Kotov}, Ivan V. and {Krabbendam}, Victor L. and {Krughoff}, K. Simon and {Kub{\'a}nek}, Petr and {Kuczewski}, John and {Kulkarni}, Shri and {Ku}, John and {Kurita}, Nadine R. and {Lage}, Craig S. and {Lambert}, Ron and {Lange}, Travis and {Langton}, J. Brian and {Le Guillou}, Laurent and {Levine}, Deborah and {Liang}, Ming and {Lim}, Kian-Tat and {Lintott}, Chris J. and {Long}, Kevin E. and {Lopez}, Margaux and {Lotz}, Paul J. and {Lupton}, Robert H. and {Lust}, Nate B. and {MacArthur}, Lauren A. and {Mahabal}, Ashish and {Mandelbaum}, Rachel and {Markiewicz}, Thomas W. and {Marsh}, Darren S. and {Marshall}, Philip J. and {Marshall}, Stuart and {May}, Morgan and {McKercher}, Robert and {McQueen}, Michelle and {Meyers}, Joshua and {Migliore}, Myriam and {Miller}, Michelle and {Mills}, David J.},
        title = "{LSST: From Science Drivers to Reference Design and Anticipated Data Products}",
      journal = {\apj},
         year = 2019,
        month = mar,
       volume = {873},
       number = {2},
          eid = {111},
        pages = {111},
          doi = {10.3847/1538-4357/ab042c},
archivePrefix = {arXiv},
       eprint = {0805.2366},
 primaryClass = {astro-ph},
       adsurl = {https://ui.adsabs.harvard.edu/abs/2019ApJ...873..111I}
}

@ARTICLE{Son25,
       author = {{Son}, Suyeon and {Kim}, Minjin and {Ho}, Luis C. and {Li}, Ruancun},
        title = "{Probing the Physical Origin of the Balmer Decrement in the Broad-line Region of Nearby Active Galactic Nuclei via Spectral Variability}",
      journal = {\apj},
         year = 2025,
        month = dec,
       volume = {995},
       number = {1},
          eid = {37},
        pages = {37},
          doi = {10.3847/1538-4357/ae1ef1},
archivePrefix = {arXiv},
       eprint = {2511.07714},
 primaryClass = {astro-ph.GA},
       adsurl = {https://ui.adsabs.harvard.edu/abs/2025ApJ...995...37S}
}

@ARTICLE{LaMura07,
       author = {{La Mura}, G. and {Popovi{\'c}}, L. {\v{C}}. and {Ciroi}, S. and {Rafanelli}, P. and {Ili{\'c}}, D.},
        title = "{Detailed Analysis of Balmer Lines in a Sloan Digital Sky Survey Sample of 90 Broad-Line Active Galactic Nuclei}",
      journal = {\apj},
         year = 2007,
        month = dec,
       volume = {671},
       number = {1},
        pages = {104-117},
          doi = {10.1086/522821},
       adsurl = {https://ui.adsabs.harvard.edu/abs/2007ApJ...671..104L}
}

@ARTICLE{Shen24,
       author = {{Shen}, Yue and {Grier}, Catherine J. and {Horne}, Keith and {Stone}, Zachary and {Li}, Jennifer I. and {Yang}, Qian and {Homayouni}, Yasaman and {Trump}, Jonathan R. and {Anderson}, Scott F. and {Brandt}, W.~N. and {Hall}, Patrick B. and {Ho}, Luis C. and {Jiang}, Linhua and {Petitjean}, Patrick and {Schneider}, Donald P. and {Tao}, Charling and {Donnan}, Fergus. R. and {AlSayyad}, Yusra and {Bershady}, Matthew A. and {Blanton}, Michael R. and {Bizyaev}, Dmitry and {Bundy}, Kevin and {Chen}, Yuguang and {Davis}, Megan C. and {Dawson}, Kyle and {Fan}, Xiaohui and {Greene}, Jenny E. and {Gr{\"o}ller}, Hannes and {Guo}, Yucheng and {Ibarra-Medel}, H{\'e}ctor and {Jiang}, Yuanzhe and {Keenan}, Ryan P. and {Kollmeier}, Juna A. and {Lejoly}, Cassandra and {Li}, Zefeng and {de la Macorra}, Axel and {Moe}, Maxwell and {Nie}, Jundan and {Rossi}, Graziano and {Smith}, Paul S. and {Tee}, Wei Leong and {Weijmans}, Anne-Marie and {Xu}, Jiachuan and {Yue}, Minghao and {Zhou}, Xu and {Zhou}, Zhimin and {Zou}, Hu},
        title = "{The Sloan Digital Sky Survey Reverberation Mapping Project: Key Results}",
      journal = {\apjs},
         year = 2024,
        month = jun,
       volume = {272},
       number = {2},
          eid = {26},
        pages = {26},
          doi = {10.3847/1538-4365/ad3936},
archivePrefix = {arXiv},
       eprint = {2305.01014},
 primaryClass = {astro-ph.GA},
       adsurl = {https://ui.adsabs.harvard.edu/abs/2024ApJS..272...26S}
}

@ARTICLE{Markov25,
       author = {{Markov}, Vladan and {Gallerani}, Simona and {Ferrara}, Andrea and {Pallottini}, Andrea and {Parlanti}, Eleonora and {Mascia}, Fabio Di and {Sommovigo}, Laura and {Kohandel}, Mahsa},
        title = "{The evolution of dust attenuation in z {\ensuremath{\approx}} 2-12 galaxies observed by JWST}",
      journal = {Nature Astronomy},
         year = 2025,
        month = mar,
       volume = {9},
        pages = {458-468},
          doi = {10.1038/s41550-024-02426-1},
archivePrefix = {arXiv},
       eprint = {2402.05996},
 primaryClass = {astro-ph.GA},
       adsurl = {https://ui.adsabs.harvard.edu/abs/2025NatAs...9..458M}
}

@ARTICLE{Phillips25,
       author = {{Phillips}, Camryn L. and {Strauss}, Michael A. and {Onoue}, Masafusa and {Ding}, Xuheng and {Silverman}, John D. and {Matsuoka}, Yoshiki and {Izumi}, Takuma and {Arita}, Junya and {Aoki}, Kentaro and {Baba}, Shunsuke and {Imanishi}, Masatoshi and {Kashikawa}, Nobunari and {Kawaguchi}, Toshihiro and {Lee}, Chien-Hsiu and {Sawamura}, Mahoshi and {Toba}, Yoshiki and {Wang}, Feige and {Yang}, Jinyi},
        title = "{SHELLQs-JWST: Revealing the Spectra of Extended Emission in 12 z > 6 Quasar Host Galaxies Using the JWST NIRSpec Fixed Slit}",
      journal = {\apj},
         year = 2026,
        month = may,
       volume = {1003},
       number = {1},
          eid = {47},
        pages = {47},
          doi = {10.3847/1538-4357/ae606d},
archivePrefix = {arXiv},
       eprint = {2510.22403},
 primaryClass = {astro-ph.GA},
       adsurl = {https://ui.adsabs.harvard.edu/abs/2026ApJ..1003...47P}
}

@ARTICLE{Jakobsen22,
       author = {{Jakobsen}, P. and {Ferruit}, P. and {Alves de Oliveira}, C. and {Arribas}, S. and {Bagnasco}, G. and {Barho}, R. and {Beck}, T.~L. and {Birkmann}, S. and {B{\"o}ker}, T. and {Bunker}, A.~J. and {Charlot}, S. and {de Jong}, P. and {de Marchi}, G. and {Ehrenwinkler}, R. and {Falcolini}, M. and {Fels}, R. and {Franx}, M. and {Franz}, D. and {Funke}, M. and {Giardino}, G. and {Gnata}, X. and {Holota}, W. and {Honnen}, K. and {Jensen}, P.~L. and {Jentsch}, M. and {Johnson}, T. and {Jollet}, D. and {Karl}, H. and {Kling}, G. and {K{\"o}hler}, J. and {Kolm}, M.-G. and {Kumari}, N. and {Lander}, M.~E. and {Lemke}, R. and {L{\'o}pez-Caniego}, M. and {L{\"u}tzgendorf}, N. and {Maiolino}, R. and {Manjavacas}, E. and {Marston}, A. and {Maschmann}, M. and {Maurer}, R. and {Messerschmidt}, B. and {Moseley}, S.~H. and {Mosner}, P. and {Mott}, D.~B. and {Muzerolle}, J. and {Pirzkal}, N. and {Pittet}, J.-F. and {Plitzke}, A. and {Posselt}, W. and {Rapp}, B. and {Rauscher}, B.~J. and {Rawle}, T. and {Rix}, H.-W. and {R{\"o}del}, A. and {Rumler}, P. and {Sabbi}, E. and {Salvignol}, J.-C. and {Schmid}, T. and {Sirianni}, M. and {Smith}, C. and {Strada}, P. and {te Plate}, M. and {Valenti}, J. and {Wettemann}, T. and {Wiehe}, T. and {Wiesmayer}, M. and {Willott}, C.~J. and {Wright}, R. and {Zeidler}, P. and {Zincke}, C.},
        title = "{The Near-Infrared Spectrograph (NIRSpec) on the James Webb Space Telescope. I. Overview of the instrument and its capabilities}",
      journal = {\aap},
         year = 2022,
        month = may,
       volume = {661},
          eid = {A80},
        pages = {A80},
          doi = {10.1051/0004-6361/202142663},
archivePrefix = {arXiv},
       eprint = {2202.03305},
 primaryClass = {astro-ph.IM},
       adsurl = {https://ui.adsabs.harvard.edu/abs/2022A&A...661A..80J}
}

@ARTICLE{Schulze10,
       author = {{Schulze}, A. and {Wisotzki}, L.},
        title = "{Low redshift AGN in the Hamburg/ESO Survey . II. The active black hole mass function and the distribution function of Eddington ratios}",
      journal = {\aap},
         year = 2010,
        month = jun,
       volume = {516},
          eid = {A87},
        pages = {A87},
          doi = {10.1051/0004-6361/201014193},
archivePrefix = {arXiv},
       eprint = {1004.2671},
 primaryClass = {astro-ph.CO},
       adsurl = {https://ui.adsabs.harvard.edu/abs/2010A&A...516A..87S}
}

\end{document}